\documentclass[12pt,a4paper,fleqn]{article}

\usepackage[textheight=23cm,textwidth=15cm]{geometry}
\usepackage{amsmath}
\usepackage{graphicx}
\usepackage{url}
\usepackage[american]{babel}
\usepackage{here}
\usepackage{overcite}
\usepackage{algpseudocode}
\usepackage{algorithm}
\usepackage{booktabs}
\usepackage{amsfonts}
\usepackage{tikz}

\usepackage{listings}
\begin{document}

\begin{center}

{\LARGE\bf \textsc{Molassembler}:\\[1ex] Molecular graph construction, modification and
  conformer generation for inorganic and organic molecules }

\vspace{1cm}

{\large Jan-Grimo Sobez\footnote{ORCID:\@0000--0002--3264--0622} and Markus
  Reiher\footnote{Corresponding author; e-mail:
  markus.reiher@phys.chem.ethz.ch; ORCID:\@0000--0002--9508--1565} }\\[4ex]

Laboratory of Physical Chemistry, ETH Zurich, \\
Vladimir-Prelog-Weg 2, 8093 Zurich, Switzerland

June 26, 2020

\vspace{.43cm}

\textbf{Abstract}
\end{center}
\vspace*{-.41cm}
{\small

We present the graph-based molecule software \textsc{Molassembler} for
building organic and inorganic molecules. \textsc{Molassembler} provides algorithms
for the construction of molecules built from any set of elements from the
periodic table. In particular, poly-nuclear transition metal complexes and
clusters can be considered. Structural information is encoded as a graph.
Stereocenter configurations are interpretable from Cartesian coordinates into an
abstract index of permutation for an extensible set of polyhedral shapes.
Substituents are distinguished through a ranking algorithm. Graph and
stereocenter representations are freely modifiable and chiral state is
propagated where possible through incurred ranking changes. Conformers are
generated with full stereoisomer control by four spatial dimension Distance
Geometry with a refinement error function including dihedral terms. Molecules
are comparable by an extended graph isomorphism and their representation is
canonicalizeable. \textsc{Molassembler} is written in C++ and provides Python
bindings.

}

\newpage
\section{Introduction}\label{sec:introduction}

Software encoding of molecules has become indispensable to many research fields.
Significant effort has been invested into the formal description of organic molecules, 
owing to their importance as small-molecule drugs amenable to a characterization by chemical concepts.
A trove of open-source cheminformatics programs exists serving various needs\cite{Pirhadi2016, Ambure2016}. Many of these 
computer programs have some support
for inorganic molecules, for which reliable and simple chemical principles are more difficult to define owing
to an increased electronic structure. For example, take
fragment-based~\cite{Foscato2014} and evolutionary~\cite{Chu2012} organometallic
compound design applications, whose primary conformer generation is facilitated by
Marvin~\cite{Chemaxon534} and Chemistry Development
Kit~\cite{Steinbeck2003}, respectively.

The set of software whose primary focus is an encoding of organometallic and
inorganic molecules is small, but steadily expanding. The high-level complex
generator molSimplify~\cite{ioannidis2016} is based on three-dimensional
manipulation of preoptimized molecular fragments. AARON~\cite{Guan2018}
automates the generation of configurations and conformations of organometallic
molecules in transition-metal catalytic explorations.
HostDesigner~\cite{Hay2002} selects ligands to complement metal ion guests.
DENOPTIM~\cite{Foscato2019} is a fragment-based transition metal compound
design tool.
The fast quantum mechanical metadynamics conformer ensemble generator
CREST~\cite{Pracht2020} and a related force-field method
GFN-FF~\cite{Spicher2020} for molecular mechanics are further recent additions.

For our efforts toward the automated exploration of chemical reaction
networks based on first-principles heuristics~\cite{Bergeler2015, Simm2017, Unsleber2019}, we
require structure-identification and -construction algorithms that are applicable to molecules composed of any element
from the periodic table (in particular, to inorganic compounds) in order not to restrict explorations
to irrelevant parts of chemical reaction space if certain reactants cannot be represented.
No existing cheminformatics implementation provides the capabilities required
for truly general mechanism exploration algorithms that identify nodes in a reaction network
as chemical species described by a graph. To be more specific, the requirements for this purpose
are: (i) The algorithm must be
able to interpret a local minimum on the Born-Oppenheimer potential energy
hypersurface as a chemical graph and capture the stereochemical configuration.
(ii) The molecular model must encompass multidentate and haptic ligands.
(iii) It must be possible to change the connectivity of atoms
and enumerate all stereoisomers without restriction. (iv) Molecules must be
comparable. (v) It must be possible to generate new conformers of molecules.
This set of features does not exist in any software supporting inorganic molecules, because 
bonding patterns and stereochemistry in inorganic molecules present significant challenges for
cheminformatics tools due to their immense variability.

This variability affects how atom-centered chirality manifests. Compared to
organic chemistry, the most complicated chiral unit will no longer be the tetrahedron
with at most two distinct spatial arrangements (i.e., R or S) if its substituents are all
deemed different by a ranking algorithm. 
Instead, the local shape can have as many as twelve vertices and most vertex counts accommodate multiple shapes.
For instance, five sites can arrange into idealized shapes of a
trigonal bipyramid, a square pyramid or a pentagon. The number of distinct
spatial arrangements for the maximally asymmetric substitution situation scales
approximately as the factorial of the number of sites. Note also that individual shape vertices
are not necessarily occupied by single atoms, but possibly by multiple atoms in a
haptic configuration.

The degree to which these ostensibly slight structural changes complicate the
design of algorithms necessarily depends on the specific aims of the 
molecule construction software. Let us consider some of the consequences of the proposed aims
individually: In order to interpret a local minimum of the potential energy
hypersurface as a chemical graph and capture the stereochemical configuration, 
the software must base its atom-connectivity representation on a graph and
find a data representation of stereocenters that is general, and in particular, suitable for inorganic molecules.
Returning to atom-centered chirality, this implies that for an arbitrary local
shape, the software must enumerate rotationally non-superimposable arrangements
of variable numbers of different substituents. This enumeration must consider
the effects of site interconnectivity for multidentate ligands and abstract away 
the additional complication of haptic binding. The algorithm must decide which
substituents are chemically different with a ranking algorithm that encompasses
stereodescriptors of inorganic stereocenters. 

Local shapes and especially non-superimposable arrangements within those shapes must be reliably
identifiable from Cartesian coordinates. An imposed rigid model of chiral
arrangements might also consider interconverting molecular dynamics such as
nitrogen inversion to avoid overstating chiral character. Given a
molecule's connectivity, deciding on the local shape at each non-terminal atom 
involves more complex electronic structures. For a reasonable degree of
accuracy, it will in general be insufficient to base this decision on simple concepts such as
Gillespie's valence shell electron
pair repulsion. If the molecular graph and the stereocenter
representations are to be freely modifiable and if they are identified
through relative substituent ranking, stereodescriptors must be propagated
through the ranking changes that arbitrary edits can incur. For editing continuity,
stereocenters should not generally lose their state upon addition or removal of
a site while transitioning between different molecular structures (shapes). 

Moreover, the comparison of molecules will require the
implementation of a graph isomorphism and optionally a graph canonicalization
algorithm to accelerate repeated comparisons. Generation of conformations of
arbitrary graphs requires a spatial model that can encompass multidentate and
haptic ligands and a methodology that reliably generates specific stereocenter
configurations in Cartesian coordinates. 

Note that model-free molecule comparison algorithms such as
G-RMSD~\cite{Tomonori2020}, and continuous representations of
molecules~\cite{Gomez2018} are certainly also viable approaches to some of the
challenges we seek to solve.
However, we do not consider direct comparisons of
Cartesian coordinates through measures such as root mean square deviations an option
here, because increasing deviations may be expected for increasing molecule sizes of
structures that actually resemble the same compound.

In this work, we present a new molecular construction software, \textsc{Molassembler}, that seeks to
encompass organometallic and inorganic bonding patterns in a more complex
graph-based model. 
Comparably few of these sub-problems of our non-exhaustive list 
require entirely new solutions. Many can be solved by implementing existing
algorithms or through careful extension of existing partial solutions, as we
will discuss in the following. For instance, we will show that through partial
abstraction of the International Union of Pure and Applied Chemistry (IUPAC)
organic ranking sequence rules~\cite{favre2013nomenclature}, a configurational
index of permutation can serve as a comparative stereodescriptor in a ranking
algorithm. Moreover, through extension, the original algorithm can serve to
differentiate not merely direct graph substituents, but binding sites. By
contrast, the canonicalization of stereodescriptors and their propagation into
new structures (shapes) or through ranking changes will require new solutions.

\section{Molecule model}\label{sec:molecule_model}

\textsc{Molassembler} applies a molecule model split into an undirected graph
and a list of data structures named stereopermutators. Vertices of the graph
represent atoms, each storing an element type. Edges of the graph represent
bonds, storing an idealized bond order (i.e., an integer number that may be extracted from
a real-valued bond order obtained from a quantum mechanical analysis of the electronic wave
function). Idealized bond types comprise bond
orders from single to sextuple and an algorithm-internal special bond called
'eta' indicating connectivity to an atom forming part of a haptic binding site
following the $\eta$-notation~\cite{mcnaught1997compendium} for such bonding
situations. The molecular graph must form a single connected component.
Accidental graph disconnects upon atom or bond removal are prevented by permitting
only the removal of non-bridge bonds or non-articulation vertices (removal of articulation
vertices would enlarge the number of connected components). Intentional
molecule splitting along bridge bonds is possible through specialized functions.

Stereopermutators manage relative spatial orientations of groups of bonded
atoms. They reduce the set of relative spatial
orientations to an abstract permutational space. Special relative spatial
orientations are reduced to an index of permutation. Collectively,
stereopermutators capture a molecule's configurational space. 
Stereopermutators are not named stereocenters because they also manage non-stereogenic cases.
\textsc{Molassembler} comprises two types of stereopermutators: Atom-centered
stereopermutators capture configurational differences between different
orientations of the direct graph adjacents of a central atom. Bond-centered
stereopermutators capture configurational differences arising due to rotational
barriers along bonds.

\begin{figure}[h]
  \centering
  \includegraphics[width=0.5\linewidth]{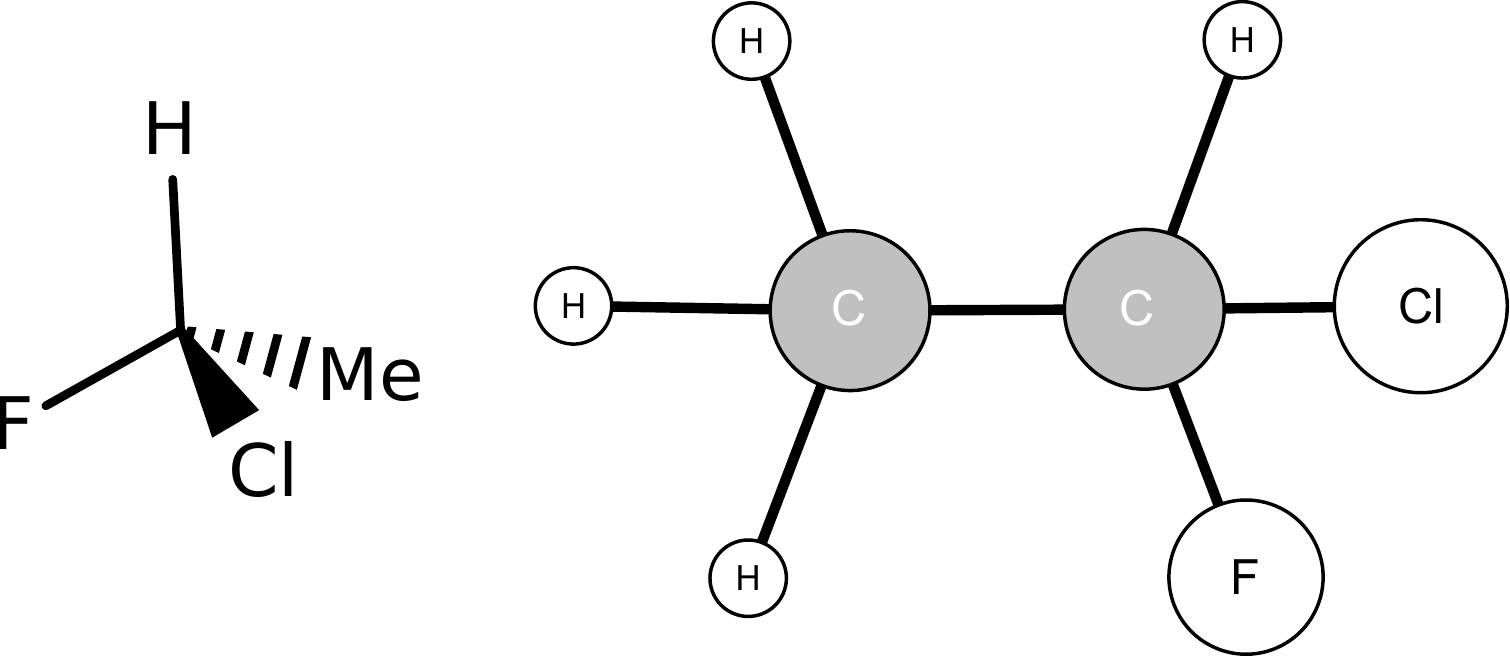}
  \caption{Illustration of spatial structure and molecule graph. \textit{Left:} A typical Lewis structure encoding also stereoinformation.
  \textit{Right:}~The connectivity of this molecule represented as a graph.
}\label{fig:model_graph} \end{figure}

Consider the example given in Figure~\ref{fig:model_graph}. The
graph of the Lewis structure captures its connectivity, but not its spatial
shape or chiral character. In addition to the graph, this requires two atom-centered
stereopermutators. One is placed at the methyl group's carbon atom, indicating a
tetrahedral local shape and an abstract binding situation \texttt{AAAB} (three
ranking-identical hydrogen atoms and one different group), for which case there
are no multiple non-superimposable arrangements. That stereopermutator is
therefore not stereogenic. The other stereopermutator is placed at the carbon
atom central in the Lewis structure. It also has a tetrahedral local shape, but
ranking-wise all its substituents are different in an \texttt{ABCD} abstract
binding situation. In that shape, there are two non-superimposable spatial
permutations of its substituents. The Lewis structure represents the S variation
of this stereocenter, and the stereopermutator represents it as an index of
permutation that parallels the IUPAC stereodescriptor.

Our molecular representation model is capable of transferring molecular structures at local
minima on the Born-Oppenheimer potential energy surface given in Cartesian coordinates to an abstract
permutational space in which configurational variations can be enumerated easily
and transformed back into Cartesian coordinates via conformer generation close
to local minima on the potential energy surface. However, the model
is not well suited for capturing structural elements of other regions of the
potential energy surface.

\textsc{Molassembler} places no structural limitations on the graph aside from
constraining it to remain a single connected component. As a consequence, it is
possible to generate graphs for which \textsc{Molassembler} cannot generate a
spatial model allowing the graph to be embedded into three dimensions, such as
the set of complete graphs (i.e., those in which all vertices are connected with
all other vertices) with more than four vertices. Deciding graph
embeddability in three dimensions is NP-hard~\cite{Thomassen1989}, and no
attempt is made to guarantee that a supplied graph is embeddable during its
construction. 

In the following sections,
we discuss the individual domains of responsibility for atom- and
bond-centered stereopermutators, the implemented ranking algorithm, manipulation
of molecules and conformer generation.

\section{Local shapes and stereopermutations}\label{sec:stereopermutations}

Atom-centered stereopermutators manage the spatial orientation of the graph
substituents of a non-terminal atom by abstracting into a permutational space in
which each permutation constitutes a mutually non-superimposable spatial
orientation. If multiple permutations are geometrically feasible, the respective
atom is a stereocenter.

\begin{figure}[h]
  \includegraphics[width=\linewidth]{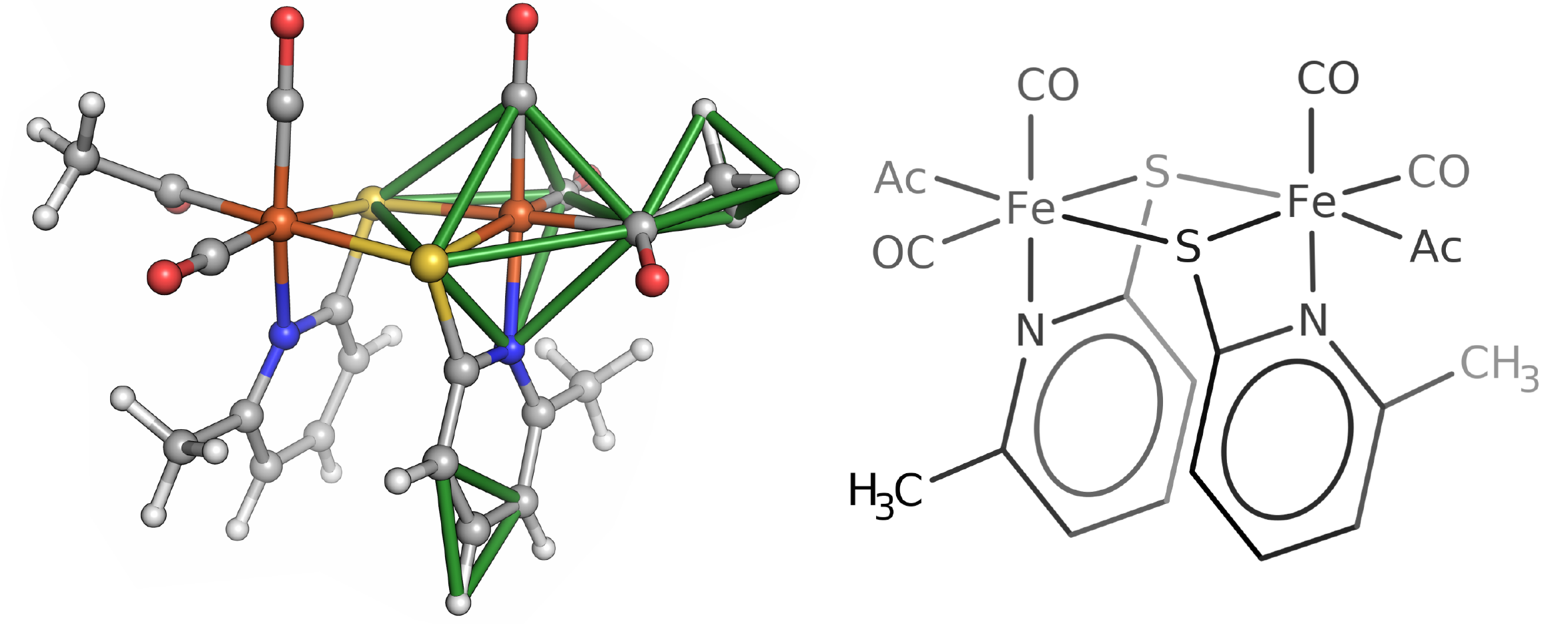}
  \caption{\textit{Left:} An example of a molecule with classified shapes shown in green for three of
  its non-terminal atoms: An equilateral triangle at the bottom, an octahedron
  in the center, and a tetrahedron at the right edge. \textit{Right:}~Simplified
  Lewis structure of the left compound. Images of molecules in Cartesian
  coordinates were generated with PyMOL~\cite{PyMOL2015}, simplified Lewis
  structures with
  MarvinSketch~\cite{MarvinSketch1926}.
  Atom coloring: Hydrogen in white, carbon in gray, oxygen in red, 
iron in orange, sulfur in yellow, and nitrogen in blue.}\label{fig:shape_classification}
\end{figure}

The local spatial arrangement of an atom's substituents is classified into a set
of idealized polyhedral shapes. \textsc{Molassembler} currently comprises thirty
different shapes ranging from two to twelve vertices. The choice of shapes aims
to reflect common geometric patterns while avoiding too fine distinctions. All
chosen shapes are spherical, i.e.\ within their coordinate frames, all vertices
of the encoded shapes have the same distance from the origin. The first class of
chosen shapes is of reduced dimensionality and contains the following shapes:
Line, bent line, equilateral triangle, T-shape, regular square, pentagon, and
hexagon~\cite{Garcon2019}. Next, four of the five Platonic solids are included: tetrahedron, octahedron,
cube, and icosahedron. Then, some other polyhedra composed of regular faces:
Uniform triangular prism, square antiprism, and cuboctahedron. Several Johnson
solids are included without modification as they are by construction
spherical: Square pyramid, pentagonal pyramid, trigonal bipyramid, and pentagonal
bipyramid. Furthermore, the hexagonal and heptagonal bipyramids are also
included. Several non-spherical Johnson solid shapes have been spherized by
projection onto a sphere and subsequent minimization on the Thomson
potential~\cite{Thomson1904} to a local minimum that retains the original point
group symmetry: Capped square antiprism, trigonal dodecahedron, and capped
trigonal prism. The shapes of the global minima for nine, ten and eleven
particles on the Thomson potential are also included: Tricapped trigonal prism,
bicapped square antiprism and edge-contracted icosahedron. Lastly, some derived
shapes are also present: Capped octahedron, a tetrahedron with one vertex vacant
and both apical- and equatorial-monovacant variations of the trigonal bipyramid.
The spatial arrangement of a non-terminal atom's immediate substituents can be
classified as one of these shapes, as demonstrated
in Figure~\ref{fig:shape_classification}.

\begin{figure}[h]
  \includegraphics[width=\linewidth]{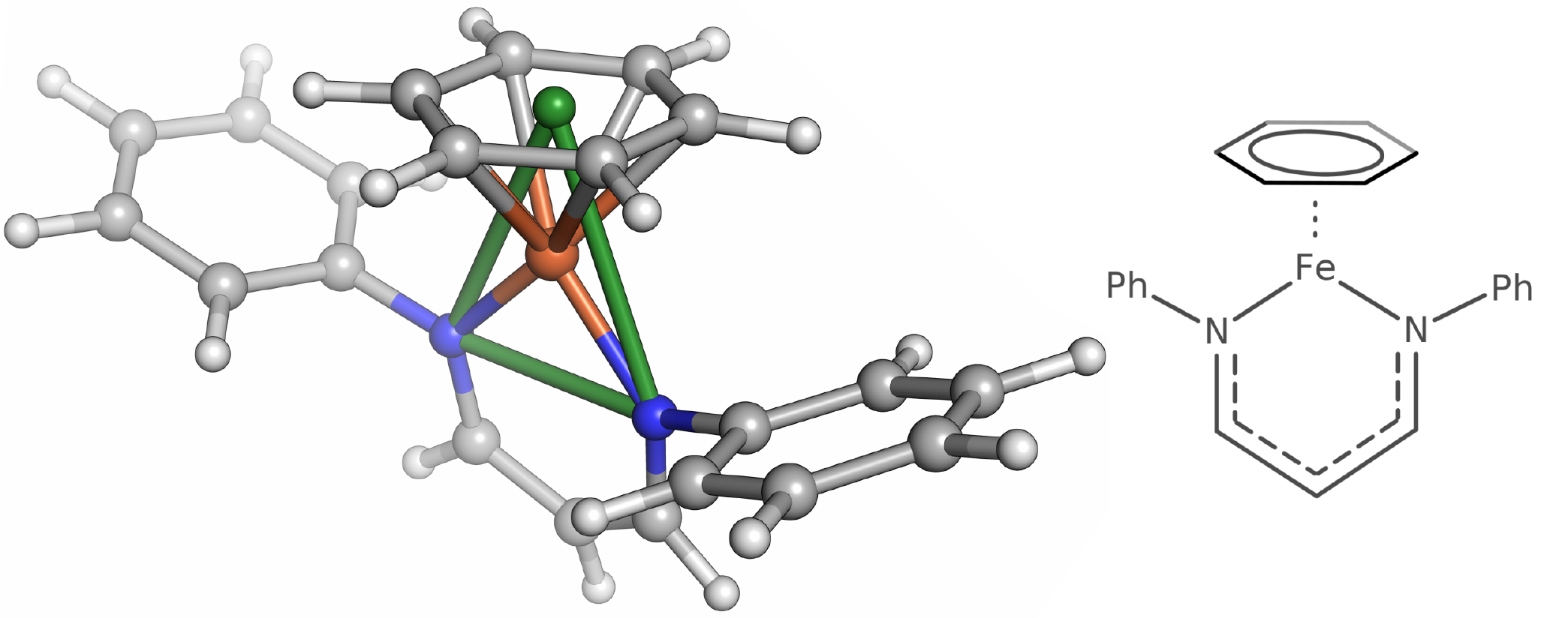}
  \caption{\textit{Left:} An example of a molecule with a haptic binding site.
  The central iron atom has two binding sites consisting of single nitrogen
  atoms. The carbon atoms of the benzene molecule above the iron atom form a
  third binding site, as they are a contiguous group of atoms bonded to the
  central iron atom. The centroid of the haptic binding site is shown as a
  pseudoatom in the center of the carbocycle. The shape of the
  substituents of the central iron atom has three vertices, one for the centroid
  of each binding site, and is classified as an equilateral triangle, shown in green.
  \textit{Right:}~Simplified Lewis structure of the left
  compound.
Atom coloring: Hydrogen in white, carbon in gray, nitrogen in 
blue, and iron in orange.}\label{fig:haptic_shape_classification}
\end{figure}

A binding site is a contiguous set of vertices adjacent to a central atom.
If a site comprises more than a single atom, the binding site is
haptic and the bonds of its edges (collected and idealized, for instance, from bond orders evaluated from
an electronic wave function) to the central atom vertex are relabeled
as eta bonds. Spatially, a binding site's vertex-set centroid is modeled to
co-locate with a vertex of the local shape as shown in
Figure~\ref{fig:haptic_shape_classification}. Haptically binding atoms are free
to rotate around the axis defined by the local shape's central atom and the
binding site's vertex-set centroid.

Binding sites are ranked by an algorithm described in
Section~\ref{sec:ranking_algorithm}. Ranking is necessary to decide which
binding sites are chemically different. Permuting chemically identical groups
does not create a new non-superimposable configuration, which must be considered
in the enumeration of stereopermutations. Links between binding sites, and by
proxy multidentate ligands, are identified with cycle perception from the
external software library RingDecomposerLib~\cite{Flachsenberg2017}.

From an idealized shape, a set of binding sites, their ranking, and their links,
\textsc{Molassembler} enumerates a set of abstract non-superimposable
permutations following a method reported in the literature~\cite{Bennett1969}. In preparation,
the ranked sites and their links are transformed into an abstract case such as
\texttt{AA(B-C)} that merely reflects whether sites are different from one
another and their connectivity. This is done by transforming the nested list of
ranked sites into a flat list where each site is represented by an incrementing
character. Within the considered shape, all permutations of
vertex occupations are generated and the rotationally non-superimposable
permutations are collected. These are the stereopermutations of the abstract case
within the considered shape.

Generating a mapping from sites to shape vertices for a particular
stereopermutation will be nontrivial if links (paths between site constituting atoms not including the central atom) 
exist and there are multiple sites
with equal ranking. In any situation, it is possible to represent the ranked
sites and their links as a colored graph: 
Vertices are sites and links are edges between sites. 
Similarly, a stereopermutation can be represented as a graph:
Vertices are shape vertices and edges are their links. Any isomorphism mapping
between the two graphs is a valid mapping from sites to shape vertices.

\begin{figure}[h]
  \centering
  \includegraphics[width=0.5\linewidth]{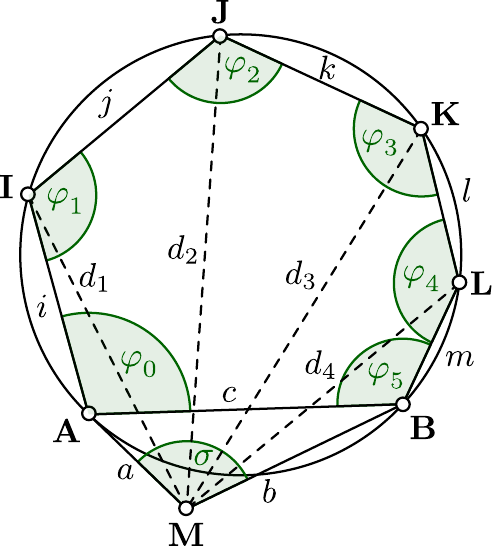}
  \caption{Geometric construction for determining the viability of an arbitrary
  bidentate cycle. The spatial distances between atoms are modeled on the basis
  of their element type and mutual bond orders. The idealized angle between
  immediate adjacents of the central atom \textbf{M} is known in the given
  stereopermutation. The resulting distance between atoms \textbf{A} and
  \textbf{B} is the closing edge to the polygon defined by the cycle path
  between \textbf{A} and \textbf{B} excluding the central atom (labeled $i$
  through $m$). This uniquely determines the cyclic polygon. From these
  quantities, the upper bound on the distances $d_i$ between members of the
  cycle and the central atom can be calculated. If, for any of the cycle
  members, the upper bound on the distance to the central atom is lower than a
  modeled distance in case they were bonded, then the stereopermutation will be considered
unfeasible since any spatial realization will contradict the graph.
}\label{fig:infeasibility_denticity} \end{figure}

Within stereopermutation enumeration, three-dimensional feasibility
of links between binding sites in each permutation is not considered. 
For instance, a bidentate oxalate ligand cannot coordinate in a trans-arrangement within an octahedron shape, 
but must be cis-coordinated: stereopermutations in which its sites are trans-arranged are not viable in three dimensions.
Such stereopermutations arise in the abstract
permutational scheme, and must be removed to avoid false classification of an
atom as a stereocenter and unexpected failures in conformer generation. We
consider it undesirable to exclude trans-ligating arrangements on principle for
specific shapes and prefer to make general geometric arguments on a case-to-case
basis. Unfeasible stereopermutations are avoided by
probing whether a cyclic polygon can be modeled for each links' cycle without
non-binding atoms of the bridge entering a bonding distance to the central atom.
This geometric algorithm is depicted in Figure~\ref{fig:infeasibility_denticity}.
The cyclic polygon was chosen because it maximizes the area of the polygon and
by proxy the distance of its vertices from the original center. This scheme will
identify individual cycles that, when modeled in a particular stereopermutation,
would contradict the graph, but miss instances where multiple individually
feasible cycles are impossible to realize jointly. For instance, in an
octahedral shape with three bidentate ligands with bridges of sufficient length
to achieve trans-ligation individually, this scheme will not remove the
stereopermutation in which all are ligated in a trans-arrangement. If the
bridges are too short for each to distort sufficiently to accommodate the
others, conformer generation will fail. It may be possible to discover a
lack of viability by modeling all bridges in a joint distance bounds matrix and
applying triangle and tetrangle inequality smoothing, but this has not yet turned out to be
fruitful. For haptic ligands, an additional feasibility criterion is applied:
\textsc{Molassembler} estimates
the conical space the site occupies and removes stereopermutations in which the
conical spaces of multiple haptic binding sites overlap.

A stereopermutator with multiple feasible stereopermutations may be left unspecified:
Although a set of feasible stereopermutations might exist, stereopermutators can be
set as none of them. In other terms, their assigned stereopermutation is
nullable. When unspecified, stereopermutators represent all of their possible
stereopermutations. For example, a conformational ensemble generated from a
compound with a single asymmetric, tetrahedral, and unspecified stereopermutator
represents a racemic mixture of both configurations.

In summary, \textsc{Molassembler} executes the following steps to identify
stereopermutations from coordinates: First, it ranks the central atom's graph
adjacents and groups the substituents into sites. Second, it combines
substituent ranking and binding site groups into a site ranking. Third, it
classifies the local shape (see next section). Fourth, \textsc{Molassembler} enumerates abstract
permutations and removes permutations deemed unfeasible. Finally, it identifies
the stereopermutation present by finding the realized stereopermutation within
the set of permutations deemed feasible.

\section{Shape information from Cartesian coordinates}\label{sec:local_shapes}

To extract which particular stereopermutation is present at a non-terminal atom
in a molecular structure, local shapes must first be classified starting from Cartesian
coordinates. This is achieved by calculating the continuous shape
measures~\cite{Pinsky1998} of all shapes with a number of vertices matching the
number of binding sites and choosing the shape for which the continuous shape
measure is minimal. The calculation of continuous shape measures is principally
of factorial complexity since the point-pairwise mapping minimizing the measure
is unknown. A faithful implementation of the suggested algorithm is
prohibitively expensive for shapes with many vertices, particularly because the
shape centroid must also be considered. The problem is not without exploitable
structure, however: Although a faithful implementation prescribes minimizing
both relative orientation and scaling over all point pairings, minimization over
scaling can be relegated to after the minimal pairing regarding relative
orientation is found, hence reducing the complexity of evaluating a single pairing. If
the shape features rotational symmetries, these can be exploited to avoid redundant
pairings. This divides the theoretical complexity by the shape's number of
superimposable rotations. A larger reduction in complexity is reached by the
application of the following heuristic: The spatial rotation is considered
converged with only a reduced, fixed number of point pairings. After performing
a rotational minimization of square distances by a quaternion fit with limited
pairings, the cost of adding individual pairings to the current set is
calculated for all remaining pairs and the minimal variation is chosen without
recalculating the spatial rotation with intermediate choices. This heuristic can
fail if multiple points lie within a small spherical area, implying the rotation
matrix is not converged by few point pairs. This failure mode is both detectable
and unlikely to occur in molecular structures. The resulting complexity in terms
of quaternion fits scales as $\mathcal{O}(N! / (N-D)!)$ where $N$ is the number of vertices
of the shape including the centroid and $D$ is the number of point pairings for
which the rotation is considered converged. \textsc{Molassembler} applies $D =
5$ and pre-sets the centroid point pair known from stereopermutator shape
fitting. We made further algorithmic attempts at finding the minimal pairing permutation
at reduced cost with simulated annealing, stochastic
tunneling~\cite{Wenzel1999Stochastic}, thermodynamic simulated
annealing~\cite{andresen1994constant}, fixed-cost greedy, and locally optimal
minimizations with reshuffling, all without bearing fruit against the heuristic
described above with respect to correctness and speed.

Initially, shapes were recognized through minimizing the sum of absolute
angular deviations between binding site centroids of the coordinates from the
idealized geometry. Although it is difficult to objectively evaluate the
agreement between shape classification algorithms and human perception, it can
be plainly stated that shape classification through angular deviations alone
quickly breaks subjective tests. Geometry indices as defined for limited sets of
shapes with four~\cite{yang2007structural, Okuniewski2015} and five~\cite{addison1984synthesis}  
vertices can improve angular deviation classification, but
are limited in applicability: \textsc{Molassembler} contains more shapes with four
vertices than the geometry index can classify. For example, the trigonal pyramid, here
denoting an axially monovacant trigonal bipyramid, not a monovacant tetrahedron,
is not considered in the value range definition of the geometry index $\tau_4$. Hence, 
the geometry index is only
applied to exclude that shape for which its value indicates the largest
distance. Overall, both the pure angular deviation and its hybrid with geometry
indices are fast and effective, but do not match visual intuition for strongly
distorted structures.

Another shape classification algorithm that we considered was the continuous symmetry
measure~\cite{Zabdrodsky1993}, which measures the degree to which a set of
points has a particular point group symmetry. Unfortunately, this is untenable
as a matching point group symmetry does not demonstrate similarity to a
particular polyhedral shape. If, for instance, a viable shape in classification
were the equatorially monovacant trigonal bipyramid (also known as disphenoid or
seesaw) of C\textsubscript{2v} point group symmetry, a collinear arrangement of
all points would minimize the continuous symmetry measure. Another thought
experiment demonstrates a further weakness: Imagine one wants to classify a shape of
four vertices, and the set of possible shapes includes the square and the
equatorially monovacant trigonal bipyramid, which have D\textsubscript{4h} and
C\textsubscript{2v} point group symmetries, respectively. The C\textsubscript{2v}
point group is a subgroup of the D\textsubscript{4h} point group. A regular square
has continuous symmetry measures $S(\textrm{D\textsubscript{4h}}) = 0$ and
$S(\textrm{C\textsubscript{2v}}) = 0$, highlighting the ambiguity of polyhedral shape matching to
point group symmetry.

In the context of continuous shape measures, the concept of minimum distortion
paths were introduced between distinct shapes~\cite{Alvarez2005}. Another,
in hindsight fruitless, shape classification algorithm studied was based on calculating the distance
of the shape to be classified from the paths separating all pairs of viable
shapes. That pair for which the distance from the path was minimal was chosen and
then the smaller shape measure of that shape pair was the classified shape.
However, this is a poor shape classification algorithm, indicated by a strong disconnect
between its classification choices and visual intuition beginning at slight
distortions.

In an attempt to quantify and visualize the behavior of experimental shape
classification algorithms, we compared their results on a uniform random
distortion scale. In this attempt, the base shape to be re-identified was distorted
by addition of fixed-length distortions of uniform random direction to each
vertex including the centroid and the resulting point cloud reclassified as a
shape. For a sense of scale of distortion vector norms relative to the shape
size, recall that \textsc{Molassembler} encompasses only spherical shapes whose
vertices are at unit distance from the coordinate origin, where the centroid is
placed. From a reductionistic perspective disregarding the distinct relationships
between particular shapes, an algorithm that reclassifies shapes correctly for
larger distortion vector norms could be considered better. We shall argue with
this metric by distinguishing algorithms with restricted or extensive
\textit{distortion tolerance}. 

\begin{figure}[h]
  \centering
  \includegraphics[width=\linewidth]{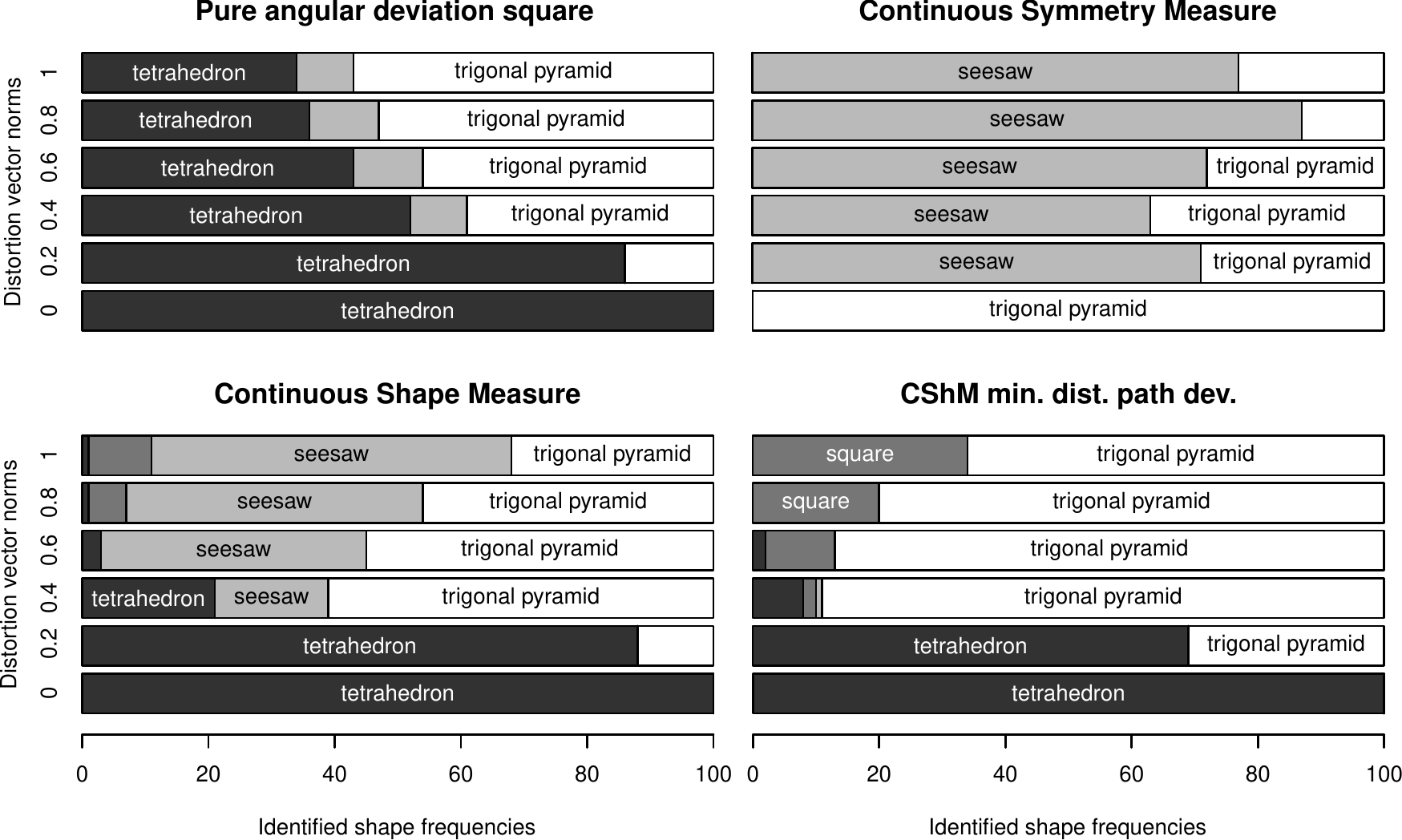}
  \caption{Shape classifications of four algorithms with varying distortion
  vector norms applied to all vertices of a regular tetrahedron. From left to
  right, top to bottom: Minimal sum of angular deviation square, minimal
  continuous symmetry measure, minimal continuous shape measure, and continuous
  shape measure (CShM) minimum distortion path deviation
  algorithms. 'Frequencies' denotes the occurrence of a shape in the set of one-hundred distorted structures generated.}\label{fig:shape_classification_tetrahedron}
\end{figure}

A comparison of shape classification algorithms at the example 
of the tetrahedron is visualized in Figure~\ref{fig:shape_classification_tetrahedron} 
(cf. also the subsequent two figures discussed below). For each algorithm, the 
idealized shape in a unit sphere is uniformly distorted one-hundred 
times by applying vectors of uniform random direction at each vertex 
point. Then, the one-hundred distorted point clouds are reclassified as shapes. This 
is repeated for increasing distortion vector norms.

In Figure~\ref{fig:shape_classification_tetrahedron}, the pure
angular deviation algorithm shows extensive distortion tolerance, but the continuous symmetry
measure based algorithm misclassifies an undistorted tetrahedron into the
C\textsubscript{3v} symmetry of the trigonal pyramid and classifies all
distorted tetrahedra as closer to C\textsubscript{2v} or C\textsubscript{3v}
point groups than the T\textsubscript{d} point group. The continuous shape
measure has a much reduced distortion tolerance, but classifies structures into more diverse
shapes at large distortions than the angular deviation algorithm. The minimum
distortion path deviation continuous shape measure algorithm has even further reduced
distortion tolerance than the pure continuous shape measure algorithm and over-represents the
trigonal pyramid at the expense of the seesaw, which is classified only once. In
this example, the discussed shortcomings of algorithms based on the continuous
symmetry measure or minimum distortion path deviation continuous shape measure
become clear.

\begin{figure}[h]
  \centering
  \includegraphics[width=\linewidth]{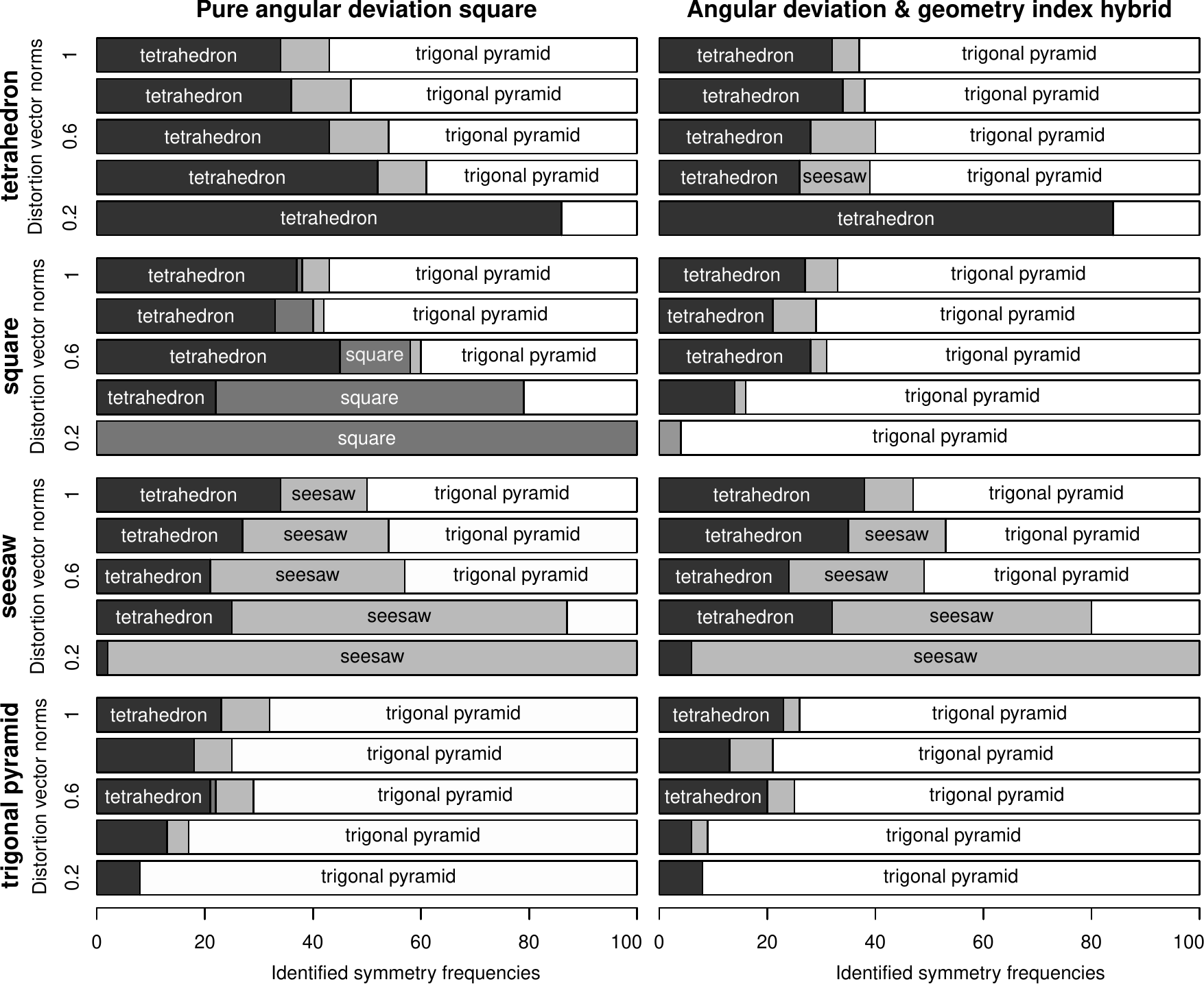}
  \caption{Shapes identified by sum of angular deviation squares algorithm (left
  column) and its hybrid algorithm with geometry indices (right column) for
  four-vertex shapes with varying distortion vector norms applied to all
  vertices. Classification of the base undistorted shape is not shown as both
  algorithms correctly identify
  it. 'Frequencies' denotes the occurrence of a shape in the set of one-hundred distorted structures generated.}\label{fig:shape_classification_size_four_first}
\end{figure}

We now compare some shape classification algorithms at the example of
multiple shapes. For all shapes with four vertices encompassed in
\textsc{Molassembler}, shape classification with angular distortion and its
hybrid with geometry indices are compared by the aforementioned metric in
Figure~\ref{fig:shape_classification_size_four_first}. It is noticeable that the
trigonal pyramid (more specifically an axially monovacant trigonal bipyramid),
despite being a rare occurrence when compared to the tetrahedron or the square,
is overrepresented compared to a random shape distribution. The addition of
geometry indices as an exclusion criterion exacerbates this, presumably because
the trigonal pyramid is never excluded by geometry index, whereas all other
shapes are. Generally, both variations of the angular distortion algorithm have
good distortion tolerance. Merely the square shape dissipates quickly in the hybrid algorithm.
Note that this is not necessarily bad as we have introduced distortion tolerance as a
measure that does not consider the relationships between shapes, but
solely as a means to compare algorithms.

\begin{figure}[h]
  \centering
  \includegraphics[width=\linewidth]{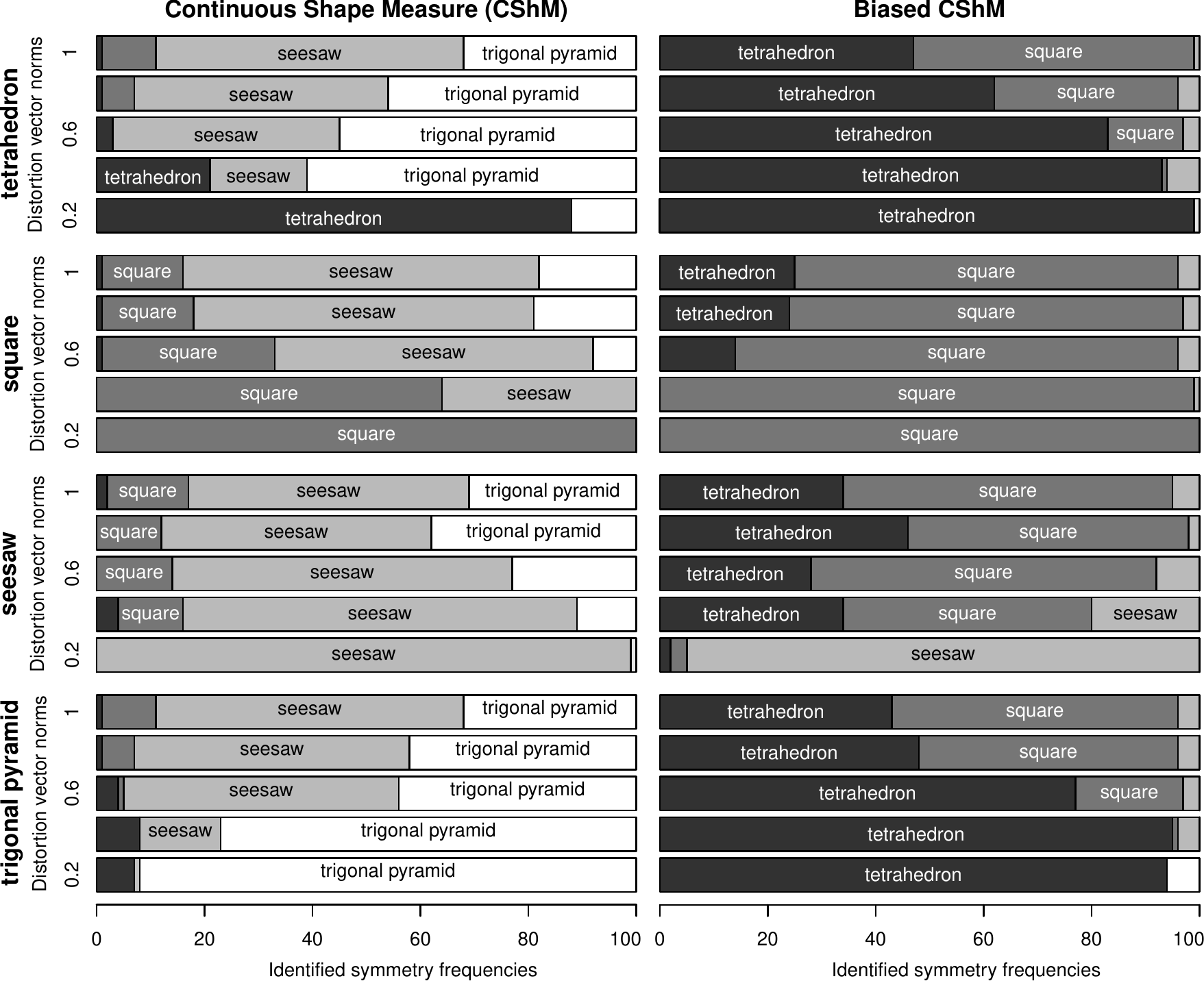}
  \caption{Shapes identified by the continuous shape measure algorithm (left
  column) and a biased variant algorithm (right column) for four-vertex shapes
  with varying distortion vector norms applied to all vertices. Classification
  of the base undistorted shape is not shown as both algorithms correctly
  identify it. 'Frequencies' denotes the occurrence of a shape in the set of one-hundred distorted structures generated.}\label{fig:shape_classification_size_four_second}
\end{figure}

Lastly, consider the continuous shape measure based algorithm and a biased
variation in Figure~\ref{fig:shape_classification_size_four_second}. For all shapes
except the tetrahedron, the unbiased algorithm exhibits good distortion tolerance
characteristics, and the set of classified shapes at large distortions is
consistently diverse when compared to the angular distortion algorithm. The
seesaw and trigonal pyramid shapes, which are equatorially or axially monovacant
trigonal bipyramids, are comparatively rare structures relative to the
tetrahedron or square. The biased variation reweights continuous shape measures
of seesaw and trigonal bipyramids to reduce their classification probability. As
a result, the algorithm exhibits extensive distortion tolerance for the tetrahedron and square
shapes and reduced distortion tolerance for the seesaw and trigonal pyramid shapes. This biased
algorithm is the shape classification algorithm chosen for \textsc{Molassembler}.

Returning to the identification of a stereopermutation from Cartesian coordinates:
By calculating the continuous shape measure in order to classify the shape, one
also obtains a minimal square distance pairwise mapping between shape vertices
and binding site centroids. This allows for direct mapping into the realized
stereopermutation, which can then be sought in a generated set of feasible
stereopermutations.

\section{Atom-centered stereopermutators}\label{sec:atom_stereocenters}

Instead of IUPAC stereodescriptors such as \textit{R} and \textit{S} for the two
stereopermutations of an asymmetric tetrahedron, \textsc{Molassembler} exposes
two indices of permutation that serve as transferable and comparable
stereodescriptors under different conditions. The first is the index of
permutation within the abstract permutational space disregarding feasibility.
This integer stereodescriptor is transferable among equivalent abstract inputs:
If the number of different binding sites (as determined by ranking), their
linking, and the local shape match between cases, the index of permutation will be
comparable. The second is the index of an abstract stereopermutation within an
ordered list of stereopermutations deemed feasible, which will be comparable only if
all previous conditions hold and all linking cycles are composed of matching
element types and bond orders. Note that the value range of the second
stereodescriptor can change upon changes to the algorithm determining
feasibility.

The idealization of orientations into local shapes enables the enumeration of
stereopermutations in an abstract space. However, two issues arise: Most
importantly, distorted geometries present challenges for shape classification
and conformer generation. It can be difficult to recognize the correct shape in
strongly distorted structures. Special care must be taken not to misinterpret
distortions owing to, e.g., small cycles in terms of a false shape, and
tolerances on internal coordinates must be expanded in spatial modeling if
distortions are necessary to form a conformation. This problem is discussed
further in Section~\ref{sec:conformer_generation}. Second, as the list of
local shapes is bounded, the particular local shape one might require to
adequately describe a molecular configuration may not be in the available set.
However, the list of available shapes is designed to be extensible with little effort to
alleviate such problems.

The applied model of spatial orientations around a central atom cannot
intrinsically distinguish in which combinations of atoms and shapes
stereopermutations can interconvert with low energy barriers (consider, for example, inversion
at a nitrogen atom). In essence, the current model treats molecular configuration
as rigid, a situation considered to be realized classically at zero Kelvin. 
Though optional, a thawing protocol is
invoked by default in which a stereopermutator managing a nitrogen atom in a monovacant
tetrahedral local shape with three different binding sites will be a stereocenter 
only if it is a member of a cycle of size three or four~\cite{Rauk1970}. Without
this protocol, any monovacant tetrahedral nitrogen atom with three
different binding sites is a stereocenter. If none of the ligands are linked in
a trigonal or pentagonal bipyramid, axial and equatorial ligands interchange at
room temperature through Berry pseudorotation~\cite{Berry1960} and the Bartell
mechanism~\cite{adams1970structure}, respectively. \textsc{Molassembler}
will therefore 'thermalize' stereopermutations in such shapes in the thawing
approximation if none of the binding sites are linked, reducing such shapes'
chiral character.

In spite of an abstract model of orientation, molecules are modifiable in close
analogy to three-dimensional editing. If an atom and its immediate surroundings
constitute a stereocenter and its stereopermutation is known, then under
specific circumstances, adding or removing a binding site will not cause loss of
information. In two situations, chiral state loss is the correct result. First,
if the resulting set of binding sites post-edit cannot constitute a stereocenter
in the target shape. Second, binding site addition can be ambiguous; from a
square shape to a square pyramid shape, the new apical position can be either
above or below the plane of existing shape vertices. By default, the algorithm
opts to discard chiral state on ambiguous transitions or on significant total
angular or chiral deviations of shape vertices. Alternatively, one can choose to
retain chiral state if the transition is unique, or to let \textsc{Molassembler}
choose at random from the best transitions. The set of best transitions between
shapes of adjacent sizes is partially computed at compile-time with certain compilers
in order to avoid runtime costs. Transitions are calculated instead of
hard-coded to keep the list of local shapes easily extensible.

Atom-centered stereopermutator propagation encompasses substituent count changes, site
count changes, ranking order changes, and shape changes, any combination of which
may occur simultaneously. To limit implementation complexity, propagation is
limited to shape size, site, and substituent count changes of at most one. Under
these circumstances, a mapping between sites can be generated on the basis of
constituting substituent set similarity quantified by an edit distance metric in
analogy to string edit distance metrics like Levenshtein
distance~\cite{Levenshtein1966}. If the shape is changed, a mapping of shape
vertices between shapes is chosen in accordance with propagation settings.
Propagation then proceeds in four steps: First, site indices are placed at the
old shape vertices with the stereopermutator's set stereopermutation. Second,
if a shape vertex mapping exists, the old site indices will be transferred into the
new shape. Third, old site indices are mapped to new site indices with the site
index mapping. From the new site indices placed at the shape vertices of the new
shape, a stereopermutation is generated that can then be sought in the set of
new stereopermutations. If the stereopermutation cannot be propagated and the
stereopermutator in its new shape and ranking is a stereocenter, it will be left
undetermined. The propagation practiced here has distinct parallels to the
transformations in reaction rules of Andersen et al.\cite{Andersen2017},
particularly with respect to shape vertex propagation.

\section{Bond-centered stereopermutators}\label{sec:bond_stereocenters}

\begin{figure}[h]
  \centering
  \includegraphics[width=0.6\linewidth]{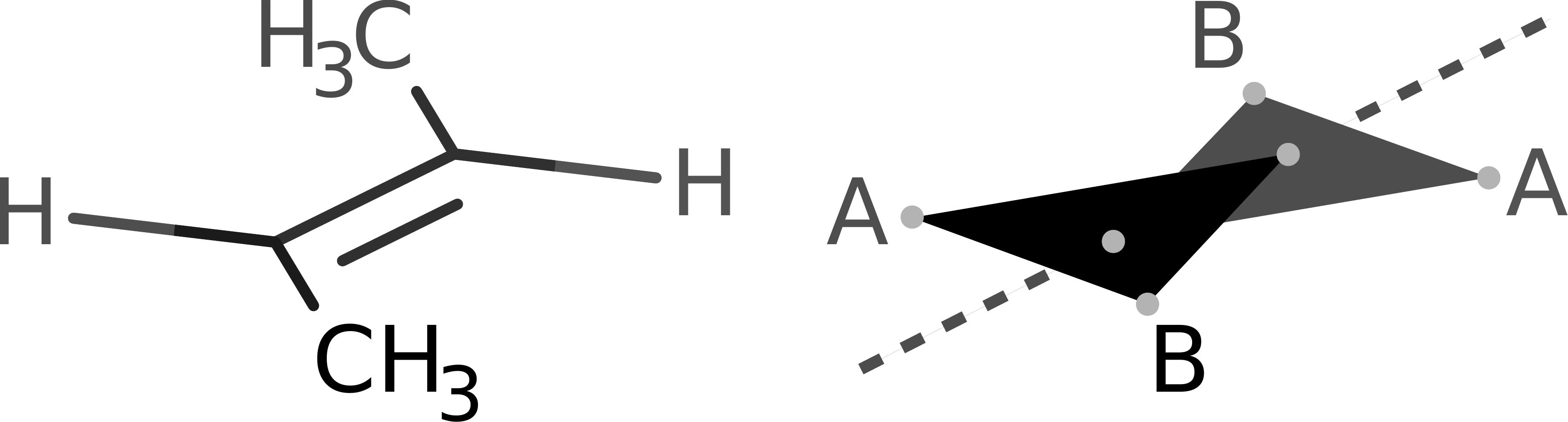}
  \caption{\textit{Left:} Lewis structure of (E)-but-2-ene. \textit{Right:}~Abstract classification of the double-bond 
   motif of (E)-but-2-ene into
  two triangle shapes and the substituents' ranking cases. Light gray circles
  denote shape vertex loci. 
  }\label{fig:double_bond_perspective}
\end{figure}

In contrast to atom-centered stereopermutators that capture non-superimposable
orientations of the immediate graph adjacents of a single atom, bond-centered
stereopermutators capture non-superimposable relative orientations of a bond's
substituents that arise due to hindered rotation around the bond axis; consider, e.g., 
substituents bonded to the atoms of a non-rotatable double bond (see Figure \ref{fig:double_bond_perspective}), where for 
each atom there exists also an atom-centered stereopermutator
characterizing the local shape. Bond-centered stereopermutators are generalized to
arbitrary combinations of shapes instead of being specialized toward the common
combinations among and between equilateral triangles and bent shapes that
constitute most rotational barriers along double bonds. 

Stereodescriptors derived from a procedure enumerating all possible rotational
orientations must be comparable and transferable among equivalent situations.
Consequently, inputs to the stereopermutation enumeration algorithm are standardized
and abstracted. 
Each side of the bond is reduced to its local shape, the fused
shape vertex within that shape (i.e., the vertex that is part of the bond), 
and the ranking characters of all non-fused vertices, as shown in Fig.~\ref{fig:double_bond_perspective}.

The stereopermutators at each end of the bond carry the information 
required for the abstraction: The ranking of their binding sites permits 
the abstraction to ranking characters, and the specified 
stereopermutation 
places ranking characters at shape vertices.

There exist degrees of freedom in the data representation of the input to 
the stereopermutation enumeration algorithm. Any shape vertex may be 
fused into a bond. Additionally, shape vertex enumeration is arbitrary. 
These degrees of freedom can be removed by rotating the fused vertex to 
the algebraically smallest shape vertex of its set of rotationally 
interconvertible vertices. For example, the octahedron has a single set 
of interconvertible vertices, but the square pyramid has two: The 
equatorial set of four vertices, and the singleton set containing the 
apical vertex. Removing representational degrees of freedom ensures the 
stereopermutation enumeration algorithm generates transferable 
stereodescriptors.

The enumeration scheme proceeds to identify the set of closest off-axis shape
vertices that will be most relevant to rotational energy barriers. If, at either
side of the bond, this shape vertex set's ranking characters are all equivalent,
then rotation around the bond is isotropic. Next, dihedral orientations are
generated by sequentially aligning off-axis shape vertices across both sets.
Alignment of off-axis shape vertices does not need to be ecliptic, but can also be
staggered, enabling bond-centered stereopermutators to also enumerate
hypothesized rotational minimum-energy structures in e.g.\ alkanes. 

This scheme will omit the trans-alignment in a combination of two bent shapes,
however, so it is explicitly added. Furthermore, the emergent ordering of
stereopermutations does not yield stereodescriptors paralleling the relative
order of the IUPAC stereodescriptors \textit{E}/\textit{Z} as defined in the
sequence rules. To ease the implementation of a ranking algorithm, the
stereopermutation sequence is modified to match the relative order of IUPAC
stereodescriptors.

Bond-centered stereopermutators can be placed at any bond where both constituting atoms
are managed by a stereopermutator whose stereopermutation is specified.
Unspecified atom-centered stereopermutators cannot form the basis of a bond-centered
stereopermutator because their mapping from binding sites to shape vertices is
unknown. Constituting shapes are not limited to the common cases constituting
multiple-order bonds in organic chemistry. The rotational energy structure of
wild combinations of shapes with many vertices is likely to be more complex than
\textsc{Molassembler} would suggest, however.

Analogously to atom-centered stereopermutators, bond-centered stereopermutations may not be
viable in small rings. For instance, altering the orientation of the double
bond in cyclopentene to an $E$ orientation yields an unembeddable graph, and this
bond should therefore not be considered for bond-centered stereopermutation. Unfeasible
bond-centered stereopermutations can be identified in analogy to unfeasible links in
atom-centered stereopermutators. 

\begin{figure}[h]
  \centering
  \includegraphics[width=\linewidth]{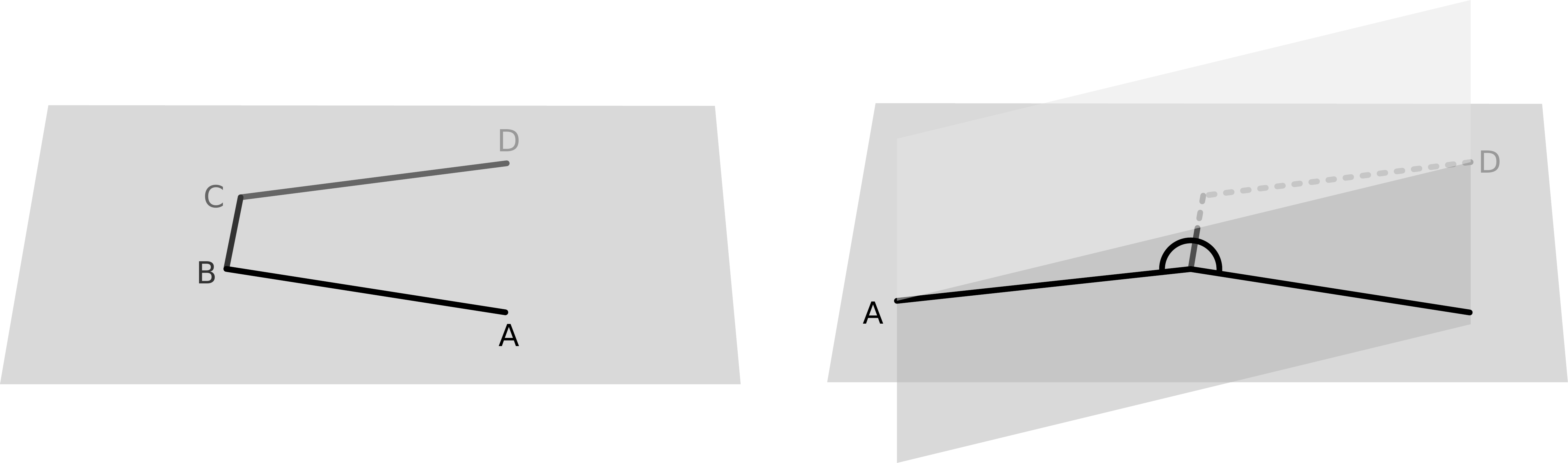}
  \caption{\textit{Left:} A dihedral sequence $ABCD$ composed of segments
  of unit length and joined at $\frac{2\pi}{3}$ angles with dihedral angle zero.
  The plane in which the cyclic polygon expands coincides with the plane
  composed of the points $BCD$ as this offers points on the cycle the
  maximal distance to the base vertices $BC$.
  \textit{Right:}~The same dihedral sequence at dihedral angle $\pi$. The cyclic
  polygon now expands in a plane perpendicular to that of $BCD$.
  }\label{fig:bond_stereopermutator_infeasibility}
\end{figure}

In atom-centered stereopermutator 
feasibility, one segment of the cyclic polygon is the segment $c$ that 
replaces the 1--3 distance modeled explicitly by two atomic distances and 
one angle in Figure~\ref{fig:infeasibility_denticity}. In bond-centered 
stereopermutator feasibility, this segment is the replacement of the 1--4 
distance, which is explicitly modeled via three atomic distances, two 
angles, and one dihedral angle. Another difference is that the cyclic 
polygon expansion plane must be modeled in three dimensions (see Supporting Information).
In a sequence of four atoms at a dihedral angle of zero, the cyclic polygon 
expands within the same plane as the dihedral-constituting atoms to 
maximize atomic distances from cycle atoms to dihedral-angle-constituting atoms. 
At dihedral angle $\pi$, the expansion plane maximizing cycle-atom 
distances to the dihedral-angle base atoms lies perpendicular to that of the 
dihedral-angle-constituting atoms, as shown in 
Figure~\ref{fig:bond_stereopermutator_infeasibility}.

Local modeling of bond-centered stereopermutator feasibility by cyclic polygons has a
necessarily limited purview. Consider a graph cycle of arbitrary length
containing two bond-centered stereopermutators each composed of two atom-centered
stereopermutators of equilateral triangle shape. Each bond-centered stereopermutator
permits a cis and trans arrangement of the cycle continuation. At small cycle
sizes, individual feasibility modeling detects that only a cis arrangement is
possible and argue that neither bond-centered stereopermutator is stereogenic. At large
cycle sizes, both arrangements are deemed feasible. It is conceivable that a
cycle size exists in which only one bond-centered stereopermutator can be cis at a time
for the graph to be embeddable. Individual bond-centered stereopermutator feasibility
modeling will not capture this.

In carbocycles of size seven and below, 
bond stereopermutators are not stereogenic, yet they do serve to enforce 
planarity of their substituents in conformers.
During molecule construction, if a cycle is detected
as approximately flat, bond-centered stereopermutators are placed at all edges of the
cycle, enforcing coplanarity of all cycle edges.

As chiral state propagation of bond-centered stereopermutators on alterations of underlying
atom-centered stereopermutators apart from ranking changes has not been implemented,
alterations of atom-centered stereopermutators forming a bond-centered
stereopermutator cause the bond-centered stereopermutator to be dropped.

\section{Ranking algorithm}\label{sec:ranking_algorithm}

Ranking is based on the IUPAC sequence rules for organic
molecules~\cite{favre2013nomenclature}. The rules for organic molecules form a
starting point for the differentiation of ligands that is generalizeable to
\textsc{Molassembler}'s model of stereocenters. Without the ambition to generate
canonical IUPAC names for molecules of either organic or inorganic character,
there is no need to implement further rules and relative preferences were laid out
for inorganic molecules in Ref.~\citenum{IUPACInorg}. Differences to organic-molecule ranking
arise mainly from generalization of the sequence rules from organic
stereodescriptors to the library model and from partial omissions of sequence
rules due to missing implementations. The stereodescriptors \textit{R} and
\textit{S} arising from maximally asymmetric tetrahedral centers directly
correspond to indices of permutation arising from the atom-centered permutational
scheme. The stereodescriptors \textit{Z/seqCis} and \textit{E/seqTrans}
similarly correspond to indices of permutation arising from the bond-centered
stereopermutation scheme. Note that helical chirality stereodescriptors \textit{M} and
\textit{P} are currently not detected by \textsc{Molassembler}.

The following changes and omissions were made with respect to the IUPAC sequence
rules: First, no algorithm to enumerate mesomeric Kekul\'{e} structures is
currently implemented, so the relevant part of sequence rule 1a in Ref.~\citenum{favre2013nomenclature}
is discarded.
This omission is planned to be fixed in a future version of our software. Second, no
distinction will be made between asymmetric and pseudoasymmetric stereogenic units,
affecting sequence rules 4a and 4c. Third, \textsc{Molassembler} considers
stereopermutators alike for sequence rule 4b if the number of stereopermutations
and the index of permutation match. This is merely an abstraction of the
sequence rule that covers all explicit cases laid out in the sequence rule
definition. Lastly, the relative priority of stereodescriptors \textit{R} before
\textit{S} and \textit{Z} before \textit{E} defined in sequence rule 5 is
abstracted to algebraic ordering of the index of permutation. Inherently, since
the vertex enumeration of shapes is arbitrary, the relative order of indices of
permutation is also arbitrary. We can therefore choose the vertex enumeration at
the tetrahedron and monovacant tetrahedron so that the integer ordering
parallels the order of the corresponding \textit{R} and \textit{S}
stereodescriptors. 

The generation of indices of permutation at bond-centered
stereopermutators is similarly altered to parallel the IUPAC stereodescriptors
\textit{E/seqCis} and \textit{Z/seqTrans}. Indices of permutation for shapes
with more vertices than the tetrahedron can distinctly depend on the abstract
binding case. For instance, indices of permutation for a trigonal bipyramid of
abstract binding case \texttt{ABCDE} are not comparable to indices of
permutation for the abstract binding case \texttt{AABCD}. Furthermore, they are
not comparable across different shapes. Consequently, relative ordering of
stereodescriptors in sequence rule 5 is established by sequential comparison of
the shape (algebraic ordering of its index in the list of all shapes),
lexicographic comparison of the abstract binding case, and finally comparison of
the index of permutation. The relative ordering established remains arbitrary,
but the sequence rule distinguishes more cases that may occur in inorganic
compounds without affecting the relative priority of the organic
stereodescriptors.

A ranking algorithm based upon these modified IUPAC sequence rules can establish
chemical differences between individual atom substituents, but not between
binding sites. For a ranking at the level of binding sites, \textsc{Molassembler} applies
two sequence rules. First, sites consisting of more atoms precede sites
consisting of fewer atoms. 
Second, sites are ordered lexicographically 
by their constituting atoms' ranking positions set. The ranking 
positions set is ordered descending by value.

Ranking implementation correctness was tested against relevant IUPAC
examples~\cite{favre2013nomenclature} and a validation suite from a proposed revision of
the sequence rules~\cite{Hanson2018}. Not surprisingly, owing to the implementation of
only a limited set of sequence rules and some omissions from their definitions (as discussed
above), \textsc{Molassembler} did
not pass all tests from this validation suite. However, additional differences arose due to
different interpretations of the sequence rules. The consequence of an incorrect
ranking can be twofold. By falsely differentiating substituents, duplicate
stereopermutations can arise that are superimposable. By falsely equating
substituents, stereopermutations can be missed. These issues point to future work
on the implementation, which will
lift limitations imposed upon sequence rules so that
sequence rule interpretation can be brought into full agreement with that of the validation
set.

Currently, the generalization of the sequence rules to \textsc{Molassembler}'s 
molecule model resulted in a ranking algorithm that can be applied in 
myriad situations. Stereodescriptors of multidentate and haptic 
ligands in arbitrary polyhedral shapes and stereodescriptors of any 
multiple bond-order combination of polyhedral shapes at any fused 
vertices are accounted for. Despite its incompleteness, the ranking 
algorithm as implemented is correct for a vast majority of typical cases.

\section{Molecule representation, equivalence, and manipulation}\label{sec:molecule_manipulation} 

A set of atom (nuclear) Cartesian coordinates is in principle free of any
information about the presence of bonds between atoms. \textsc{Molassembler}
allows the translation of molecules given as a set of nuclear coordinates and a
bond order matrix into graphs. This proceeds by discretization of the bond order
matrix, subsequent connected component identification, and finally
classification of local shapes and identification of stereopermutations of a
specific molecule. 

Crucial for the correct molecule interpretation is that bonds are identified
correctly. A vast number of methods may be applied to calculate bond orders for
systems under study. Generally, we do not recommend a particular method that
will always work for any kind of system, simply because of the fact that the
assignment of chemical bonds will be affected by subjective ratings in regions
where bond orders are not close to an integer number. Instead, we chose not to
integrate any particular scheme into the process, requiring a floating-point
bond order matrix as input instead. Additionally, \textsc{Molassembler}
explicitly supports two variants of bond order discretizations. The first mode
interprets bonds between atoms according to a bond order matrix in a binary
fashion: Bonds either exist or they do not exist, and no differentiation is made
regarding their order, (with the exception of 'eta' bonds). Little detail in
modeling of molecules is lost in this mode because bond stereopermutators are
checked in a phenomenological fashion: If a dihedral angle is aligned in
ecliptic fashion, it is assumed that there is an energetic reason for it.
Similarly, flat cycles are detected and preserved by bond stereopermutators,
ensuring at least a structural capture of aromaticity. Generally, this method of
molecule interpretation is robust and can be combined with low-effort bond order
calculation methods such as the comparison of atom-pair spatial distances
against sums of covalent radii. The other mode of discretization of the
floating-point bond order matrix is rounding into the nearest integer bond
order. Bond stereopermutators can be checked either at all bonds or only at
those bonds above a specified real-number bond order threshold to reduce the
likelihood that a spuriously eclipsed dihedral angle is misconstrued as an
energetically favorable effect. Recall that \textsc{Molassembler} expects
structures at a local potential energy surface minimum as input. Methods
yielding floating-point bond orders suitable for discretization in this mode may
be system-appropriate force fields or quantum mechanical methods, possibly with
Natural Bond Order~\citenum{Glendening2012} or Intrinsic Bond
Order~\citenum{Knizia2013} approaches. The generated conformational ensembles of
molecules interpreted by binary bond discretization are generally poorer in
quality than those of the nearest-integer approach, because atom-pair distances
are then universally estimated as if they were of order one, losing fidelity
from their input. Note, however, that \textsc{Molassembler} does not aim to
generate conformational ensembles of thermodynamic quality. A generated
conformer merely has to be close enough to a local potential energy surface
minimum so that subsequent refinement by an appropriate method (e.g., quantum
chemical structure optimization) leads into the intended minimum, and this
property is preserved. By supporting both manners of bond order discretization,
\textsc{Molassembler} allows a trade-off between reliability of interpretation
against fidelity of modeling and quality of generated conformational ensembles
if the system under study imposes it. In future work, a possible extension of
the bond discretization procedure may attempt to maximize the number of
low-distortion recognized local shapes either to correct the discretization
itself or serve to highlight problematic cases.

\begin{figure}[h]
  \centering
  \includegraphics[width=0.7\linewidth]{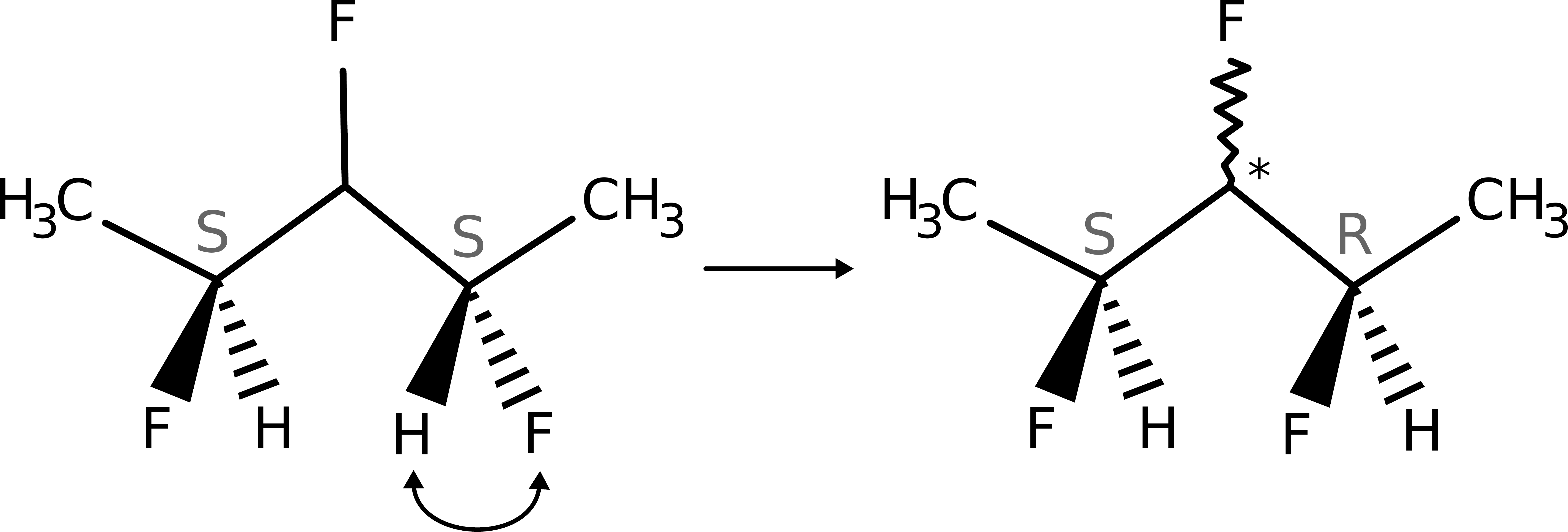}
  \caption{\textit{Left:} This example shows a molecule with two substituents at the
  central carbon atom that are constitutionally isomeric and carry the same
  stereodescriptor, yielding an identical ranking. The central carbon atom is
  not stereogenic. If a molecule edit changes the stereopermutation of one of
  these substituents, akin to swapping the hydrogen and fluorine atom as indicated,
  the stereodescriptor flips. \textit{Right:}~The desired result of the
  molecular manipulation. Although the change is local to the right arm of the
  molecule, it incurs a differentiation between the two arms at the central
  carbon atom, which is stereogenic and unspecified after re-ranking.}\label{fig:why_rerank} 
\end{figure}

Small manipulations of molecules, as trivial as they may seem, impose a
significant amount of nontrivial work. Stereodescriptors are propagated at
binding site additions and removals to the target shape 
as discussed above. 
Since stereodescriptors are chosen based on substituent ranking, which may have been
altered by the edit, it is necessary to re-rank all binding sites of non-terminal atoms.
For an example see Figure~\ref{fig:why_rerank}. Should a binding site
ranking change by the edit, assigned stereopermutations are propagated to the
new ranking. Consequently, 
to ensure continuity, molecule edits scale at least linearly in the number of atoms.

In large-scale manipulations of molecules such as connecting multiple molecules we may
avoid increasing complexity by delaying chiral state
propagations due to ranking changes until all
graph modifications are finished. 
Hence, specialized functions for large-scale
manipulations such as connects, disconnects, substitutions, and chelating ligand
additions are made available in our software.

Equivalence between two molecules is based on a colored graph
isomorphism algorithm with modular vertex coloring so that molecules can be
compared for varying definitions of molecular equality. The information that
can be exploited to color vertices are element types, bond orders, local shapes, and
stereopermutations. To expedite molecular-graph comparisons in large sets, it is helpful to standardize all involved
graphs. If both graphs in a comparison are in canonical form, it is no
longer necessary to search for an index permutation matching vertices across the
graphs; instead, it is possible to base comparisons on an identity vertex mapping.
Such canonicalization is modular, as is the colored graph isomorphism. If two
molecular graphs have been brought into standardized form with the same level of information as the
comparison is tasked with, equivalence comparison of molecule instances is
a direct identity mapping comparison instead of a colored graph isomorphism.

\begin{figure}[h]
  \centering
  \includegraphics[width=\linewidth]{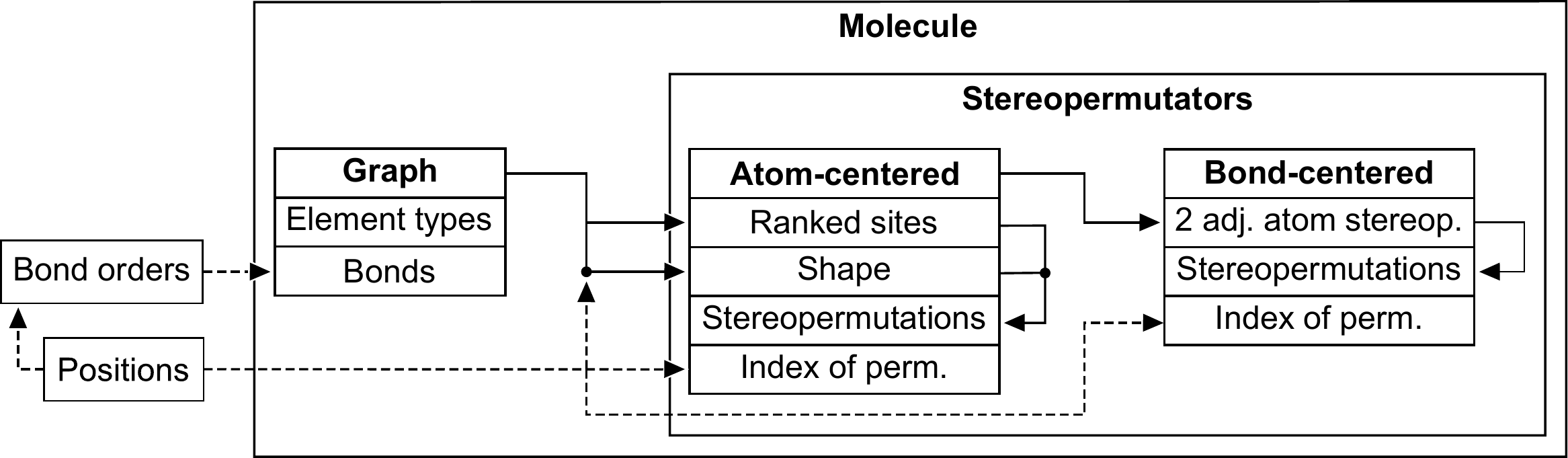}
  \caption{Simplified data flow in construction of a molecule representation
  from various primitive data types. Solid paths represent the default flow of
  data, dashed paths represent optional data flow. For instance, by default, the
  local shape of non-terminal atoms is inferred from the graph in the
  construction of atom-centered stereopermutators. Optionally and preferably, if
  positions are available, local shapes can be classified from atom (nuclear) positions.
  }\label{fig:data_flow}
\end{figure}

The components of the molecule model and some of their dependencies are
illustrated in Figure~\ref{fig:data_flow}. Nuclear Cartesian coordinates and
bond orders are optional primitive data types in the construction of a molecule.
Minimally, a molecule representation consists of a graph storing element types
and bonds. Atom-centered stereopermutators are constructed at non-terminal atoms
of the graph from a ranking of its substituents and a local shape either
inferred from the graph or classified from coordinates. These parts are combined
into the set of stereopermutations. Two adjacent atom-centered stereopermutators
can form the foundation of a bond-centered stereopermutator, which determines its
stereopermutations from the information embedded in both atom-centered
stereopermutators. The indices of permutation of both atom-centered and
bond-centered stereopermutators can be inferred from positions.

Worth mentioning are a number of further available features: 
(i) Molecule instances are serializeable into plain-text JavaScript Object Notation and its
binary encodings for database or file storage. 
(ii) It is possible to assert whether
molecules are enantiomers of one another. 
(iii) McGregor's maximum common subgraph
algorithm~\cite{McGregor1982} with custom vertex comparators enables subgraph
matching. 
(iv) \textsc{Molassembler} also contains an
openSMILES~\cite{james2015opensmiles} compliant SMILES parser notably
implementing stereoconfiguration for square, trigonal bipyramid and octahedron
shapes in addition to the tetrahedron.

\section{Conformer generation}\label{sec:conformer_generation}

Conformer generation in \textsc{Molassembler} is achieved with Distance
Geometry~\cite{Blaney1994}. 
In the context of molecular conformer generation, 
Distance Geometry enables a multi-step transformation from an atom-pairwise 
distance bounds matrix to atom coordinates. A distance bounds matrix 
is filled with lower and upper bounds on atom-pair distances as desired in the
resulting conformer.

\begin{figure}[h]
  \centering\includegraphics[width=0.7\linewidth]{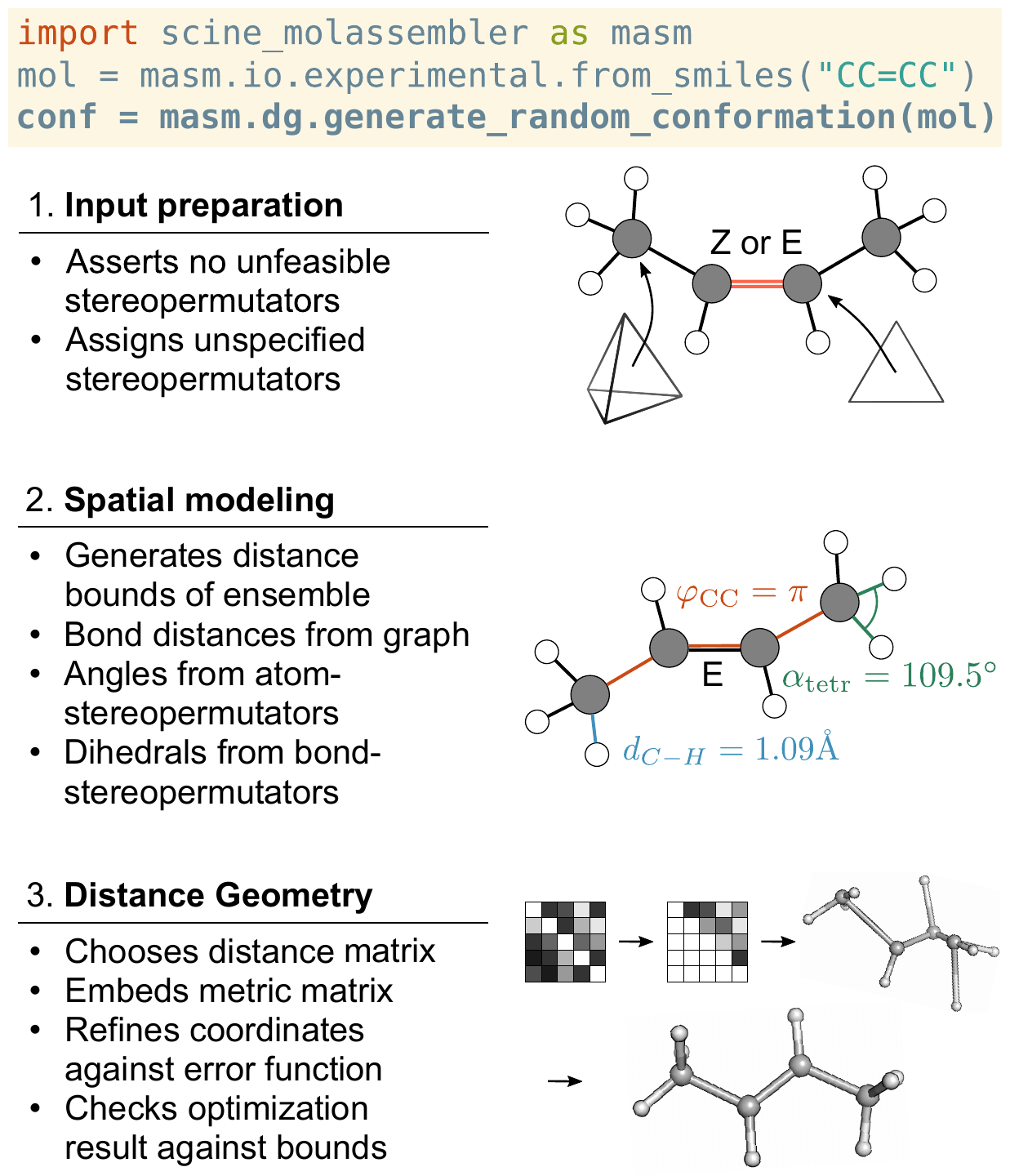}
  \caption{Condensed steps of the conformer generation procedure at the
  example of but-2-ene with an unspecified bond stereopermutator at the double
  bond. Input to conformer generation is a modelled molecule. After the
  unspecified bond stereopermutator is randomly assigned, spatial modelling
  proceeds by generating a distance bounds matrix. This distance bounds matrix
  is then finally converted into spatial coordinates by Distance Geometry.
  }\label{fig:cg_steps}
\end{figure}

Conformer generation proceeds in three major steps (refer to
Figure~\ref{fig:cg_steps}): The first stage is preparatory,
where we exclude some graphs that cannot be modeled and assign unspecified
stereopermutators. Second is spatial modeling, where the molecular model is
translated into a distance bounds matrix. In the third step, we apply Distance
Geometry and obtain Cartesian atom coordinates.

It is necessary for spatial modeling that all
stereopermutators of a molecule must have at least one feasible
stereopermutation and may not be unspecified. In the first preparatory stage,
unspecified stereopermutators are set iteratively by choosing a random
unspecified stereopermutator and setting its stereopermutation at random until
no unspecified stereopermutators remain. For atom-centered stereopermutators,
each stereopermutation is weighted by its relative statistical occurrence. 

In the spatial modeling stage, we leverage the molecular model to collect
atom-pairwise distance bounds for a distance bounds matrix. Distance bounds are estimated between all pairs of atoms
up to a graph distance of three. The distance bounds of immediate graph
adjacents are populated from estimations of atom distances for the given bond
order. Each atom-centered stereopermutator provides angular information between
its substituents, which allows the modeling of the distance bounds of 1--3
bonded atoms when combined with 1--2 distance bounds. Each bond-centered
stereopermutator provides dihedral information between substituents, which
allows the modeling of the distance bounds of 1--4 bonded atoms when combined
with 1--2 distance bounds and the angular information of the two atom-centered
stereopermutators constituting the bond. If no bond-centered stereopermutator is
present on a bond, default full definition range dihedral bounds are assumed
instead in the modeling of the 1--4 bonded distance. 

\begin{figure}[h]
  \centering \includegraphics[width=\linewidth]{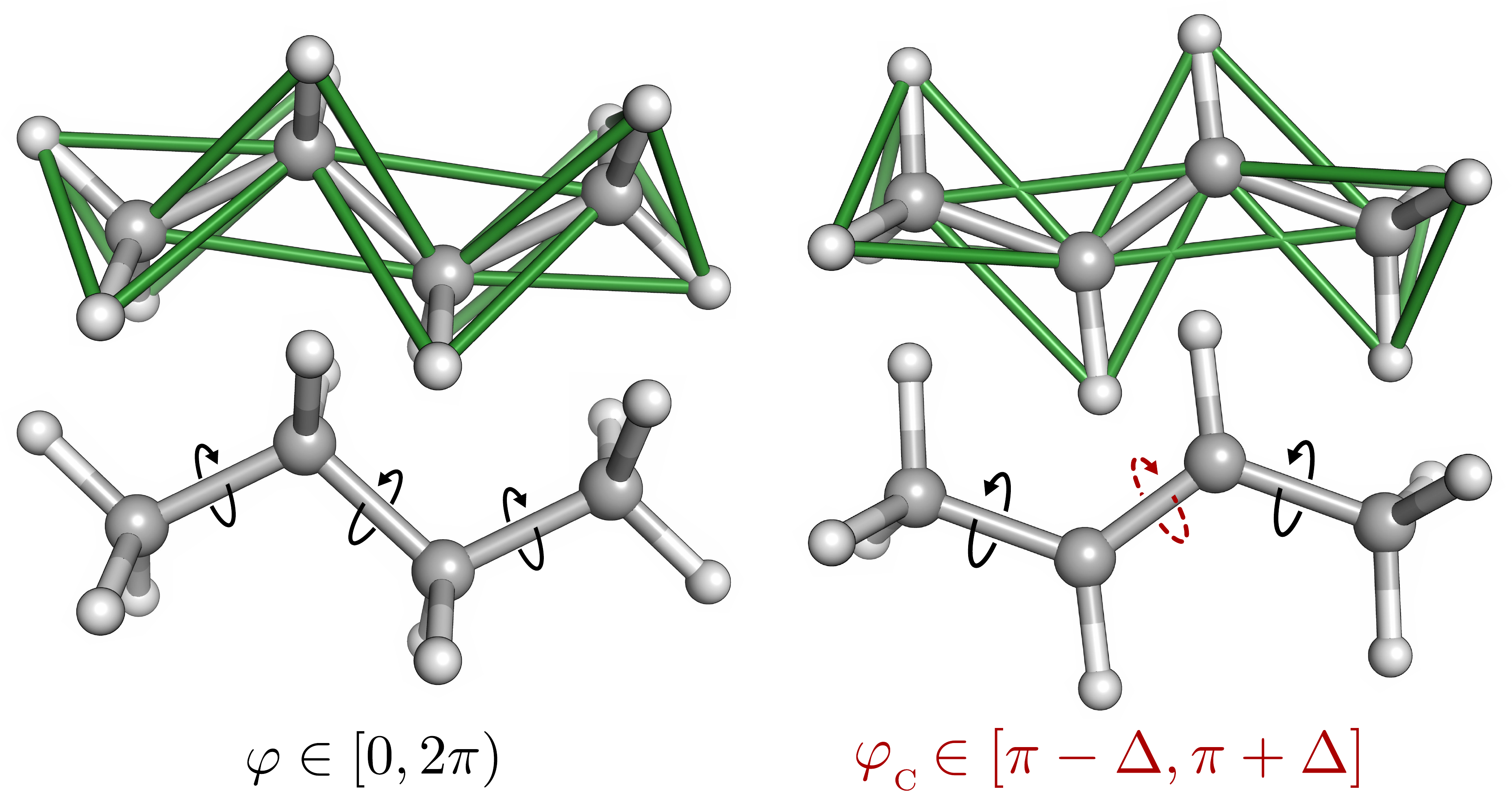}
  \caption{Representations of the molecular model of butane (left) and
  but-2E-ene (right) and differences in conformer generation spatial modeling.
  While butane has an atom-centered stereopermutator at each carbon atom
  indicating a local tetrahedral shape, the central two atom stereopermutators
  of but-2E-ene indicate a local equilateral-triangle shape. When modeling
  butane, all dihedrals along the carbon atoms are freely rotatable. In
  contrast, but-2E-ene has a bond-centered stereopermutator placed at the
  central carbon bond whose assigned stereopermutation matching the
  stereodescriptor 'E' restricts e.g.\ the carbon dihedral angle to a range of
  width $\Delta$ centered around the trans-dihedral angle.
}\label{fig:dg_spatial_modeling} 
\end{figure}

For example, a butane molecule will have atom-centered stereopermutators placed
at each carbon atom indicating the local shape of a tetrahedron, as shown in
Figure~\ref{fig:dg_spatial_modeling}. All dihedral angles are freely rotatable
as no bond-centered stereopermutators are present. 
In contrast,
atom-centered stereopermutators of but-2-ene at the double bond indicate the
local shape of an equilateral triangle. The single bond-centered
stereopermutator placed at the double bond indicates whether a $Z$- or
$E$-ethene substructure is modeled by its assigned stereopermutation.  The
distance bounds among the atoms adjacent to the double bond are affected by the
selected bond stereopermutation.

Spatial modeling modifies the collected bounds to encompass distorted
structures. Angle bounds of local shapes are expanded if they are a member of a
cycle with fewer than six members. Angles between substituents of spirocenters
with small cycles are explicitly modeled. It is possible to specify fixed
positions for atoms, albeit with the condition that none, one or all of its
binding sites' atom positions must be fixed as well.

Having obtained a distance bounds matrix from spatial modeling, we proceed with
the final stage, the application of Distance Geometry. 
A distance matrix is generated by triangle inequality bounds smoothing the
distance bounds matrix and iteratively choosing pairwise distances between the
resulting bounds. To avoid $O(N^5)$ complexity for full metrization, the
triangle inequality smoothing problem is transferred to a shortest-path graph
problem~\cite{Dress1988}. The resulting single-source shortest-path problem on a
graph with negative edge weights is solved by the simplified GOR1
algorithm~\cite{Cherkassky1996} as it performs well on such graphs.
\textsc{Molassembler} implements three metrization~\cite{havel1984distance}
variants: Full metrization, 10\% metrization, and four-atom metrization.

The distance matrix is then transformed into the metric matrix and embedded into
Cartesian coordinate space. These coordinates typically poorly reflect their
distance bound inputs and require optimization against an error function.
Such error functions typically contain two
types of terms: Distance errors mediating particle pair distances, and chiral
errors mediating four-particle relative orientation by a signed tetrahedron
volume, similar to an improper dihedral.

In refinement, distance bound violation error terms can prevent the tetrahedron
volume spanned by a chiral constraint from inverting. The more chiral
constraints occur, the more problematic this becomes, in particular if chiral
constraints have overlapping particle domains. The more vertices the local shape of an atom-centered stereopermutator has, the
more chiral constraints are emitted to ensure capture of its chirality,
exacerbating the problem. To ensure proper inversion of incorrectly embedded
chiral constraints, we adopted multiple-stage four spatial coordinates
refinement: Distance errors are applied on four spatial coordinates while chiral
errors are applied on three. In the first stage of refinement, the structure is
free to expand into the fourth spatial dimension to minimize chiral errors. When
all chiral constraints have the correct sign, an additional potential is applied
on the fourth spatial dimension to compress structures back into three
dimensional space.

The refinement error function in our implementation features two further changes:
Bond-centered stereopermutators emit dihedral bounds and hence the refinement
error function contains dihedral error terms. In contrast to enforcing
zero-dihedral angle orientations of four particles with chiral constraints of
zero volume, dihedral error terms offer more flexibility in the specificity of
refined structures. Additionally, for haptic ligands, in which the centroids of
the set of vertices forming the binding site occupy a shape vertex, chiral and
dihedral errors are calculated with centroids of particle sets instead of with
positions of individual particles.

The refinement error function depends on a set of $N$ particles with positions
$\vec{r}_i$. The symmetric matrices $\mathbf{L}$ and $\mathbf{U}$ contain the
lower and upper distance bounds. Each chiral constraint of the set $C$ consists
of four particle sets $S_\alpha, S_\beta, S_\gamma, S_\delta$, and a lower and an
upper signed volume bound $L_V$ and $U_V$, respectively. Each dihedral constraint of the set
$D$ consists of four particle sets, a lower and an upper signed dihedral bound
$L_\phi$ and $U_\phi$, respectively.

The signed tetrahedron volume of a chiral constraint is calculated from the centroids of
its constituting particle sets $\vec{s}_j$ as:
\begin{equation}
  \vec{s}_j = \frac{1}{|S_j|}\sum_{i=1}^{|S_j|} \vec{r}_{S_{j, i}}
\end{equation}
\begin{equation}
  V_{\alpha\beta\gamma\delta} =
    \frac{1}{6}\left(\vec{s}_{\alpha}-\vec{s}_{\delta}\right)^{T}
    \cdot
    \left[
      \left(\vec{s}_{\beta}-\vec{s}_{\delta}\right)
      \times \left(\vec{s}_{\gamma}-\vec{s}_{\delta}\right)
    \right].
\end{equation}
The distance error $d_{ij}$, the chiral error $C_{\alpha\beta\gamma\delta}$, and
the dihedral error $D_{\alpha\beta\gamma\delta}$ are defined as:
\begin{equation}
  d_{ij} = \mathrm{max}^2\left(0,
  \frac{\left(\vec{r}_{j}-\vec{r}_{i}\right)^{2}}{U_{ij}^{2}}-1\right) 
  + \textrm{max}^{2} \left(0, \frac{2 L_{ij}^{2}}{L_{ij}^{2} +
  \left(\vec{r}_{j}-\vec{r}_{i}\right)^{2}}-1\right)
\end{equation}
\begin{equation}
  C_{\alpha\beta\gamma\delta} = \mathrm{max}^{2} \left(0, 
    V_{\alpha\beta\gamma\delta}-U_{V}
  \right) + \mathrm{max}^{2} \left(0,
    L_{V}-V_{\alpha\beta\gamma\delta}
  \right)
\end{equation}
\begin{equation}
  D_{\alpha\beta\gamma\delta} = \mathrm{max}^2 \left(0,
    \left\lvert\phi_{\alpha\beta\gamma\delta}+\left\{
        \begin{array}{r | r}
        2\pi & \phi < \frac{U_\phi + L_\phi - 2\pi}{2}\\
        0 & \\
        -2\pi & \phi > \frac{U_\phi + L_\phi + 2\pi}{2}\\
        \end{array}
      \right\}-\frac{U_\phi{}+L_\phi}{2}
    \right\rvert-\frac{U_\phi{}-L_\phi}{2}
  \right)
\end{equation}
where $\mathrm{max}^{2}$ denotes the operation 'square the largest element out of the two given in parenthesis'.
The total error function then reads
\begin{equation}
  \mathrm{errf}\left(\left\{\vec{r}_i\right\}\right) 
  = \sum_{i < j}^{N} d_{ij}
  + \sum_{\left(S_\alpha, S_\beta, S_\gamma, S_\delta,
  U_{V}, L_{V}\right) \in{} C} C_{\alpha\beta\gamma\delta}
  + \sum_{\left(S_\alpha, S_\beta, S_\gamma, S_\delta,
  U_\phi, L_\phi\right) \in{} D} D_{\alpha\beta\gamma\delta}
\end{equation}
(note the doubled letter 'r' introduced in the abbreviation of the function
to distinguish it from the standard error function).
The error function $\mathrm{errf}$ is minimized with analytical gradients by
L-BFGS~\cite{Liu1989} in three stages: 
First, without application of a potential on the 
fourth spatial dimension and dihedral errors, the tetrahedra modeled by 
chiral errors are allowed to invert. Second, a potential is applied to 
the fourth dimension and components along the fourth spatial dimension 
are eliminated. Lastly, dihedral errors are enabled. Conformer 
generation can then be trivially parallelized.

Directed conformer generation has been implemented by placing bond stereopermutators
enumerating staggered arrangements at bonds whose rotations are non-isotropic.
\textsc{Molassembler} implements a trie data structure to track and ease
generation of rotational combinations among the set of staggered bond
stereopermutators. 
These combinations can then be passed to a special conformer generation function. 
Until feasibility criteria for combinations of dihedral
arrangements in cycles are devised and implemented, bonds in cycles of any size
are excluded from consideration in directed conformer generation, however.

\section{Shortcomings}\label{sec:shortcomings}

As mentioned before, the general model of stereopermutators treats molecular
configuration as frozen with the exceptions of (i) nitrogen atom inversion, (ii) the
Berry pseudorotation, and (iii) the Bartell mechanism, for all of which the
thawing of stereopermutations has been implemented as an option.
Most likely, these are not the only cases in which
stereopermutations can easily interconvert. In all unhandled cases,
\textsc{Molassembler} may overstate chiral character.

Feasibility determination of stereopermutations involving haptic binding sites
based on conical-space overlap avoidance is limited by the algorithm estimating
the required conical space. This algorithm currently estimates the
required conical space for biatomic or cyclic topologies; apart from this, no
feasibility checks are made. The potential consequences are overstatements of
chiral character and failures to generate conformations.

Our effort to ensure that modeled conformers are close to
local minima of the potential energy hypersurface is still rather limited and improved
structure prediction can be envisioned for future improvement of our software. In the present version, atom pair distances in
discrete bond orders are simply estimated by Universal Force
Field~\cite{rappe1992uff} parameters. Systematic distortions of common local
shapes such as those due to Jahn-Teller distortion~\cite{jahn1937stability} in
an octahedron have not been addressed. It is therefore inadvisable to directly
analyze generated conformational ensembles without subjecting the generated
structures to suitable structure optimization or molecular mechanics methods.

The fundamentally factorial complexity of shape recognition and
stereopermutation enumeration will limit the applicability of
\textsc{Molassembler} for compounds with very high coordination numbers.
Additionally, for such compounds, the approach of classifying shapes into a
strict set of defined shapes may be impractical. The Thomson
Potential~\cite{Thomson1904} demonstrates effectively that the more point
charges are introduced, the more local minima the potential surface exhibits.
Similarly, the set of shapes with numerous vertices would have to expand
significantly to correctly capture various exotic shapes and point group
symmetries. In particular for many vertices, it might be preferable to cut shape
classification from the library entirely and instead try to identify rotational
symmetry elements that the stereopermutation algorithm could apply directly.

\textsc{Molassembler} does not pass the validation test set proposed by 
Hanson et.\ al.~\cite{Hanson2018}. The library ranking algorithm 
predates their work, which laid bare the possibility of alternate 
interpretations of the existing sequence rules. The validation set also 
includes test cases for a proposed additional sequence rule which is not 
implemented in \textsc{Molassembler}. In future work, the library ranking 
algorithm will be brought into closer alignment with their proposed 
changes. A ranking algorithm capable of exact differentiation is of 
paramount importance as stereogenicity of atom centers may be 
misrepresented. Furthermore, no axial or helical chirality is yet 
identified by out implementation and certain molecular features may be missed.

\section{Demonstration examples}\label{sec:demonstration}

\begin{figure}[h]
  \includegraphics[width=\linewidth]{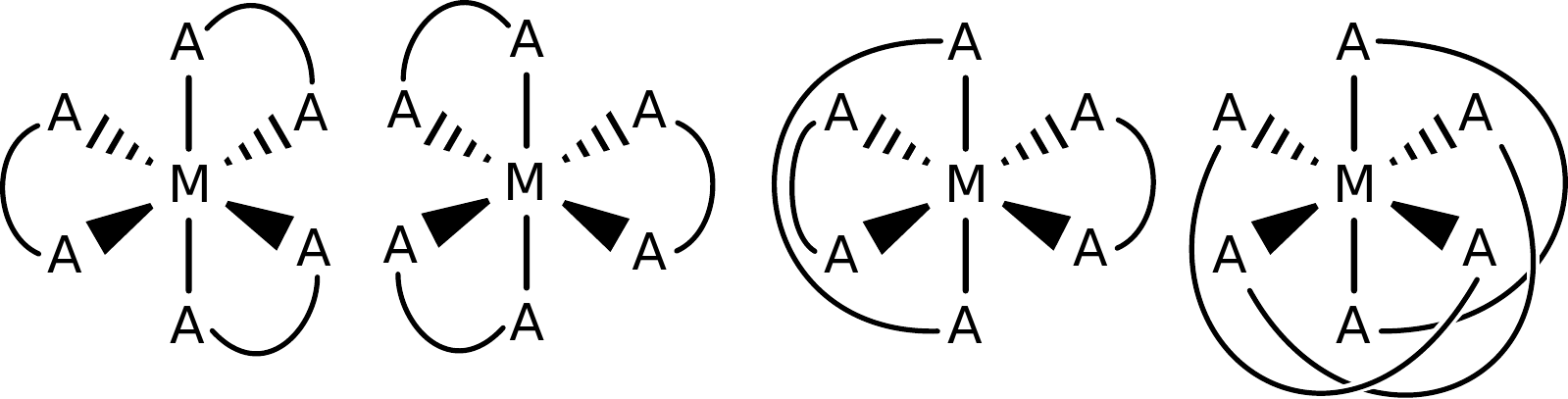}
  \caption{The four abstract stereopermutations of the binding case
  \texttt{(A-A)\textsubscript{3}} in an octahedron, from left to right: The
  $\Lambda$ and $\Delta$ isomers, in which all bidentate ligands are
  cis-arranged, a stereopermutation in which a single ligand is trans-arranged,
  and lastly a stereopermutation in which all ligands are
  trans-arranged.}\label{fig:demonstration_feasibility} 
\end{figure}

\textsc{Molassembler} has already been employed in the context of automated
exploration of chemical reaction networks~\cite{grimmel2019}, in which it served
as molecular graph interpreter, equivalence oracle, and conformer generator.
Therefore, we begin with a demonstration of two of its core features:

We shall consider an example where feasibility checks are important: An
octahedral center with three short-bridge bidentate ligands. For the abstract
binding case \texttt{(A-A)\textsubscript{3}}, there are four stereopermutations,
as shown in Figure~\ref{fig:demonstration_feasibility}. In the particular case of
\texttt{[Fe($\mu$\textsuperscript{2}-Oxalate)\textsubscript{3}]\textsuperscript{3+}},
the feasibility algorithm rules out all stereopermutations in which an oxalate
is trans-arranged due to its short bridges. \textsc{Molassembler} reports that
the stereopermutator has four stereopermutations but only two of these are
feasible.

\begin{figure}[h]
  \includegraphics[width=\linewidth]{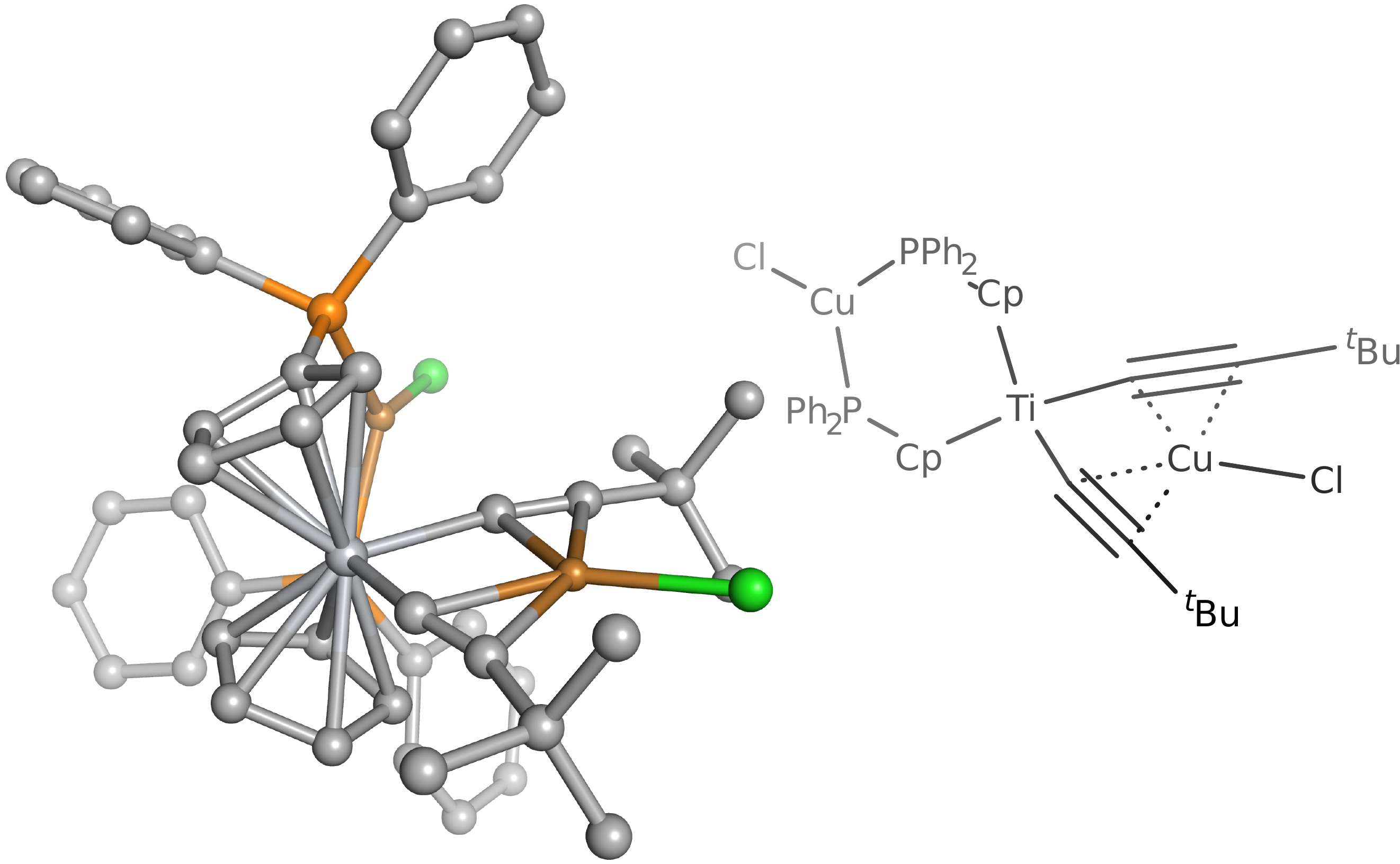}
  \caption{\textit{Left:} The heavy-atom skeleton of a molecule with two types
  of haptic ligands: The bridged cyclopentadienyl groups bonded to a titanium
  atom in the center, and the triple bonds bonded to the copper atom towards the
  right. \textit{Right:}~Simplified Lewis structure of the left
  compound.
Atom coloring: Carbon and titanium in gray, phosphorus in 
orange, copper in brown, and chlorine in green.}\label{fig:demonstration_haptic} 
\end{figure}

Next, we demonstrate that the abstraction \textsc{Molassembler} introduces
regarding binding sites is effective for haptic ligands. In
Figure~\ref{fig:demonstration_haptic}, an example molecule is shown with a
relatively complex case for shape classification, ranking, and stereopermutation
-- all at once. The assumption that the centroid of a set of contiguous binding
atoms is placed at a vertex of an underlying shape is well-applicable here,
reducing the five-atom copper atom to a three-site equilateral triangle. Ranking
correctly deduces that its haptic ligands are identical in an \texttt{(A-A)B}
abstract binding case. Similarly, the twelve-atom titanium atom is reduced to a
four-site tetrahedron with a \texttt{(A-A)(B-B)} binding case.
Stereopermutation concludes that neither center is stereogenic.

\begin{figure}[h]
  \includegraphics[width=\linewidth]{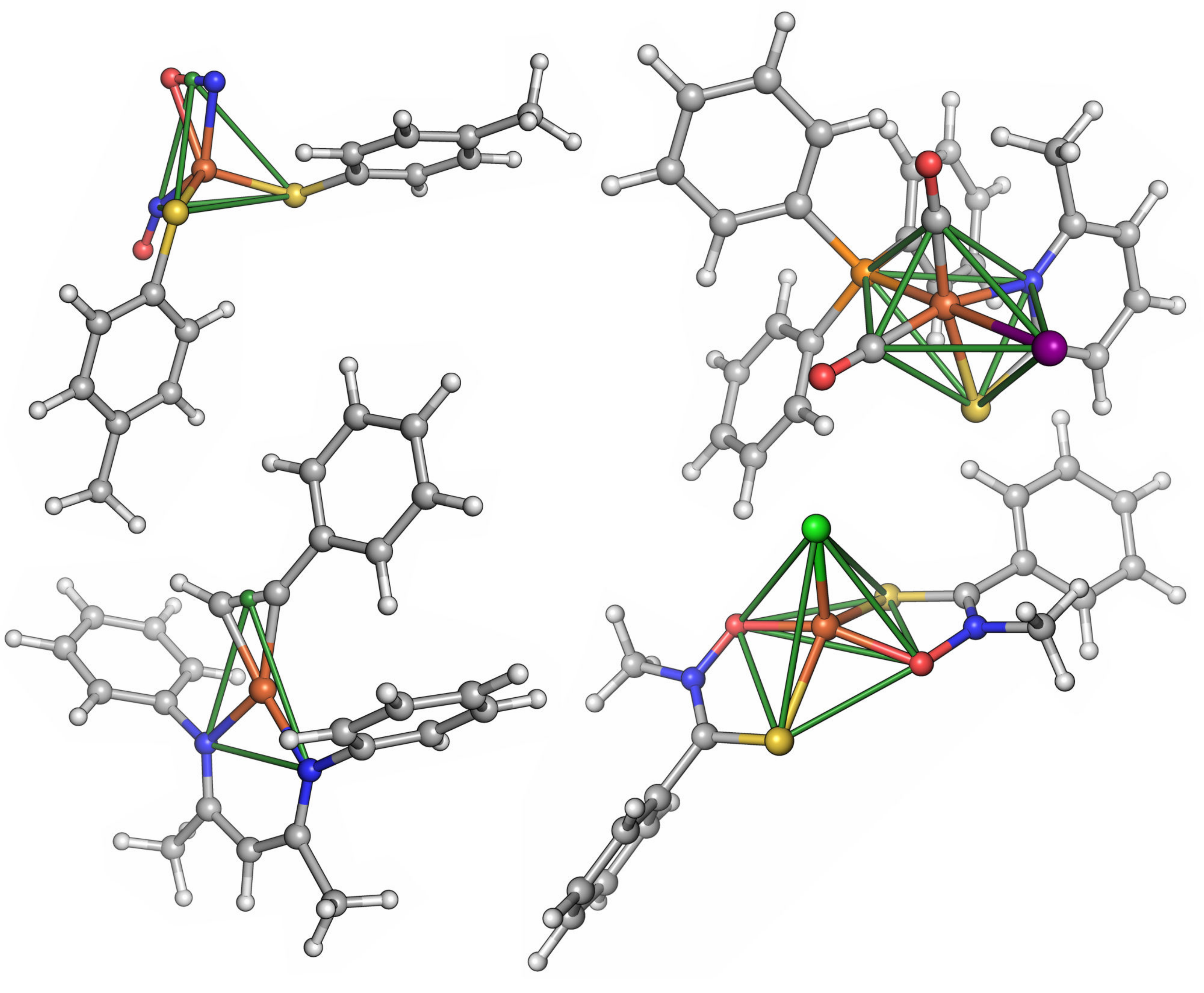}
  \caption{Clockwise from top left: Compounds \#28, \#34, \#26, and \#14 from
  the dataset of Ref.~\cite{Proppe2017}, with the interpreted
  polyhedral shape of iron atoms highlighted in green.
  }\label{fig:demonstration_proppe}
\end{figure}

In order to show that \textsc{Molassembler} can accurately capture
stereochemistry, we have interpreted the full set of compounds collected by
Proppe and Reiher~\cite{Proppe2017}, which comprises 44 iron compounds with
diverse bonding patterns, into the presented molecular model. Some selected
examples are shown in Figure~\ref{fig:demonstration_proppe} and the full
compound set is arrayed in Figures S2--S9. This is a significant stress test
since the test set contains a varied set of coordination polyhedra, some with
significant distortions. In our judgment, only in the case of compound \#26
(displayed bottom right in Figure~\ref{fig:demonstration_proppe}) is the shape
classification result dubious: A trigonal bipyramid is classified instead of a
square pyramid. Apart from generating molecular models, we have also generated
conformers of stereoisomers for a selected compound to exhibit more of the
program interface. For further details, see the Supporting Information.

\section{Conclusions}\label{sec:conclusion}

The molecular model applied in \textsc{Molassembler} is capable of accurately
representing many inorganic molecular complexes. The data representation of
stereoisomerism generalizes to complex shapes and ligands, permitting an
abstract high-level approach to molecular structure. Transferable
stereodescriptors are devised for any combination of shape and abstract binding
case, including haptic and multidentate ligands. Molecule instances are
interpretable from Cartesian coordinates, constructible from various
cheminformatics formats, editable, canonicalizeable, serializeable, and
comparable. Both stochastic and directed conformer generation are implemented.
Despite the few shortcomings in its first version, \textsc{Molassembler}
is already a feature-rich, general-purpose program that enables programs to
expand into parts of organometallic and inorganic chemical space. In future
releases, we will address the existing shortcomings, particularly with regard
to ranking and handling of aromaticity. \textsc{Molassembler} is open-source
under the BSD 3-clause license and available in the SCINE project~\cite{Brunken2019}.

\section*{Acknowledgments}\label{sec:acknowledgments}

This work has been financially supported by ETH Z\"urich (ETH-38 17-1). 

\begin{appendix}
\section{Library architecture and dependencies}\label{sec:code_architecture}

\textsc{Molassembler} is a C++ library written in the C++14 language standard. It
has multiple dependencies: The SCINE Utilities~\cite{Brunken2019} form the data
exchange formats prevalent throughout our software. Linear algebra computations are
mediated by the Eigen~\cite{Guennebaud2010} library.
RingDecomposerLib~\cite{Flachsenberg2017, Kolodzik2012} provides graph
cycle perception with a chemical perspective, nauty~\cite{McKay2014} serves
graph canonicalization purposes, JSON for Modern C++~\cite{Lohmann2013} offers
serialization and deserialization to JavaScript Object Notation (JSON), and
  several
Boost libraries~\cite{BoostAlgorithms2017, BoostBimap2017, BoostFilesystem2017, BoostFunctional2017, BoostGraph2002, BoostInteger2017, BoostMath2017, BoostOptional2017, BoostOutcome2019, BoostOutcome2019, BoostPhoenix2017, BoostProcess2017, BoostRange2017, BoostSpirit2017, BoostVariant2017} form the basis of the molecular graph
representation and offer relevant algorithms and data structures.

Optional dependencies offer functionality if available: If one wishes to compile
the available Python bindings, the C++ library pybind11~\cite{Wenzel2017} will
be required. Parallelization of core library algorithms will be available if
OpenMP~\cite{Dagum1998} is supported by the available compiler and linker.
\textsc{Molassembler} will accept SMILES~\cite{Weininger1988} and InChI~\cite{Stein2003}
input if OpenBabel's~\cite{openbabel2011} chemical file format conversion
utility is found at runtime.

Configuration of optional components, compiler-agnostic building and
installation are supported cross-platform by modern CMake scripts.

\textsc{Molassembler} draws from two sub-libraries, each serving singular
purposes. One is responsible for defining the local shapes and calculating
properties and transition mappings between them. The other is responsible for
the symbolic permutation required in stereopermutation. Further 
sub-libraries model cyclic polygons, implement the simplified GOR1 shortest path
algorithm for Boost Graph Library data structures, and provide data structures
and algorithms for compile-time computation. The main body of
\textsc{Molassembler} itself is split into the application programming interface
and the private implementations thereof. Overall, \textsc{Molassembler} consists
of roughly sixty thousand lines of code and twenty thousand lines of comments
and documentation. With the open-source release we provide an installation and user's manual as well as separate technical
documentations for the C++ library and the Python module (see https://scine.ethz.ch/download/molassembler).

\newpage
\section{Supporting Information}

\subsection{Shape classification algorithms}

Algorithms to classify point clouds into idealized polyhedral shapes that we investigated for our purposes
are listed in Table~\ref{table:shapealgs}.

\begin{table}[h]
  \caption{List of shape classification algorithms}\label{table:shapealgs}
  \begin{tabular}{l l l}
    \toprule
    Name & Listing & Reference\\
    \midrule
    Continuous shape measure &\ref{alg:cshm} &\cite{Pinsky1998}\\
    Continuous shape measure with heuristics &\ref{alg:cshm_heuristics} &\\
    Continuous shape measure minimum distortion path deviation
    &\ref{alg:cshmmindistpath} &\cite{Alvarez2005}\\
    Continuous symmetry measure &\ref{alg:csm_p1},\,\ref{alg:csm_p2}
    &\cite{Zabdrodsky1993}\\
    Angular square deviation &\ref{alg:angsqdev} & \\
    Angular deviation and geometry index hybrid &\ref{alg:angdevgeoindhybrid}
    &\cite{yang2007structural, Okuniewski2015}\\
    \bottomrule
  \end{tabular}
\end{table}

Note, however, that several
algorithms in this table refer to undefined functions. These are briefly explained in the
accompanying text. All algorithms listed here have been implemented in
\textsc{Molassembler}. For further details, we refer
to the source code and its accompanying documentation.

\subsubsection{Continuous shape measure-derived algorithms}

\begin{algorithm}
  \caption{Continuous shape measure}\label{alg:cshm}
  \begin{algorithmic}[1]
    \Function{Normalize}{$\mathbf{R}$}\Comment{$\mathbf{R}$ is $N$ Cartesian
    coordinates $\vec{r}_i$}
    \State$\bar{r} = \frac{1}{N} \sum_i^N \vec{r}_i$\Comment{$\bar{r}$ centroid}
    \State$\mathbf{S}\gets \mathbf{R} - \bar{r}$
    \State$l\gets0$
    \ForAll{$\vec{s}_i\in\mathbf{S}$}
      \State$l\gets \max(l,\ |\vec{s}_i|)$
    \EndFor{}
    \State\textbf{return} $\frac{1}{l}\cdot\mathbf{S}$\Comment{Rescale so largest
    vector is unit length}
    \EndFunction{}\\

    \Function{ShapeMeasure}{$\mathbf{P}, s$}\Comment{$\mathbf{P}$ is $N$
    Cartesian coordinates, $s$ a shape}
      \State\textbf{require} $N=$\ \Call{VertexCount}{$s$}
      \State$\mathbf{C}\gets$ \Call{Normalize}{$\mathbf{P}$}\Comment{Point cloud
      coordinates}
      \State$\mathbf{S}\gets$ \Call{Coordinates}{$s$}\label{alg:cshm:coords}\Comment{Idealized shape coordinates}
      \State$s_{\min}\gets \infty$\Comment{Minimal shape measure}
      \ForAll{$\sigma\in\textsc{Permutations}(N)$}
        \State$\mathbf{S}_\sigma\gets \sigma\mathbf{S}$\Comment{Apply permutation}
        \State$\mathbf{R}\gets$ \Call{QuaternionFit}{$\mathbf{C}, \mathbf{S}_\sigma$}
        \State$\mathbf{S}_{R}\gets\mathbf{R}\mathbf{S}_\sigma$\Comment{Apply rotation}
        \State$s\gets\underset{a}{\min}\sum_i^N |\vec{c}_i-a\vec{s}_{R,
        i}|^2$\Comment{Minimize over isotropic scaling}
        \State$s_{\min}\gets\min(s_{\min}, s)$
      \EndFor{}
      \State\textbf{return} $100 \cdot s_{\min} \div \sum_i^N
      |\vec{c}_i|^2$\Comment{Normalized shape measure}
    \EndFunction{}\\

    \Function{ClassifyShapeByCShM}{$\mathbf{P}$}
      \State\textbf{return}
      $\underset{s\in\textsc{Shapes}(N)}{\mathrm{argmin}}$ \Call{ShapeMeasure}{$\mathbf{P},
      s$}\label{alg:cshm:shapes}
    \EndFunction{}
  \end{algorithmic}
\end{algorithm}

The continuous shape measure calculation is presented in Alg.~\ref{alg:cshm}.
For each possible mapping of the $N$ vertices of the point cloud to the vertices
of the reference polyhedron, this algorithm involves separate minimizations over
a rotation matrix and an isotropic scaling parameter. The minimization
over the rotation matrix is here explicitly carried out by a quaternion fit,
but is in principle open to other optimization methods. There are three
undefined functions in this listing. \texttt{Coordinates} yields the idealized
coordinates of its shape parameter. The function \texttt{Shapes} returns the set
of all shapes with a number of vertices equal to its integer parameter. Finally,
the \texttt{VertexCount} function returns the number of vertices of a shape. In
brief, this algorithm classifies a point cloud as that shape of matching number
of vertices for which the continuous shape measure is minimal.

\begin{algorithm}
  \caption{Continuous shape measure, with heuristics}\label{alg:cshm_heuristics}
  \begin{algorithmic}[1]
    \Function{ShapeMeasureHeuristics}{$\mathbf{P}, s$}\Comment{As in Alg.~\ref{alg:cshm}}
      \State\textbf{require} $N=$\ \Call{VertexCount}{$s$}
      \State$\mathbf{C}\gets$ \Call{Normalize}{$\mathbf{P}$}\Comment{Point cloud
      coordinates}
      \State$\mathbf{S}\gets$ \Call{Coordinates}{$s$}\Comment{Idealized shape coordinates}
      \State$s_{\min}\gets \infty$\Comment{Minimal shape measure}
      \State$\sigma_{\min}\gets\emptyset$
      \ForAll{$i\neq j\neq k\neq l\neq m\ | \ \{i, j, k, l, m\} \in \{1,
      2, \ldots, N\}$}
        \State$\sigma_p\gets\begin{pmatrix}1&2&3&4&5\\i&j&k&l&m\end{pmatrix}$\Comment{Partial
          point mapping from $\mathbf{C}$ to $\mathbf{S}$}
        \State$\mathbf{R}\gets$ \Call{PartialQuaternionFit}{$\mathbf{C},
        \mathbf{S}, \sigma_p$}
        \State$\mathbf{S}_p\gets\mathbf{R}\mathbf{S}$
        \State$s_p\gets\sum_i^5 |\vec{c}_i-\vec{s}_{p, \sigma_p(i)}|^2$
        \If{$s_p > s_{\min}$}\Comment{Partial fit penalty greater than best?}
          \State\textbf{continue}\Comment{Skip this permutation}
        \EndIf{}
        \State$U\gets\{6, 7, \ldots, N\}$\Comment{Points of $\mathbf{C}$ not
        mapped by $\sigma_p$}
        \State$V\gets\{1, 2, \ldots, N\} \cap \sigma_p$\Comment{Points of
        $\mathbf{S}$ not mapped by $\sigma_p$}
        \State$a_{i,\,j} \gets |\vec{c}_{U_i} - \vec{s}_{p, V_j}|^2$\Comment{Cost matrix
        of adding a mapping}
        \State$\sigma_{c, \min}\gets\underset{\sigma_c}{\mathrm{argmin}}
        \sum_i^{N-5}a_{i,\,\sigma_c(i)}\ |\ \sigma_c \in
        \textsc{Permutations}(N-5)$
        \State$\sigma\gets\sigma_p\cup\begin{pmatrix}U_1&U_2&\cdots&U_{N-5}\\V_{\sigma_c(1)}&V_{\sigma_c(2)}&\cdots&V_{\sigma_c(N-5)}\end{pmatrix}$\Comment{Merge
          cost min.\ perm.}
        \State$\mathbf{R}\gets$ \Call{QuaternionFit}{$\mathbf{C}, \sigma\mathbf{S}$}
        \State$\mathbf{S}_R\gets\mathbf{R}\sigma\mathbf{S}$
        \State$s\gets\sum_i^N |\vec{c}_i - \vec{s}_{R, i}|^2$
        \If{$s < s_{\min}$}
          \State$s_{\min}\gets s$
          \State$\sigma_{\min}\gets\sigma$
        \EndIf{}
      \EndFor{}
      \State$\mathbf{R}\gets$ \Call{QuaternionFit}{$\mathbf{C}, \sigma_{\min}\mathbf{S}$}
      \State$\mathbf{S}_{R}\gets\mathbf{R}\sigma_{\min}\mathbf{S}$\Comment{Apply rotation}
      \State$s\gets\underset{a}{\min}\sum_i^N |\vec{c}_i-a\vec{s}_{R,
      i}|^2$\Comment{Minimize over isotropic scaling}
      \State\textbf{return} $100\cdot s\div\sum_i^N
      |\vec{c}_i|^2$\Comment{Normalized shape measure}
    \EndFunction{}
  \end{algorithmic}
\end{algorithm}

A number of heuristics can be added, of which the authors themselves suggest
several, such as pre-pairing vertices. Our benchmarks indicate that the
quaternion fit is the most time intensive step within the loop. We chose
several changes, presented in Alg.~\ref{alg:cshm_heuristics}. For one, we
performed the minimization over isotropic scaling only on the best permutation,
outside the main loop. Another change was that we consider quaternion fits
converged after five vertex pairs have been mapped. For this partial mapping, a
quaternion fit is performed. Afterwards a cost matrix is set up for all
remaining possible vertex mappings, and the minimal cost permutation is merged
with the initial five vertex pairs. Only then are a full quaternion fit
performed and the square deviations calculated. Shape classification from a
point cloud proceeds as before, except that the shape measure calculation
algorithm is replaced with the presented \texttt{ShapeMeasureHeuristics} method.

\begin{algorithm}
  \caption{Continuous shape measure minimum distortion path
  deviation}\label{alg:cshmmindistpath}
  \begin{algorithmic}[1]
    \Function{PathDeviation}{$\mathbf{X}, s_1, s_2$}\Comment{$\mathbf{X}$
    is $N$ Cart.\ coord., $s_i$ are shapes}
      \State$\theta\gets\arcsin\left(\sqrt{\textsc{ShapeMeasure}(\textsc{Coordinates}(s_1),
      s_2)}\div 10\right)$
      \State$C\gets\textsc{Normalize}(\mathbf{X})$
      \State$\alpha\gets\arcsin\left(\sqrt{\textsc{ShapeMeasure}(C, s_1)}\div
      10\right)$
      \State$\beta\gets\arcsin\left(\sqrt{\textsc{ShapeMeasure}(C, s_2)}\div
      10\right)$
      \State\textbf{return} $\left(\alpha+\beta\right)\div\left(\theta-1\right)$
    \EndFunction\\
    \Function{ClassifyShapeByPathDev}{$\mathbf{X}$}\Comment{$\mathbf{X}$
    is $N$ Cart.\ coord.}
      \State$p\gets\underset{s_1 \neq s_2\in\textsc{Shapes}(N)}{\mathrm{argmin}}$ 
        \Call{PathDeviation}{$\mathbf{X}, s_1,
      s_2$}\Comment{$p$ is a pair of shapes}
      \If{\Call{ShapeMeasure}{$\mathbf{X}, p_1$} $<$
      \Call{ShapeMeasure}{$\mathbf{X}, p_2$}}
        \State\textbf{return} $p_1$
      \Else{}
        \State\textbf{return} $p_2$
      \EndIf{}
    \EndFunction{}
  \end{algorithmic}
\end{algorithm} 

Lastly, the minimum distortion path deviation shape classification algorithm is
based on the concept of minimum distortion paths between two shapes. It proceeds
by calculating the distance of the point cloud from the minimum distortion path
between all pairs of viable shapes. For the pair of shapes for which the
distance from the distortion path is minimal, the shape with the lower shape
measure is chosen.

\subsubsection{Continuous symmetry measure}

\begin{algorithm}
  \caption{Continuous symmetry measure}\label{alg:csm_p1}
  \begin{algorithmic}[1]
    \Function{ElementPartitions}{$S$}\Comment{$S$ set of symmetry
    elements}
      \State$G\gets\emptyset$\Comment{Set of element partitions}
      \State$P\gets\emptyset$\Comment{Set of probe points}
      \Procedure{ProcessPoint}{$\vec{p}$}
        \If{$\vec{p}\in P$}
          \State\textbf{return}
        \EndIf{}
        \State$G\gets G\,\cup\,$\Call{Partition}{$S,
        \vec{p}$}\Comment{Partition $S$ by result of application
        to $\vec{p}$}
        \State$P\gets P\cup \vec{p}$
      \EndProcedure{}

      \State\Call{ProcessPoint}{$\vec{e}_z$}
      \State\Call{ProcessPoint}{$\vec{e}_z + \Delta\vec{e}_x$}\Comment{With $\Delta <
      1$}
      \State\Call{ProcessPoint}{$\vec{e}_x$}
      \State\Call{ProcessPoint}{$\vec{e}_y$}
      \ForAll{$\vec{v}\in\{\textsc{Axis}(\mathbf{E})\ |\ \mathbf{E}\in S\}$}\Comment{\textsc{Axis}
      yields unaltered line, if exists}
        \State\Call{ProcessPoint}{$\vec{v}$}
      \EndFor{}
      \State\textbf{return} $P$
    \EndFunction{}\\

    \Function{CSM}{$\mathbf{X}, P, E$}
      \State\textbf{require} $(|P| = |E|) \land \forall(|P_i| = |E_i|)$
      \State$M\gets|P|$
      \State$N\gets|P_1|$
      \State$v_{\min}\gets\infty$
      \ForAll{$\sigma\in\textsc{Permutations}(M)$}
        \State$\bar{x}\gets\frac{1}{MN}\sum_i^M\sum_j^N\mathbf{E}_{i,j}\vec{x}_{\sigma(P_i)}$\Comment{Fold
        points by symmetry elements}
        \State$v\gets\sum_i|\mathbf{E}_{i,1}^{-1}\bar{x}-\vec{x}_{\sigma(P_i)}|^2$\Comment{Unfold
        and sum square difference norms}
        \State$v_{\min}\gets\min(v_{\min}, v)$
      \EndFor{}
      \State\textbf{return} $100\cdot v_{\min}\div M$
    \EndFunction{}
      \State\Comment{Continued in Alg.~\ref{alg:csm_p2}}
      \algstore{alg:csm}
  \end{algorithmic}
\end{algorithm}

\begin{algorithm}
  \caption{Continuous symmetry measure (continued from Alg.~\ref{alg:csm_p1})}\label{alg:csm_p2}
  \begin{algorithmic}[1]
    \algrestore{alg:csm}
    \Function{AlignedSymmetryMeasure}{$\mathbf{X}, g$}
      \State$E\gets$ \Call{Elements}{$g$}
      \State$P\gets$ \Call{ElementPartitions}{$E$}
      \State$a_k\gets|P_k|$\Comment{$a_k$ coefficients of diophantine eq.}
      \If{$N > |E|$}
        \State$a\gets a^\frown|E|$\Comment{Extend
        coefficients by order of point group}
        \State$P\gets P\cup\{E\}$\Comment{Extend
        partitions by set of all elements}
      \EndIf{}
      \State$v_{\min}\gets\infty$
      \ForAll{sequences $x\ |\ \sum_i a_i x_i = N, \forall x_i\in\mathbb{N}_0$}\label{alg:csm:diophantine}\Comment{Solve diophantine}
        \ForAll{partitions $P_X$ of $N$ points into $x_i$ sets of size
        $a_i$}
          \State$v\gets\frac{1}{|P_X|}\sum_i\textsc{CSM}(\mathbf{X}, P_{X, i}, P_i)$
          \State\Comment{$P_i$ a partition of into sets of size
          equal to $P_{X, i}$}
          \State$v_{\min}\gets\min(v_{\min}, v)$
        \EndFor{}
      \EndFor{}
      \State\textbf{return} $v_{\min}$
    \EndFunction{}\\

    \Function{SymmetryMeasure}{$\mathbf{X}, g$}\Comment{$\mathbf{X}$ is $N$
    Cart.\ coord., $g$ a point group}
      \State\textbf{return}
      $\underset{\mathbf{R}\in\mathbf{SO(3)}}{\min}$
      \Call{AlignedSymmetryMeasure}{$\mathbf{R}\mathbf{X}$}
    \EndFunction{}\\

    \Function{ClassifyShapeByCSM}{$\mathbf{X}$}\Comment{$\mathbf{X}$ is $N$
    Cartesian coordinates}
      \State$\mathbf{C}\gets$ \Call{Normalize}{$\mathbf{X}$}
      \State\textbf{return} $\underset{s\in\textsc{Shapes}(N)}{\mathrm{argmin}}$
      \Call{SymmetryMeasure}{$\mathbf{C}$, \textsc{PointGroup}$(s)$}
    \EndFunction{}
  \end{algorithmic}
\end{algorithm}

An algorithm for the calculation of the continuous symmetry measure is presented
in Algs.~\ref{alg:csm_p1} and \ref{alg:csm_p2}. There are several undefined
functions here. The function \texttt{PointGroup} returns the point group of its
shape parameter. \texttt{Elements} returns the full set of symmetry elements of
a point group. \texttt{Partition}, which accepts a set of symmetry elements and
a vector, groups symmetry elements by identical results of applying the 
transformation matrix of the symmetry elements to the vector argument.

The continuous symmetry measure calculation implementation presented in
Algs.~\ref{alg:csm_p1} and \ref{alg:csm_p2} is admittedly clumsy. The
minimization over all rotation matrices of a function, which is as costly to
evaluate as \texttt{AlignedSymmetryMeasure}, is difficult to streamline. The
algorithm presented here is noticeably more complex than the sketch laid out for
the measure's calculation from the original authors, because the sketch lays out
the optimal path through a full algorithm. The additional complexities, such as
exploring all possible solutions of the diophantine equation
at~\algref{alg:csm_p2}{alg:csm:diophantine} and all possible partitions of the
$N$ points according to the current solution of the diophantine immediately
thereafter, are neccessary. Perhaps the complexity they incur can be
significantly reduced through application of heuristics. This is not explored
here, however.

\subsubsection{Angular deviation variations}

\begin{algorithm}
  \caption{Angular square deviation}\label{alg:angsqdev}
  \begin{algorithmic}[1]
    \Function{AngularDeviation}{$\mathbf{X}, s$}
      \State\textbf{return}
      $\underset{\sigma\in\textsc{Rotations}(s)}{\min}\sum_{i<j}\left[\textsc{Angle}(\mathbf{X},
      i, j) - \textsc{IdealAngle}(s, \sigma(i),
      \sigma(j))\right]^2$
    \EndFunction{}\\

    \Function{ClassifyShapeByAngularDeviation}{$\mathbf{X}$}\Comment{$\mathbf{X}$
    is $N$ Cart.\ coord.}
      \State$\mathbf{C}\gets$ \Call{Normalize}{$\mathbf{X}$}
      \State\textbf{return} $\underset{s\in\textsc{Shapes}(N)}{\mathrm{argmin}}$
      \Call{AngularDeviation}{$\mathbf{C}, s$}
    \EndFunction{}
  \end{algorithmic}
\end{algorithm}

The pure angular deviation shape classification algorithm is based only on the
central angles of vertices within their polyhedral shape. It requires knowing
which vertex of the point cloud is the central vertex of the polyhedron since
that is the base point with which the angles are calculated. The algorithm is
shown in Alg.~\ref{alg:angsqdev}. The undefined functions are \texttt{Angle} and
\texttt{IdealAngle}, the latter of which returns the angle between the vertices
from the idealized coordinates of the shape.

\begin{algorithm}
  \caption{Angular deviation and geometry index hybrid}\label{alg:angdevgeoindhybrid}
  \begin{algorithmic}[1]
    \Function{ClassifyShapeByHybrid}{$\mathbf{X}$}\Comment{$\mathbf{X}$
    is $N$ Cartesian coordinates}
      \State$\mathbf{C}\gets$ \Call{Normalize}{$\mathbf{X}$}
      \State$S\gets\emptyset$\Comment{Set of excluded shapes}
      \If{$N = 4$}
        \State$\tau\gets$ \Call{Tau4Prime}{$\mathbf{C}$}
        \If{$\tau < 0.12$}
          \State$S\gets\{\textrm{Seesaw}, \textrm{Tetrahedron}\}$
        \ElsIf{$0.12\leq\tau<0.62$}
          \State$S\gets\{\textrm{Square}, \textrm{Tetrahedron}\}$
        \ElsIf{$0.62\leq\tau$}
          \State$S\gets\{\textrm{Square}, \textrm{Seesaw}\}$
        \EndIf{}
      \ElsIf{$N = 5$}
        \State$\tau\gets$ \Call{Tau5}{$\mathbf{C}$}
        \If{$\tau < 0.5$}
          \State$S\gets\{\textrm{Trigonal bipyramid}\}$
        \ElsIf{$\tau > 0.5$}
          \State$S\gets\{\textrm{Square pyramid}\}$
        \EndIf{}
      \EndIf{}
      \State$V\gets\textsc{Shapes}(N)\cap S$\Comment{Set of viable shapes}
      \State\textbf{return} $\underset{s\in V}{\mathrm{argmin}}$
      \Call{AngularDeviation}{$\mathbf{C}, s$}
    \EndFunction{}
  \end{algorithmic}
\end{algorithm}

Finally, the hybrid algorithm with geometry indices $\tau_4^{\prime}$ and
$\tau_5$ as shown in Alg.~\ref{alg:angdevgeoindhybrid} leverages the geometry
indices to exclude the most unlikely shapes according to their value range
definitions.

\subsection{Stereopermutations}

\subsubsection{Atom stereopermutation enumeration}

An individual atom stereopermutation is represented as the product type of a
sequence of ranking characters $c_k$ and a set of ordered pairs $L$ representing
links between shape vertices. The sequence of ranking characters has length
equal to the number of vertices of the shape and represents a mapping from a
vertex position to a ranking character. Relational ordering of
stereopermutations is defined via sequential lexicographical comparison of
ranking characters and vertex links.

The enumeration of atom stereopermutations with tracking of relative statistical
occurrence weights proceeds as laid out in Alg.~\ref{alg:stereop_enum}. Note
that a stereopermutation is represented as a tuple.

\begin{algorithm}
  \caption{Atom stereopermutation enumeration}\label{alg:stereop_enum}
  \begin{algorithmic}[1]
    \Function{ApplyPermutation}{$(c_k, L), \sigma$}\Comment{$\sigma$ a
    permutation}
      \State\textbf{return} $(\sigma c_k, \{(\inf \{\sigma(l_1), \sigma(l_2)\}, \sup
      \{\sigma(l_1), \sigma(l_2)\})\ |\ l\in L\})$
    \EndFunction\\

    \Function{AllRotations}{$(c_k, L), s$}\Comment{$s$ a shape}
      \State\textbf{return} $\{\textsc{ApplyPermutation}((c_k, L),
      \sigma)\ |\ \sigma\in\textsc{Rotations}(s) \}$
    \EndFunction\\

    \Function{Enumerate}{$c_k, L, s$}\Comment{$s$ a shape}
      \State$u_1\gets\inf \textsc{AllRotations}((c_k, L), s)$\Comment{Sequence of stereopermutations}
      \State$v_1\gets1$\Comment{Sequence of occurrence counts}
      \ForAll{$\sigma\in\textsc{Permutations}(\textsc{VertexCount}(s))$}
        \State$r\gets\inf\textsc{AllRotations}(\textsc{ApplyPermutation}((c_k,
        L), \sigma))$
        \If{$r\in u_k$}
          \State$i\gets u_k^{-1}(r)$\Comment{Get index of $r$ in sequence $u$}
          \State$v_i\gets v_i + 1$\Comment{Increment occurrence count}
        \Else{}
          \State$u\gets u^{\frown}r$\Comment{Extend sequence of permutations}
          \State$v\gets v^{\frown}1$\Comment{Extend sequence of counts}
        \EndIf{}
      \EndFor{}
      \State\textbf{return} $\{(u_i, v_i\div\textsc{Gcd}(v_k))\ |\ i\in\{1, \ldots, |u|\}\}$
    \EndFunction{}
  \end{algorithmic}
\end{algorithm}

\subsubsection{Bond stereopermutation}

Rotational isomerism along multiple bond-order bonds is modeled with bond
stereopermutations. Each side of the bond must have an atom stereopermutator
with an assigned stereopermutation. Indeterminate atom stereopermutators cannot
map their graph substituents to shape vertices, which is necessary for bond
stereopermutation enumeration. The two atom stereopermutators each contribute a
shape, a mapping of graph substituents to shape vertices, and their local
substituent rankings. 

First, representational degrees of freedom are removed to ensure that the
stereopermutation enumeration algorithm generates transferable
stereodescriptors. Any shape vertex may fuse, but shape vertices that are
rotationally interconvertible yield the same set of permutation, but in a
different order. Additionally, shape vertex enumeration is ordered arbitrarily.
These degrees of freedom are removed by rotating the fused vertex to the
algebraically smallest shape vertex of its set of rotationally interconvertible
vertices. For example, the octahedron has a single set of interconvertible
vertices, but the square pyramid has two: The equatorial set of four vertices,
and the singleton set containing the apical vertex.

The degree of freedom reduction is laid out in
Algs.~\ref{alg:bstereop_dof_aux},\,\ref{alg:bstereop_dof}. \texttt{Orbits}
collects sets of vertices that interconvert in a permutation.
\texttt{MergePermutations} is an undefined function that merges overlapping
orbits. \texttt{VertexGroups} yields the vertex sets that are interconvert
rotationally in a given shape. The last auxiliary function,
\texttt{IndexOfPermutation}, establishes ordering of permutations. With these
auxiliaries, the representational freedom at each side of the double bond,
consisting of a shape, a fused vertex and the ranking characters at each vertex
of the shape, can be reduced as shown in Alg.~\ref{alg:bstereop_dof}.

\begin{algorithm}
  \caption{Bond stereopermutation degree of freedom reduction
  auxiliaries}\label{alg:bstereop_dof_aux}
  \begin{algorithmic}[1]
    \Function{Orbits}{$\sigma$}
      \State$O\gets\emptyset$
      \ForAll{$i\in\{1, \ldots, |\sigma|\}$}
        \If{$i\in O_j$ for any $j$}
          \State$S\gets\{1\}$
          \State$j\gets\sigma(i)$
          \While{$j\neq i$}
            \State$S\gets S\cup\{j\}$
            \State$j\gets\sigma(j)$
          \EndWhile{}
          \State$O\gets O\cup \{S\}$
        \EndIf{}
      \EndFor{}
      \State\textbf{return} $O$
    \EndFunction\\

    \Function{VertexGroups}{$s$}
      \State\textbf{return} $\textsc{Orbits}(\textsc{MergePermutations}(\textsc{Rotations}(s)))$
    \EndFunction\\

    \Function{IndexOfPermutation}{$\sigma$}
      \State$x\gets0$
      \State$a\gets1$\Comment{Factor}
      \State$b\gets2$\Comment{Position}
      \ForAll{$i\in\{N-1, N-2, \ldots, 1\}$}
        \State$l\gets0$\Comment{Number of larger successors}
        \ForAll{$j\in\{i+1, i+2, \ldots, N\}$}
          \If{$\sigma(j) < \sigma(i)$}
            \State$l\gets l+1$
          \EndIf{}
        \EndFor{}

        \State$x\gets x+l\cdot a$
        \State$a\gets a\cdot b$
        \State$b\gets b+1$
      \EndFor{}
      \State\textbf{return} $x$
    \EndFunction{}
  \end{algorithmic}
\end{algorithm}

\begin{algorithm}
  \caption{Bond stereopermutation degree of freedom
  reduction}\label{alg:bstereop_dof}
  \begin{algorithmic}[1]
    \Function{ReduceDOF}{$s, v, c_k$}\Comment{shape $s$, fused vertex $v$,
    ranking chars. $c_k$}
      \State$v_{\min}\gets\inf\ \{X\ |\ X\in\textsc{VertexGroups}(s)\land v\in
      X\}$\Comment{Smallest in orbit}
      \State$P\gets\{\sigma\ |\
      \sigma\in\textsc{Rotations}(s)\land\sigma(v)=v_{\min}\}$
      \State$\sigma_{\min}\gets\underset{\sigma\in
      P}{\mathrm{argmin}}\ \textsc{IndexOfPermutation}(\sigma)$
      \State\textbf{return} $(\sigma_{\min}(v), \sigma_{\min}c_k)$
    \EndFunction{}
  \end{algorithmic}
\end{algorithm}

The enumeration of bond stereopermutations is less amenable to concise
expression in mathematical terms than existing algorithm listings, but can be
sketched well in words. After reducing the representational degrees of freedom
of the state at each side of the bond, the group of shape vertices with minimal
angle to the fused vertices is selected at each side. These groups' mutual
interaction is the first-order approximation to the dihedral potential and the
interaction of vertices further back is neglected entirely. Next, the vertices
of each set are ordered by their ranking characters and their value.

If the ranking characters of the set of shape vertices closest to the fused
vertex at either side is a singleton set, i.e.\ all of these vertices have the
same ranking character, then rotation around the bond being modeled will be isotropic
and there will be only one stereopermutation.

Otherwise, the dihedral angles are explicitly modeled by arranging both shapes
along the fused vertices in a joint coordinate system and sequentially aligning
pairs of vertices across both vertex sets in the previously established order
and measuring all pairs of dihedral angles at each alignment.

\subsubsection{Bond stereopermutation feasibility}

The plane that the cyclic polygon expands into is modeled as follows: We define
the dihedral point sequence $\vec{A}, \vec{B}, \vec{C}, \vec{D}$ with angles
$\alpha = \angle ABC, \beta = \angle BCD$ and dihedral $\varphi = \angle ABCD$
with $\left|AB\right| = a, \left|BC\right| = b, \left|CD\right| = c$ and
$\alpha,\beta \in \left[0, \pi\right], \varphi \in \left[-\pi, \pi\right]$.

We choose the following positions for the points without loss of generality:
\begin{align}
  &\vec{A} = \mathbf{R}_{z}(\alpha) a \vec{e}_x\\
  &\vec{B} = \vec{0}\\
  &\vec{C} = b \vec{e}_x\\
  &\vec{D} = \mathbf{R}_{x}(\varphi)\left[\vec{C} + \mathbf{R}_z(\pi -
  \beta)c\vec{e}_x\right],
\end{align}
where $\vec{e}_x$ is the unit vector along the $x$-axis and $\mathbf{R}_i$ is the
rotation matrix along the axis $i$. The dihedral distance is the length of the
line segment $\overline{AD}$. In order to find the shortest distance between the
line segments $\overline{AD}$ and $\overline{BC}$, we define:
\begin{align}
  \overline{BC}: \vec{r}(\lambda) 
  &= \vec{B} + \lambda\left(\vec{C}-\vec{B}\right)
  = \lambda\vec{C},
  \lambda \in \left[0, 1\right]\\
  \overline{AD}: \vec{s}(\mu) 
  &= \vec{A} + \mu\left(\vec{D}-\vec{A}\right),
  \mu \in \left[0, 1\right]
\end{align}
The shortest distance between the line segments must be orthogonal to both line
segments' direction vectors:
\begin{align}
  \left[\lambda\vec{C} - \left(\vec{A} + \mu(\vec{D}-\vec{A})\right)\right]
  &\cdot\vec{C} = 0\\
  \left[\lambda\vec{C} - \left(\vec{A} +
  \mu(\vec{D}-\vec{A})\right)\right]
  &\cdot\left(\vec{D}-\vec{A}\right) = 0
\end{align}
Solving this system of equations yields:
\begin{align}
  \mu_0 &= - \frac{A_y S_y + A_z S_z}{S_y^2 + S_z^2}\\
  \lambda_0 &= \frac{A_x + \mu_0 S_x}{b}
\end{align}
with $\vec{S} = \vec{D} - \vec{A}$.
These solutions do not satisfy their conditions yet, and must be adjusted
slightly:
\begin{align}
  \lambda_m &= \min\left(\max(\lambda_0, 0), 1\right)\\
  \mu_m &= 
  \begin{cases}
    -\frac{\displaystyle A_x}{\displaystyle S_x} & \lambda_0 \leq 0\\
    \frac{\displaystyle b - A_x}{\displaystyle S_x} & \lambda_0 \geq 1\\
    \lambda_0 & \textrm{else}
  \end{cases}.
\end{align}
The plane in which the cyclic polygon expands can then be defined through the three
points $\vec{A}, \vec{D}$, and $\vec{r}(\lambda_m)$.
For $\varphi = \pm \pi$, these points become collinear and the plane definition
is not uniquely defined. Under these circumstances, the plane can instead be
defined using the point $\vec{A}$ and the two vectors
$\left(\vec{D}-\vec{A}\right)$ and $\vec{e}_z$.

\subsection{Isomorphism and Canonicalization}

The graph data structures and several related algorithms in
\textsc{Molassembler} are provided by the Boost Graph Library from the Boost
family of C++ libraries~\cite{BoostGraph2002}. Molecule comparison is implemented as
a colored graph isomorphism that is modular in vertex coloring. The following
types of information can be collected to color a vertex: the element type of the
atom, its bond orders to adjacent vertices, the shape of its atom
stereopermutator (if present), and the stereopermutation of its stereopermutator
(if present). Note that indeterminate atom stereopermutators will have a
different color than a stereopermutator with a set stereopermutation. As a
consequence, molecules with indeterminate stereopermutators will not compare
equal to an otherwise equivalent molecule with assigned stereopermutators. The
varying levels of information for graph coloring are collected into a 128-bit 
hash, but in a bijective manner, i.e.\ without the possibility of hash
collision. These hashes then represent the color of each vertex and are passed
as input to the Boost isomorphism function.

Canonicalization of the molecule representation is similarly modular. First,
modular hashes of vertices are calculated according to algorithm input. Next,
the canonical automorphism is determined by processing the graph and its vertex
hashes by the nauty~\cite{McKay2014} library. Specifically, the
\textsc{sparsenauty} function is called, specifying distance vertex invariants
for canonical labeling. Lastly, the canonical automorphism is applied to the
graph and its stereopermutators.

\subsection{Distance Geometry error function derivative}

\newcommand{\avgVecDiff}[2]
{\left(\vec{s}_{#1}-\vec{s}_{#2}\right)}

\newcommand{\vecDiff}[2]
{\left(\vec{#1}-\vec{#2}\,\right)}

\newcommand{\vecCross}[2]
{\left(\vec{#1}\times\vec{#2}\,\right)}

\newcommand{\posDependence}
{\left(\left\{\vec{r}_i \right\}\right)}

\newcommand{\distanceErrorFirstTermPart}
{\frac{\vec{r}_{ij}^{2}}{U_{ij}^{2}}-1}

\newcommand{\distanceErrorFirstTerm}[1]
{\mathrm{max}^{#1}\left(0, \distanceErrorFirstTermPart\right)}

\newcommand{\distanceErrorSecondTermPart}
{\frac{2 L_{ij}^{2}}{L_{ij}^{2} + \vec{r}_{ij}^{2}}-1}

\newcommand{\distanceErrorSecondTerm}[1]
{\mathrm{max}^{#1} \left(0, \distanceErrorSecondTermPart\right)}

\newcommand{\chiralErrorFirstTerm}[1]
{\mathrm{max}^{#1} \left(0, 
    V_{\alpha\beta\gamma\delta}\posDependence{}-U_{V}
  \right)
}

\newcommand{\chiralErrorSecondTerm}[1]
{\mathrm{max}^{#1} \left(0,
    L_{V}-V_{\alpha\beta\gamma\delta}\posDependence{} 
  \right)
}

\newcommand{\iNSum}{\sum_{i = 1}^{N}}
\newcommand{\ijNSum}{\sum_{i < j}^{N}}
\newcommand{\chiralSum}
{\sum_{\left(S_\alpha, S_\beta, S_\gamma, S_\delta,
  U_{V}, L_{V}\right) \in{} C} 
}
\newcommand{\dihedralSum}
{\sum_{\left(S_\alpha, S_\beta, S_\gamma, S_\delta,
  U_\phi, L_\phi\right) \in{} D}
}
\newcommand{\phiSymbol}
{\phi_{\alpha\beta\gamma\delta}\posDependence{}}

\newcommand{\dihedralTerm}[1]
{\mathrm{max}^{#1} \left(0,
    \left\lvert\phi{} + \bar{\phi} -\frac{U_\phi{}+L_\phi}{2}
    \right\rvert-\frac{U_\phi{}-L_\phi}{2}
  \right)
}

\newcommand{\volumeCalculationSets}[4]
{\avgVecDiff{#1}{#4}^{T}
  \cdot \left[
    \avgVecDiff{#2}{#4}
    \times\avgVecDiff{#3}{#4}
  \right]
}

\newcommand{\volumeCalculation}[4] {\left(\vec{#1}-\vec{#4} \right)^{T} \cdot
\left[ \left(\vec{#2}-\vec{#4} \right) \times\left(\vec{#3}-\vec{#4} \right)
\right] }

\newcommand{\partialPosVec}[1]
{\frac{\partial}{\partial\vec{r}_{#1}}}

\newcommand{\partialVec}[1]
{\frac{\partial}{\partial\vec{#1}\,}}

\newcommand{\errf}{\mathrm{errf}\posDependence}

\newcommand*\circledn[1]{
  \tikz[baseline=(char.base)]{
    \node[shape=circle,draw,inner sep=2pt] (char) {#1};
  }
}

For a given set of $N$ particles with positions $\vec{r}_i$, the Distance
Geometry error function applied is the sum of distance errors $a$, chiral errors
$b$ and dihedral errors $c$.
\begin{equation}
  a = \ijNSum\left[
      \distanceErrorFirstTerm{2} + \distanceErrorSecondTerm{2}
    \right]
\end{equation}

The symbols $U_{ij}$ and $L_{ij}$ are the upper and lower distance bounds for
the atoms $i$ and $j$.
\begin{align}\begin{aligned}
  b = \chiralSum &\chiralErrorFirstTerm{2}\\
  +&\chiralErrorSecondTerm{2}
\end{aligned}\end{align}

Within the chiral errors, $C$ is a set of chiral constraint tuples consisting of
the particle sets $S_\alpha, S_\beta, S_\gamma$ and $S_\delta$. These contain
mutually disjoint particle indices and contain at least one element. In further
notation, $\vec{s}$ denotes the average spatial position of all elements of a
set, e.g.\ for $S_\alpha$:
\begin{equation}
  \vec{s}_\alpha = \frac{1}{|S_\alpha|}\sum_{i=1}^{|S_\alpha|}
  \vec{r}_{S_{\alpha, i}},
\end{equation}

where $|S_\alpha|$ denotes the number of elements in the set and $S_{\alpha,
i}$ is the $i$-th element in the set.\\

The constraint tuple further consists of the scalars $U_{V}$ and
$L_{V}$, which are upper and lower bounds on
the volume spanned by the average positions of the sets $S_\alpha$, $S_\beta$,
$S_\gamma$ and $S_\delta$. This volume is
calculated in the symbol $V_{\alpha\beta\gamma\delta}$, which is the signed
tetrahedron volume spanned by $\vec{s}_\alpha, \vec{s}_\beta, \vec{s}_\gamma$
and $\vec{s}_\delta$:
\begin{equation}
  V_{\alpha\beta\gamma\delta}\left( \left\{\vec{r}_i \right\} \right) =
    \volumeCalculationSets{\alpha}{\beta}{\gamma}{\delta}.
\end{equation}

It is important to note that tetrahedron volumes such as
$U_{V}$ are signed values. On odd permutations of constituting indices, these
quantities change sign.
\begin{equation}
  c = \dihedralSum\dihedralTerm{2}
\end{equation}

Within the dihedral errors, $D$ is a set of dihedral constraint tuples. Each
tuple consists of four particle index sets $S_\alpha, S_\beta, S_\gamma$ and
$S_\delta$ and upper and lower bounds on the dihedral angle $U_\phi$ and
$L_\phi$. Exactly as for chiral errors, the particle index sets do not
intersect and each contain at least a single element. The dihedral angle $\phi$
is defined via the particle index sets, where $\vec{s}_{ij} = \vec{s}_j - \vec{s}_i$:
\begin{align}\begin{aligned}
  \phi = \mathrm{atan2} \Big(\left(
      \vec{s}_{\alpha\beta} \times \vec{s}_{\beta\gamma}
    \right) \times \left(
      \vec{s}_{\beta\gamma} \times \vec{s}_{\gamma\delta}
    \right) \cdot 
    \frac{\vec{s}_{\beta\gamma}}{\lvert \vec{s}_{\beta\gamma}\rvert},
     \left(
      \vec{s}_{\alpha\beta} \times \vec{s}_{\beta\gamma}
    \right) \cdot \left(
      \vec{s}_{\beta\gamma} \times \vec{s}_{\gamma\delta}
    \right)
  \Big)
\end{aligned}\end{align}

Here we have used a three-vector dihedral angle definition and inserted
the index set average-position differences that constitute the dihedral angle.
Another definition of the dihedral angle with merely the inverse cosine is:
\begin{align}\begin{aligned}
  \phi = \arccos
    \frac{\left(
      \vec{s}_{\beta\alpha} \times \vec{s}_{\gamma\beta}
    \right)\left(
      \vec{s}_{\gamma\beta} \times \vec{s}_{\gamma\delta}
    \right)}{\lvert
      \vec{s}_{\beta\alpha} \times \vec{s}_{\gamma\beta}
    \rvert\lvert
      \vec{s}_{\gamma\beta} \times \vec{s}_{\gamma\delta}
    \rvert
    },
\end{aligned}\end{align}

whose derivatives w.r.t.\ the constituting position vectors are considerably
easier to evaluate. $\bar{\phi}$ is defined as:
\begin{equation}
  \bar{\phi} = \begin{cases}
    2\pi & \text{if } \phi < \frac{U_\phi + L_\phi - 2\pi}{2}\\
    -2\pi & \text{if } \phi > \frac{U_\phi + L_\phi + 2\pi}{2}\\
    0 & \text{else}\\
  \end{cases}
\end{equation}

For a scalar-valued function $E$, the columnar gradient of the error function is
composed of partial derivatives to the individual position vectors $\vec{r}_i$:
\begin{align*}
  \nabla E &= \left(
    \frac{\partial E}{\partial \vec{r}_1},
    \frac{\partial E}{\partial \vec{r}_2},
    \ldots,
    \frac{\partial E}{\partial \vec{r}_N}
  \right)
\end{align*}

Each individual component $\left( \partial E / \partial \vec{r}_\xi \right)$
is a vector whose components are the scalar derivatives:
\begin{align*}
  \frac{\partial E}{\partial \vec{r}_\xi} &= \begin{pmatrix}
    \partial E / \partial \vec{r}_{\xi , x} \\
    \partial E / \partial \vec{r}_{\xi , y} \\
    \partial E / \partial \vec{r}_{\xi , z} 
  \end{pmatrix}
\end{align*}

We split the problem into five main terms:
\begin{align*}
  \partialPosVec{\xi} \errf 
  &= \underbrace{\partialPosVec{\xi} \left(
      \ijNSum \distanceErrorFirstTerm{2}
    \right)
  }_{\circledn{1}}\\
  &+ \underbrace{\partialPosVec{\xi} \left(
      \ijNSum \distanceErrorSecondTerm{2}
    \right)
  }_{\circledn{2}}\\
  &+ \underbrace{\partialPosVec{\xi} \chiralSum \chiralErrorFirstTerm{2}
  }_{\circledn{3}}\\
  &+ \underbrace{\partialPosVec{\xi} \chiralSum \chiralErrorSecondTerm{2}
  }_{\circledn{4}}\\
  &+ \underbrace{\partialPosVec{\xi} \dihedralSum \dihedralTerm{2}
  }_{\circledn{5}}.
\end{align*}

\subsection{Distance error terms}

We begin with \circledn{1}, applying the chain rule:
\begin{align}\begin{aligned}
  &\partialPosVec{\xi} \ijNSum \distanceErrorFirstTerm{2}\\
  = &2 \ijNSum \distanceErrorFirstTerm{} \partialPosVec{\xi}
  \distanceErrorFirstTerm{}.
\end{aligned}\end{align}

If $\xi = i$, then:
\begin{equation}
  \partialPosVec{i} \distanceErrorFirstTermPart
  = - \frac{2}{U_{ij}^{2}} r_{ij}
\end{equation}

and likewise, but positive, for $\xi = j$. Consequently:
\begin{align}\begin{aligned}
  \partialPosVec{\xi} \left( \distanceErrorFirstTerm{} \right)
  &= \frac{2}{U_{ij}^{2}} \vec{r}_{ij} \begin{cases}
    -1 & \text{if } \xi = i\\
    1 & \text{if } \xi = j\\
    0 & \text{else}
  \end{cases}\\
  &= \frac{2}{U_{ij}^{2}} \vec{r}_{ij} \left(\delta_{\xi j}-\delta_{\xi i} 
  \right),
\end{aligned}\end{align}

where we have discarded the possibility of $\distanceErrorFirstTermPart < 0$
since this case is adequately covered by the first maximum function. $\delta$ is
the Kronecker delta. So, in total we have:
\begin{equation}\label{eq_breakdown_deltas}
  \text{\circledn{1}}
  = \ijNSum \left(
    \delta_{\xi j} - \delta_{\xi i} 
  \right) \underbrace{\frac{4}{U_{ij}^{2}} 
    \vec{r}_{ij} \distanceErrorFirstTerm{}
  }_{f(i, j)}
\end{equation}

We can transform the summation further with Kronecker deltas:
\begin{align}
  &\ijNSum \left( \delta_{\xi j} - \delta_{\xi i} \right) f(i, j)\\
  = &\sum_{i = 1}^{\xi - 1} f(i, \xi) 
    - \sum_{j = \xi + 1}^{N} f(\xi, j)\\
  = &\sum_{i = 1}^{\xi - 1} f(i, \xi) 
    + \sum_{i = \xi + 1}^{N} f(i, \xi)\\
  = &\sum_{i = 1}^{N} \left(1 - \delta_{i\xi} \right) f(i, \xi),
\end{align}

where we have used that $f(i, j) = -f(j, i)$ (see Eq.~\ref{eq_breakdown_deltas}).
All in all:
\begin{equation}
  \text{\circledn{1}}
  = \iNSum \left(1 - \delta_{i\xi}\right) \frac{4}{U_{i\xi}^{2}}
    \vec{r}_{i\xi}\max\left(0,
      \frac{\vec{r}_{i\xi}^{2}}{U_{i\xi}^{2}}-1
    \right)
\end{equation}

Let us continue with \circledn{2}. Once again, we apply the chain rule:
\begin{align}\begin{aligned}
  &\partialPosVec{\xi} \ijNSum \distanceErrorSecondTerm{2}\\
  = &2 \ijNSum \distanceErrorSecondTerm{} \partialPosVec{\xi}
    \distanceErrorSecondTerm{}
\end{aligned}\end{align}

For $\xi = i$,
\begin{equation}
  \partialPosVec{i} \left( \distanceErrorSecondTermPart \right)
  = \frac{4 L_{ij}^{2} \vec{r}_{ij}}
    {\left(L_{ij}^{2} + \vec{r}_{ij}^2\right)^2},
\end{equation}

and likewise, but negative, for $\xi = j$. Therefore,
\begin{align*}
  \partialPosVec{\xi} \left( \distanceErrorSecondTerm{} \right)
  &=\frac{4 L_{ij}^{2} \vec{r}_{ij}}{\left(L_{ij}^{2} + \vec{r}_{ij}^2\right)^2} \begin{cases}
    1 & \text{if } \xi = i\\
    -1 & \text{if } \xi = j\\
    0 & \text{else}
  \end{cases}\\
  &= \frac{4 L_{ij}^{2} \vec{r}_{ij}}{\left(L_{ij}^{2} + \vec{r}_{ij}^2\right)^2} \left(
    \delta_{\xi i} - \delta_{\xi j} 
  \right),
\end{align*}

where we have excluded the possibility of $\distanceErrorSecondTermPart < 0$
since this case is covered by the first maximum function. Altogether:
\begin{equation}\label{eq_breakdown_deltas_second}
  \text{\circledn{2}}
  = \ijNSum \left(
    \delta_{\xi i} - \delta_{\xi j} 
  \right) \underbrace{\frac{8 L_{ij}^{2} \vec{r}_{ij}}{\left(L_{ij}^{2} + \vec{r}_{ij}^2\right)^2} \distanceErrorSecondTerm{}
  }_{g(i, j)}.
\end{equation}

Transforming the summation:
\begin{align}
  \ijNSum \left( \delta_{\xi i} - \delta_{\xi j} \right) g(i, j)
  &= - \sum_{i = 1}^{\xi - 1} g(i, \xi) 
    + \sum_{j = \xi + 1}^{N} g(\xi, j)\\
  &= \sum_{i = 1}^{\xi - 1} g(\xi, i) 
    + \sum_{i = \xi + 1}^{N} g(\xi, i)\\
  &= \sum_{i = 1}^{N} \left(1 - \delta_{i\xi} \right) g(\xi, i),
\end{align}

Where we have used $g(i, j) = -g(j, i)$ (see
Eq.~\ref{eq_breakdown_deltas_second}). All in all:
\begin{equation}
  \text{\circledn{2}}
  = \iNSum \left(1 - \delta_{i\xi}\right) 
    \frac{8 L_{\xi i}^{2} \vec{r}_{i\xi}}{\left(L_{\xi i}^{2} + \vec{r}_{i\xi}^2\right)^2} \max \left(0, \frac{2 L_{\xi i}^{2}}{L_{\xi i}^{2} + \vec{r}_{i\xi}^{2}} - 1\right)
\end{equation}

\subsubsection{Chiral error terms}

Next, we consider \circledn{3}. Applying the chain rule yields:
\begin{align*}
  &\quad\partialPosVec{\xi} \chiralSum \chiralErrorFirstTerm{2}\\
  &= 2 \sum_{\left(\ldots\right) \in{} C} \chiralErrorFirstTerm{} \partialPosVec{\xi} \chiralErrorFirstTerm{}
\end{align*}

The partial derivatives of $V_{\alpha\beta\gamma\delta}\posDependence$ with
respect to $\vec{r}_\xi$ are split into five cases. The index $\xi$ can be an
element of one of the four sets $S_\alpha, S_\beta, S_\gamma, S_\delta$ (and
only one, since they are mutually disjoint) or not. The derivative for the last
case is zero. In the remaining cases, one average vector $\vec{s}$ is a function
of $\vec{r}_\xi$ but the rest are not. The partial derivative of any average
vector $\vec{s}$ is:
\begin{equation}
  \partialPosVec{\xi} \vec{s}_\alpha 
  = \partialPosVec{\xi} \frac{1}{|S_\alpha|}\sum_{i=1}^{|S_\alpha|}
  \vec{r}_{S_{\alpha, i}}
  = \begin{cases}
    \frac{1}{|S_\alpha|} \mathbf{I}_3 & \text{if } \xi \in S_\alpha\\
    0 & \text{else}
  \end{cases}
\end{equation}

where $\mathbf{I}_3$ denotes the three dimensional identity matrix. The individual set
membership cases are thus as follows:
\begin{align*}
  \text{\circledn{$S_\alpha$}} &\quad \partialPosVec{\xi} \left\{ 
    \avgVecDiff{\alpha}{\delta}^{T}
    \cdot \left[
      \avgVecDiff{\beta}{\delta}
      \times \avgVecDiff{\gamma}{\delta}
    \right]
  \right\} \\
  &= \partialPosVec{\xi} \left\{ 
    \vec{s}_\alpha^{\,T}
    \cdot \left[
      \avgVecDiff{\beta}{\delta}
      \times \avgVecDiff{\gamma}{\delta}
    \right]
  \right\} - \vec{0} \\
  &= \frac{1}{|S_\alpha|} \left[
    \avgVecDiff{\beta}{\delta} \times \avgVecDiff{\gamma}{\delta} 
  \right]\\
  &= \frac{\vec{s}_{\delta\beta} \times \vec{s}_{\delta\gamma}}{|S_\alpha|}
\end{align*}

In shorthand notation, in which all $\vec{s}$ symbols are replaced by their subscripts:

\begin{align*}
  \text{\circledn{$S_\beta$}} &\quad \partialPosVec{\xi} 
    \vecDiff{\alpha}{\delta}^{T}
    \cdot \left[
      \vecDiff{\beta}{\delta}
      \times \vecDiff{\gamma}{\delta}
    \right] \\
  &= \partialPosVec{\xi}
    \vecDiff{\alpha}{\delta}^{T}
    \cdot \left[
      \vec{\beta} \times \vec{\gamma} - \vec{\beta} \times \vec{\delta} - \vec{\delta} \times \vec{\gamma} + \underbrace{\vec{\delta} \times \vec{\delta}}_{=0}
    \right] \\
  &= \partialPosVec{\xi} \left\{
    \vec{\beta}^{\,T} \cdot \left[
      \vec{\gamma} \times \vecDiff{\alpha}{\delta}
    \right] 
    - \vec{\beta}^{\,T} \cdot \left[
      \vec{\delta} \times \vecDiff{\alpha}{\delta}
    \right] 
  \right\} - \vec{0} \\
  &= \frac{1}{|S_\beta|} \vecDiff{\gamma}{\delta} \times \vecDiff{\alpha}{\delta} \\
  &= - \frac{1}{|S_\beta|}\vecDiff{\alpha}{\delta} \times \vecDiff{\gamma}{\delta} \\
  &= - \frac{\vec{s}_{\delta\alpha} \times \vec{s}_{\delta\gamma}}{|S_\beta|}
\end{align*}

\begin{align*}
  \text{\circledn{$S_\gamma$}} &\quad \partialPosVec{\xi} 
    \vecDiff{\alpha}{\delta}^{T}
    \cdot \left[
      \vec{\beta} \times \vec{\gamma} - \vec{\beta} \times \vec{\delta} - \vec{\delta} \times \vec{\gamma} 
    \right] \\
  &= \partialPosVec{\xi} \left\{
    \vec{\gamma}^{\,T} \cdot \left[
      \vecDiff{\alpha}{\delta} \times \vec{\beta}
    \right] 
    - \vec{\gamma}^{\,T} \cdot \left[
      \vecDiff{\alpha}{\delta} \times \vec{\delta}\,
    \right] 
  \right\} - \vec{0} \\
  &= \frac{1}{|S_\gamma|} \vecDiff{\alpha}{\delta} \times \vecDiff{\beta}{\delta} \\
  &= \frac{\vec{s}_{\delta\alpha} \times \vec{s}_{\delta\beta}}{|S_\gamma|}
\end{align*}

\begin{align*}
  \text{\circledn{$S_\delta$}} &\quad \partialPosVec{\xi}
    \vecDiff{\alpha}{\delta}^{T}
    \left[
      \vec{\beta} \times \vec{\gamma} - \vec{\beta} \times \vec{\delta} - \vec{\delta} \times \vec{\gamma} 
    \right] \\
  &= \partialPosVec{\xi} \vec{\alpha}^{\,T} \left[
      \vec{\beta} \times \vec{\gamma} - \vec{\beta} \times \vec{\delta} - \vec{\delta} \times \vec{\gamma} 
    \right] - \partialPosVec{\xi} \vec{\delta}^{\,\,T} \left[
      \vec{\beta} \times \vec{\gamma} - \vec{\beta} \times \vec{\delta} - \vec{\delta} \times \vec{\gamma}
    \right]\\
  &= \underbrace{\partialPosVec{\xi} \vec{\alpha}^{\,T} \vecCross{\beta}{\gamma}}_{=0} 
    - \partialPosVec{\xi} \vec{\alpha}^{\,T} \left( 
      \vec{\beta} \times \vec{\delta}\,
    \right)
    - \partialPosVec{\xi} \vec{\alpha}^{\,T} \vecCross{\delta}{\gamma} \\
    &\quad- \partialPosVec{\xi} \vec{\delta}^{\,\,T} \vecCross{\beta}{\gamma}
    + \partialPosVec{\xi} \underbrace{\vec{\delta}^{\,\,T} \vecCross{\delta}{\gamma}}_{=0}
    + \partialPosVec{\xi} \underbrace{\vec{\delta}^{\,\,T} \vecCross{\beta}{\delta}}_{=0} \\
  &= - \partialPosVec{\xi} \vec{\delta}^{\,\,T} \vecCross{\gamma}{\alpha}
    - \partialPosVec{\xi} \vec{\delta}^{\,\,T} \vecCross{\alpha}{\beta}
    - \partialPosVec{\xi} \vec{\delta}^{\,\,T} \vecCross{\beta}{\gamma} \\
  &= - \frac{1}{|S_\delta|} \vec{\gamma} \times \vec{\alpha} -
  \frac{1}{|S_\delta|} \vec{\alpha} \times \vec{\beta} - \frac{1}{|S_\delta|} \vec{\beta} \times \vec{\gamma} \\
  &= - \frac{1}{|S_\delta|} \vecDiff{\alpha}{\gamma} \times \vecDiff{\beta}{\gamma}\\
  &= - \frac{\vec{s}_{\gamma\alpha} \times \vec{s}_{\gamma\beta}}{|S_\delta|}
\end{align*}

All set membership cases can be summed up in a symbol:

\begin{equation}
  V_C = \begin{cases}
    \frac{\vec{s}_{\delta\beta} \times \vec{s}_{\delta\gamma}}{|S_\alpha|} & \text{if } \xi \in S_\alpha\\
    - \frac{\vec{s}_{\delta\alpha} \times \vec{s}_{\delta\gamma}}{|S_\beta|} & \text{if } \xi \in S_\beta\\
    \frac{\vec{s}_{\delta\alpha} \times \vec{s}_{\delta\beta}}{|S_\gamma|} & \text{if } \xi \in S_\gamma\\
    - \frac{\vec{s}_{\gamma\alpha} \times \vec{s}_{\gamma\beta}}{|S_\delta|} & \text{if } \xi \in S_\delta\\
    0 & \text{else}
  \end{cases}.
\end{equation}

So, overall:
\begin{equation}
  \text{\circledn{3}} = \chiralSum 2\ \chiralErrorFirstTerm{} V_C.
\end{equation}

For \circledn{4}, the derivation is analog save for the sign of the great amount
of cases, which we extrude from the sum:
\begin{equation}
  \text{\circledn{4}} = - \chiralSum 2\ \chiralErrorSecondTerm{} V_C.
\end{equation}

Both terms concerning the chiral error can be summarized as follows:
\begin{align*}
  &\text{\circledn{3}} + \text{\circledn{4}} \\
  = & \sum_{\left(\ldots\right) \in{} C} 2\ \chiralErrorFirstTerm{} V_C\\
  & \quad -\sum_{\left(\ldots\right) \in{} C} 2\ \chiralErrorSecondTerm{} V_C\\
  = & \sum_{\left(\ldots\right) \in{} C} 2\ V_C \Big[ 
    \chiralErrorFirstTerm{} - \chiralErrorSecondTerm{} 
  \Big].
\end{align*}

\subsubsection{Dihedral error terms}

Finally, we consider $\circledn{5}$. The chain rule yields:
\begin{equation}
  \partialPosVec{\xi} \sum_{(\ldots) \in D} \max^2 \left(0, h(\phi)\right)
  = 2\ \sum_{(\ldots) \in D} \max\left(0, h(\phi)\right) 
    \partialPosVec{\xi} \max\left(0, h(\phi)\right),
\end{equation}

where we have substituted the dihedral expression in the maximum function
with $h(\phi)$. We can drop the wrapping maximum function in the derivative
since the first maximum function adequately covers the case
$h(\phi) < 0$. Employing the chain rule again, we note 
$\partialPosVec{\xi} h(\phi) = \frac{\partial h}{\partial \phi}
\frac{\partial \phi}{\partial \vec{r}_{\xi}}$. The first part is:
\begin{equation}
  \frac{\partial h}{\partial \phi} = \frac{\partial}{\partial \phi}
    \Bigg\lvert
      \underbrace{\phiSymbol + \bar{\phi} -\frac{U_\phi+ L_\phi}{2}
      }_{w(\phi)}
    \Bigg\rvert - \frac{U_\phi - L_\phi}{2}
  = \mathrm{sgn}\left(w(\phi)\right).
\end{equation}

The individual derivatives $\frac{\partial\phi}{\partial\vec{s}_{\alpha}}, 
\frac{\partial\phi}{\partial\vec{s}_{\beta}}, 
\frac{\partial\phi}{\partial\vec{s}_{\gamma}}, 
\frac{\partial\phi}{\partial\vec{s}_{\delta}}$ are given as:
\begin{equation}
  \vec{f} = \avgVecDiff{\alpha}{\beta},
  \quad\vec{g} = \avgVecDiff{\beta}{\gamma},
  \quad\vec{h} = \avgVecDiff{\delta}{\gamma},
\end{equation}
\begin{equation}
  \vec{a} = \vec{f} \times \vec{g},
  \quad\vec{b} = \vec{h} \times \vec{g}
\end{equation}

\begin{align}\begin{aligned}
  \frac{\partial\phi}{\partial\vec{s}_{\alpha}} &= 
    -\frac{\lvert\vec{g}\rvert}{\vec{a}\,^2}\vec{a}\\
  \frac{\partial\phi}{\partial\vec{s}_{\beta}}  &= 
    \frac{\lvert\vec{g}\rvert}{\vec{a}\,^2}\vec{a}
    + \frac{\vec{f}\vec{g}}{\vec{a}\,^2\lvert\vec{g}\rvert}\vec{a}
    - \frac{\vec{g}\vec{h}}{\vec{b}\,^2\lvert\vec{g}\rvert}\vec{b}\\
  \frac{\partial\phi}{\partial\vec{s}_{\gamma}} &= 
    - \frac{\lvert\vec{g}\rvert}{\vec{b}\,^2}\vec{b}
    + \frac{\vec{g}\vec{h}}{\vec{b}\,^2\lvert\vec{g}\rvert}\vec{b}
    - \frac{\vec{f}\vec{g}}{\vec{a}\,^2\lvert\vec{g}\rvert}\vec{a}\\
  \frac{\partial\phi}{\partial\vec{s}_{\delta}} &= 
    \frac{\lvert\vec{g}\rvert}{\vec{b}\,^2}\vec{b}\\
\end{aligned}\end{align}

The individual membership cases are therefore:
\begin{align}
  \text{\circledn{$S_i$}} &\quad\frac{\partial\phi}{\partial\vec{r}_{\xi}} 
    = \frac{\partial\phi}{\partial\vec{s}_{i}}
      \frac{\partial\vec{s_{i}}}{\partial\vec{r}_{\xi}}
    = \frac{1}{\lvert S_i\rvert}
      \frac{\partial\phi}{\partial\vec{s}_{i}}
\end{align}

And altogether:
\begin{equation}
  \text{\circledn{5}} = 2 \sum_{(\ldots) \in D} \max(0, h(\phi))\ \mathrm{sgn}
  (w(\phi)) \begin{cases}
    \frac{1}{|S_\alpha|} \frac{\partial\phi}{\partial\vec{s_{\alpha}}} & \text{if } \xi \in S_\alpha\\
    \frac{1}{|S_\beta|} \frac{\partial\phi}{\partial\vec{s_{\beta}}} & \text{if } \xi \in S_\beta\\
    \frac{1}{|S_\gamma|} \frac{\partial\phi}{\partial\vec{s_{\gamma}}} & \text{if } \xi \in S_\gamma\\
    \frac{1}{|S_\delta|} \frac{\partial\phi}{\partial\vec{s_{\delta}}} & \text{if } \xi \in S_\delta\\
    0 & \text{else}
  \end{cases}
\end{equation}

\subsection{Demonstration details}

\begin{figure}
  \centering
  \includegraphics[width=\linewidth]{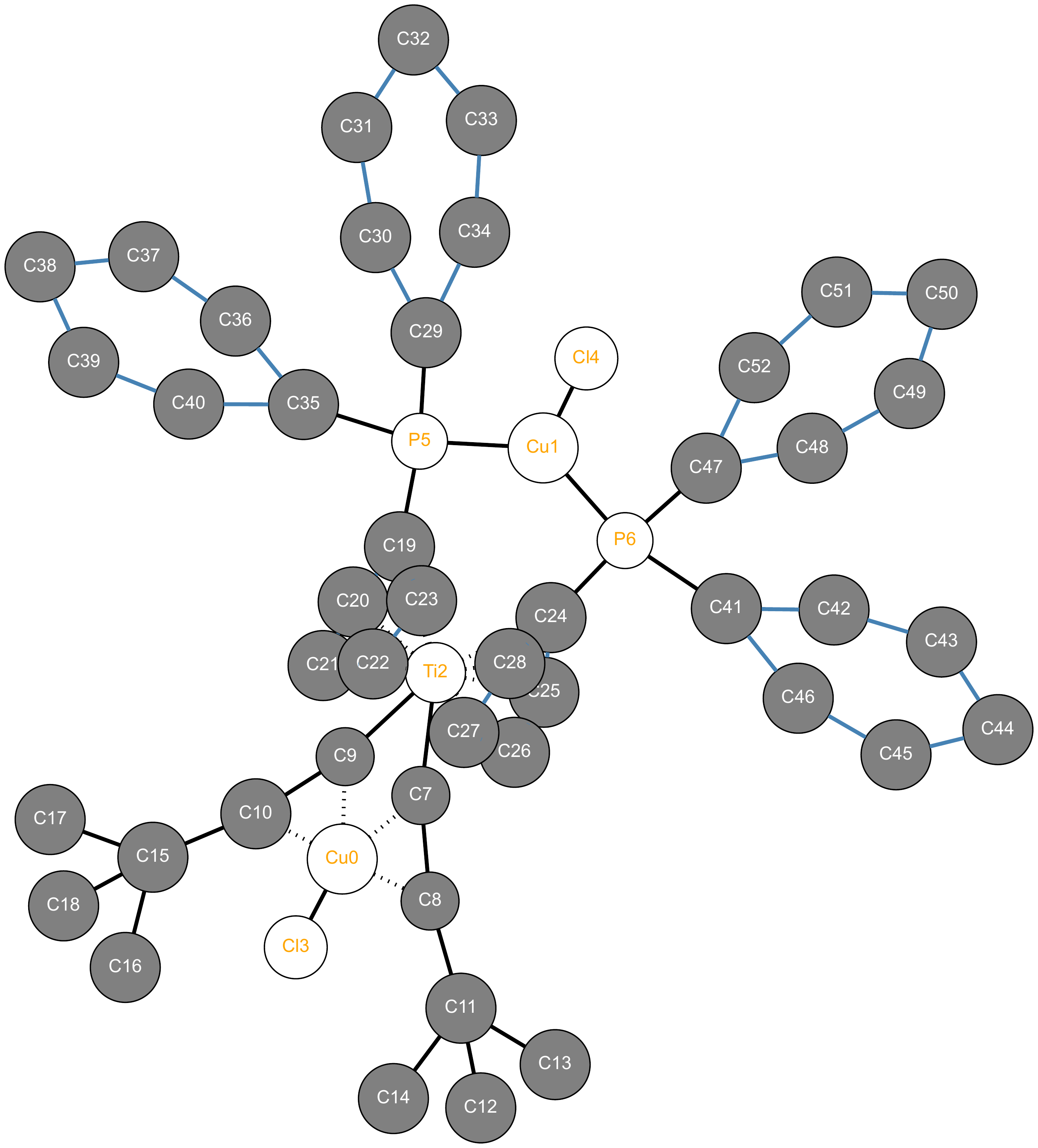}
  \caption{Graphviz output of the \texttt{BIBTAH} Cambridge Crystallographic
  Database entry after processing by \textsc{Molassembler}.
  }\label{fig:demonstration_graphviz}
\end{figure}

Besides the library itself, Python bindings, documentation and tests, the code
distribution of \textsc{Molassembler} contains a set of analysis binaries that
explore the functionality and behavior of particular aspects of the library.
Among these is the \texttt{Stereopermutations} binary. To reproduce the four
stereopermutations arising from the abstract octahedral
\texttt{(A-A)\textsubscript{3}} binding case, call the binary as follows:
\begin{lstlisting}[caption=Bidentate octahedron stereopermutations]
./Stereopermutations -s 12 -c AAAAAA -l "0,1,2,3,4,5"
Shape: octahedron
Characters: AAAAAA
Links: {{0, 1}, {2, 3}, {4, 5}}

Weight 6: chars {A, A, A, A, A, A}, links {[0, 1], [2, 3], [4, 5]}, link angles: 90 90 180
Weight 4: chars {A, A, A, A, A, A}, links {[0, 1], [2, 4], [3, 5]}, link angles: 90 90 90
Weight 4: chars {A, A, A, A, A, A}, links {[0, 1], [2, 5], [3, 4]}, link angles: 90 90 90
Weight 1: chars {A, A, A, A, A, A}, links {[0, 2], [1, 3], [4, 5]}, link angles: 180 180 180
4 stereopermutations
\end{lstlisting}

The other demonstration reinforcing that the molecular model chosen here is
suitable for the representation of inorganic molecules with haptic bonding can
be reproduced with the Python bindings in just a few lines of code:

\begin{lstlisting}[caption=Python bindings molecule construction]
import sys
import scine_molassembler as masm

name = sys.argv[1]
mols = masm.io.split(name)
for i, mol in enumerate(mols):
    for p in mol.stereopermutators.atom_stereopermutators():
      print(p)

    outname = name.replace(".mol", "-{}.dot".format(i))
    masm.io.write(outname, mol)
\end{lstlisting} 

This generates some terminal output and a Graphviz~\cite{ellson2001graphviz}
representation for all molecules in a MOLFile passed as argument. The molecule
in question is Cambridge Crystallographic Database~\cite{groom2016cambridge}
entry \texttt{BIBTAH}~\cite{Delgado1999}. The terminal output contains the
following lines of interest:

\begin{lstlisting}[caption=Python terminal output]
[...]
2: tetrahedron, AABB, A-A, B-B, 0 (1)
[...]
0: triangle, AAB, A-A, 0 (1)
[...]
\end{lstlisting}

For reference, an SVG render of the graphviz output is shown in
Fig.~\ref{fig:demonstration_graphviz}.

\subsubsection{Iron compound dataset}

The iron compound dataset collected by Proppe and Reiher~\cite{Proppe2017} with
the interpreted shapes of iron atoms, their substituent rankings, number of
abstract stereopermutations and number of feasible stereopermutations is shown
in
Tables~\ref{table:proppe_interpretation_1},\,\ref{table:proppe_interpretation_2}.
In nearly all cases, shapes and ranking results are produced as expected. In the
case of compound \#17, the classified shape is a trigonal bipyramid, yet we
would subjectively judge this a square pyramid. The coordination polyhedron is
certainly strongly distorted in this case.

In order to reproduce these results, we must first determine bond orders.
Bonds were detected in a binary fashion by summing the covalent radii of
each pair of atoms with a tolerance of 0.4\,\AA, and comparing against the
spatial distance. This procedure misclassifies six bonds as present in the
dataset (one each in structures \#13, 15, 22, 26, 33 and 35) that must be
corrected manually. The bond order determination can be repeated with the
following Python code:

\begin{lstlisting}[caption=Dataset bond order determination]
import sys
import scine_utilities as su

for filename in sys.argv[1:]:
    (coordinates, _) = su.IO.read(filename)
    bonds = su.BondDetector.detect_bonds(coordinates)
    su.IO.write_topology(filename.replace("xyz", "mol"), coordinates, bonds)
\end{lstlisting}

Next, the interpretation of molecules can be carried out:

\begin{lstlisting}[caption=Dataset molecule interpretation]
import sys
import scine_molassembler as masm

for filename in sys.argv[1:]:
    mol = masm.io.read(filename)
    # The first atom of each structure is an iron atom:
    print(mol.stereopermutators[0])
\end{lstlisting}

\begin{table}
  \caption{Interpretation of iron compounds \#1--28 from Ref.~\cite{Proppe2017}.
  This table collects information from the atom-centered stereopermutator placed
  on an iron atom in the compound (if there are multiple, they are symmetric in
  this dataset). The polyhedron column collects the IUPAC symbols for the
  coordination polyhedron of the iron atom. Ranking is the ligand binding case,
  which abstracts over haptic ligands. Links denotes the connectivity between
  characters of the ligand binding case. $F$ is the number of feasible
  stereopermutations and $A$ is the number of abstract
  stereopermutations.}\label{table:proppe_interpretation_1}
  \begin{center}
  \begin{tabular}{r l l l r r}
    \toprule
    \# & Polyhedron & Ranking & Linking & $F$ & $A$\\
    \midrule
    1 & T-4 & AAAA & & 1 & 1\\
    2 & OC-6 & AAAABB & A-A, A-A & 3 & 5\\
    3 & T-4 & AABB & B-B & 1 & 1\\
    4 & T-4 & AAAB & A-A, A-A, A-A & 1 & 1\\
    5 & TP-3 & AAB & A-A & 1 & 1\\
    6 & T-4 & AAAA & A-A, A-A & 1 & 1\\
    7 & SPY-5 & AAAAB & A-A, A-A, A-A, A-A & 2 & 5\\
    8 & T-4 & AAAA & & 1 & 1\\
    9 & SP-4 & AABC & A-B, A-C, A-B, A-C & 1 & 2\\
    10 & T-4 & AAAB & A-A, A-A, A-A & 1 & 1\\
    11 & SP-4 & AAAA & A-A, A-A, A-A, A-A & 1 & 2\\
    12 & T-4 & AAAB & A-A, A-A, A-A & 1 & 1\\
    13 & OC-6 & AAABCD & A-A, A-A, A-A, B-D, C-D & 2 & 5\\
    14 & TP-3 & AAB & A-A & 1 & 1\\
    15 & OC-6 & AAABCD & A-A, A-A, A-A, B-D, C-D & 2 & 5\\
    16 & OC-6 & AABBCD & A-B, A-C, A-B, A-C & 7 & 16\\
    17 & TBPY-5 & AABBC & A-B, A-B & 10 & 12\\
    18 & TBPY-5 & AAABC & A-C, A-C, A-C & 3 & 4\\
    19 & OC-6 & AABBCD & A-B, A-C, A-B, A-C & 7 & 16\\
    20 & OC-6 & AABBCD & A-B, A-C, A-B, A-C & 7 & 16\\
    21 & OC-6 & AABBCC & A-A, A-B, A-B, B-B & 3 & 11\\
    22 & SP-4 & ABCD & A-B, A-D, B-C & 2 & 3\\
    23 & SPY-5 & AAAAB & A-A, A-A, A-A, A-A & 2 & 5\\
    24 & T-4 & AAAA & & 1 & 1\\
    25 & TBPY-5 & AAABC & A-C, A-C, A-C & 3 & 4\\
    26 & SP-4 & ABCD & A-B, A-C, B-D & 2 & 3\\
    27 & SPY-5 & AABBC & A-A, A-B, A-B, B-B & 5 & 16\\
    28 & T-4 & AABC & & 1 & 1\\
    \bottomrule
  \end{tabular}
  \end{center}
\end{table}

\begin{table}
  \caption{Interpretation of iron compounds \#29--44 from
  Ref.~\cite{Proppe2017}. For explanations of column headers, refer to
  Table~\ref{table:proppe_interpretation_1}.}\label{table:proppe_interpretation_2}
  \begin{center}
  \begin{tabular}{r l l l r r}
    \toprule
    \# & Polyhedron & Ranking & Linking & $F$ & $A$\\
    \midrule
    29 & OC-6 & AAAABB & A-A, A-A & 3 & 5\\
    30 & OC-6 & AABBCD & A-B, A-B, A-B, A-B & 3 & 8\\
    31 & SPY-5 & AAAAB & A-A, A-A & 3 & 5\\
    32 & T-4 & AAAB & & 1 & 1\\
    33 & OC-6 & AABCDE & B-C & 12 & 15\\
    34 & OC-6 & AABCDE & B-D & 12 & 15\\
    35 & OC-6 & AABCDE & B-C & 12 & 15\\
    36 & OC-6 & AABCDE & C-D, D-E & 9 & 15\\
    37 & OC-6 & AABBCD & A-B, A-C, A-B, A-C & 7 & 16\\
    38 & OC-6 & AABCDE & C-D & 12 & 15\\
    39 & OC-6 & AABCDE & D-E & 12 & 15\\
    40 & OC-6 & AABCDE & C-E & 12 & 15\\
    41 & OC-6 & AABCDE & D-E & 12 & 15\\
    42 & SPY-5 & AABBC & A-A, A-B, A-B, B-B & 5 & 16\\
    43 & TBPY-5 & AAAAA & & 1 & 1\\
    44 & SPY-5 & AABBC & A-A, A-B, A-B, B-B & 5 & 16\\
    \bottomrule
  \end{tabular}
  \end{center}
\end{table}

All compounds with their interpreted shapes at iron atoms are shown in Figures
18-25.

\begin{figure}
  \centering
  \includegraphics[width=\linewidth]{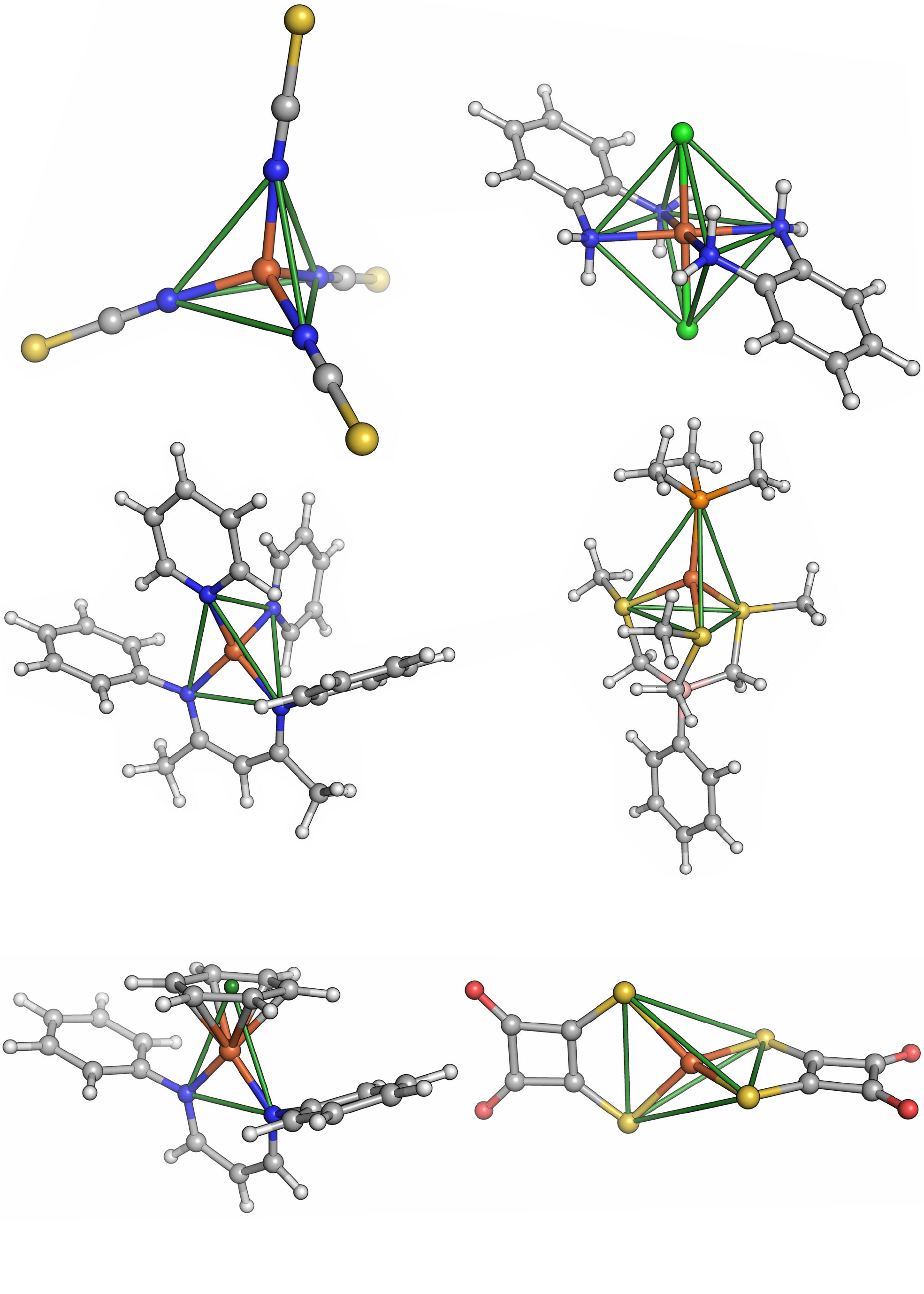}
  \caption{Iron compounds \#1--6 arranged left-to-right, then row-wise, with
  interpreted polyhedral shapes at iron atoms highlighted in green.
  }\label{fig:proppe_compounds_1}
\end{figure}
\begin{figure}
  \centering
  \includegraphics[width=\linewidth]{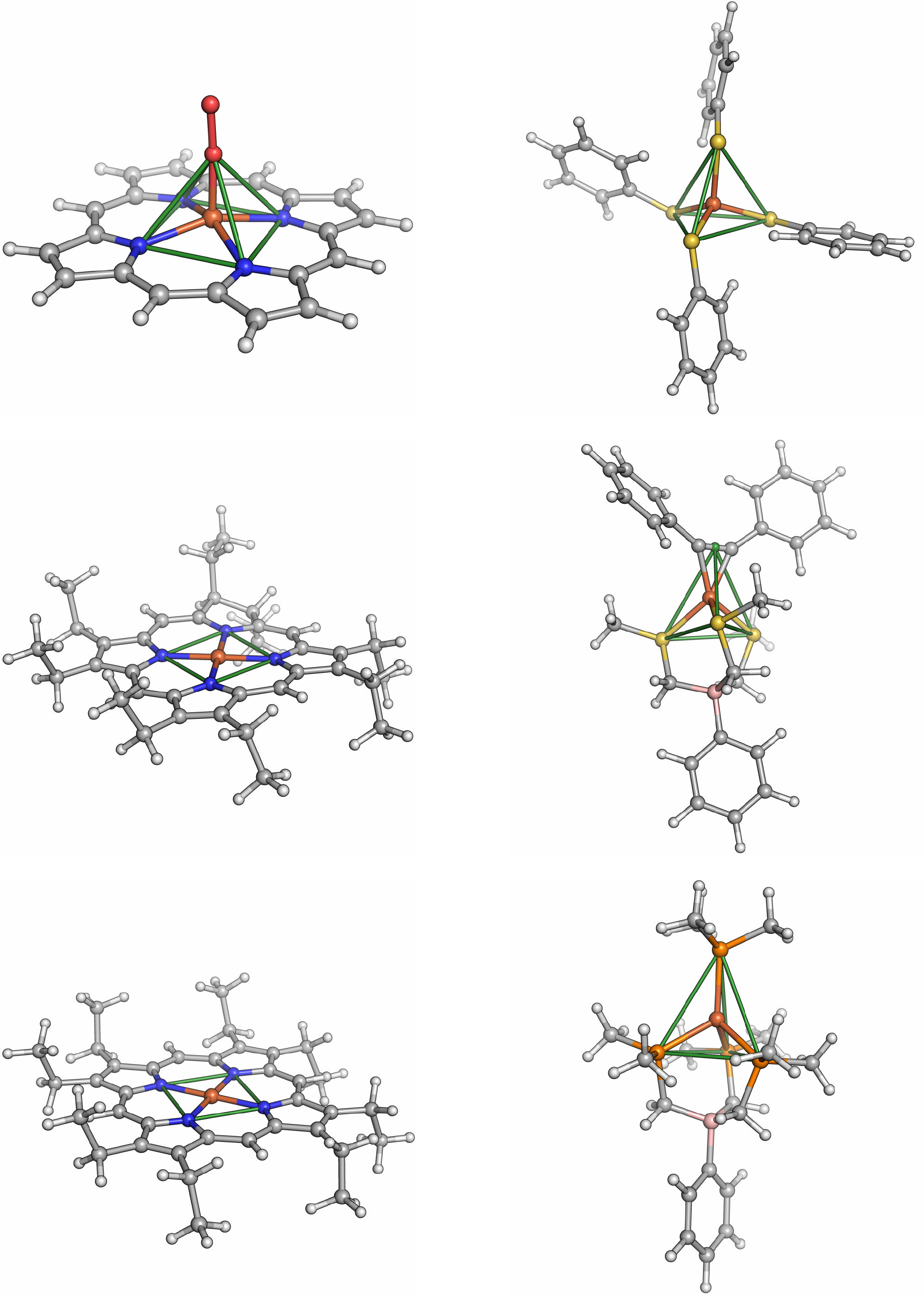}
  \caption{Iron compounds \#7--12 arranged left-to-right, then row-wise, with
  interpreted polyhedral shapes at iron atoms highlighted in green.
  }\label{fig:proppe_compounds_2}
\end{figure}
\begin{figure}
  \centering
  \includegraphics[width=\linewidth]{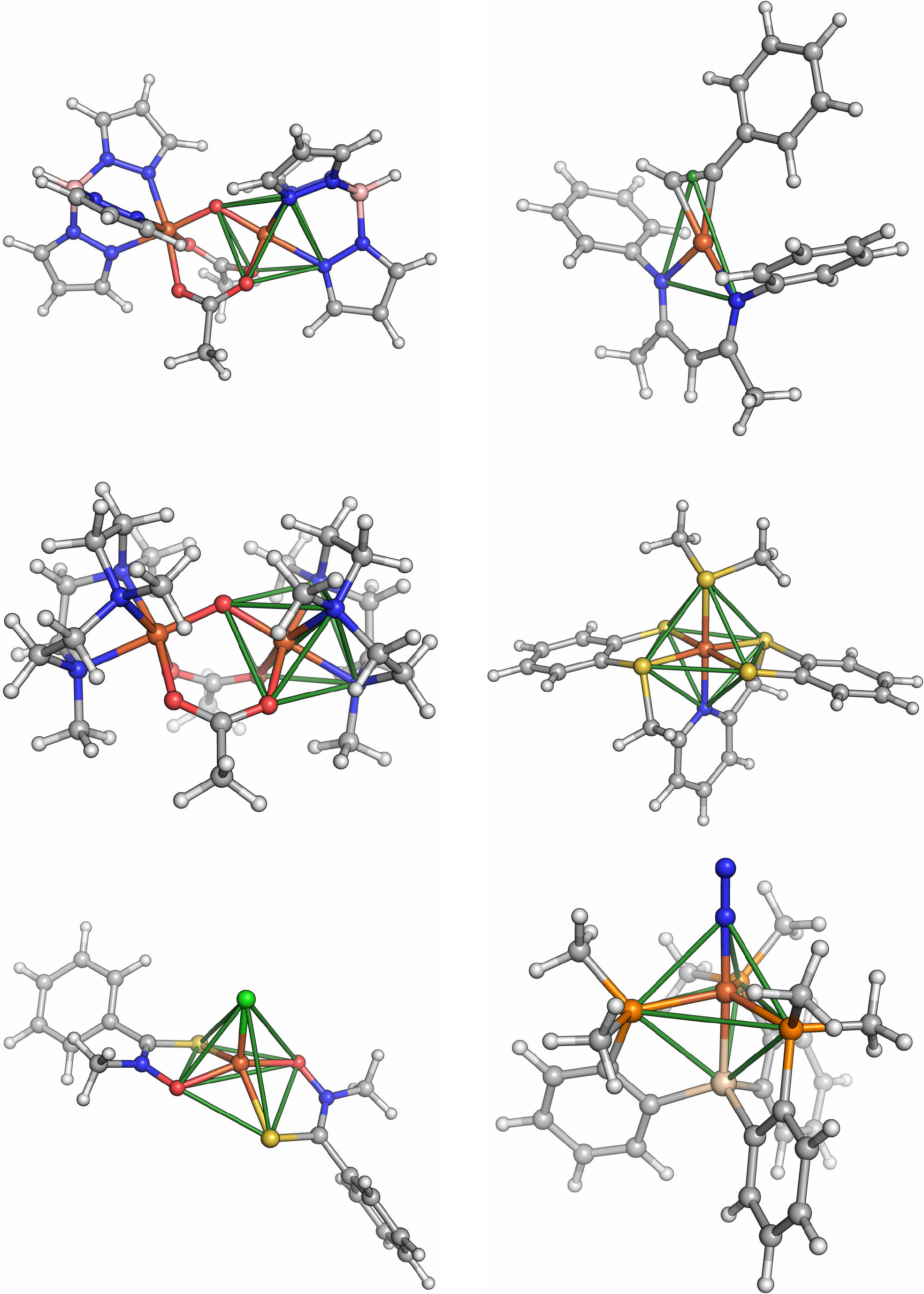}
  \caption{Iron compounds \#13--18 arranged left-to-right, then row-wise, with
  interpreted polyhedral shapes at iron atoms highlighted in green.
  }\label{fig:proppe_compounds_3}
\end{figure}
\begin{figure}
  \centering
  \includegraphics[width=\linewidth]{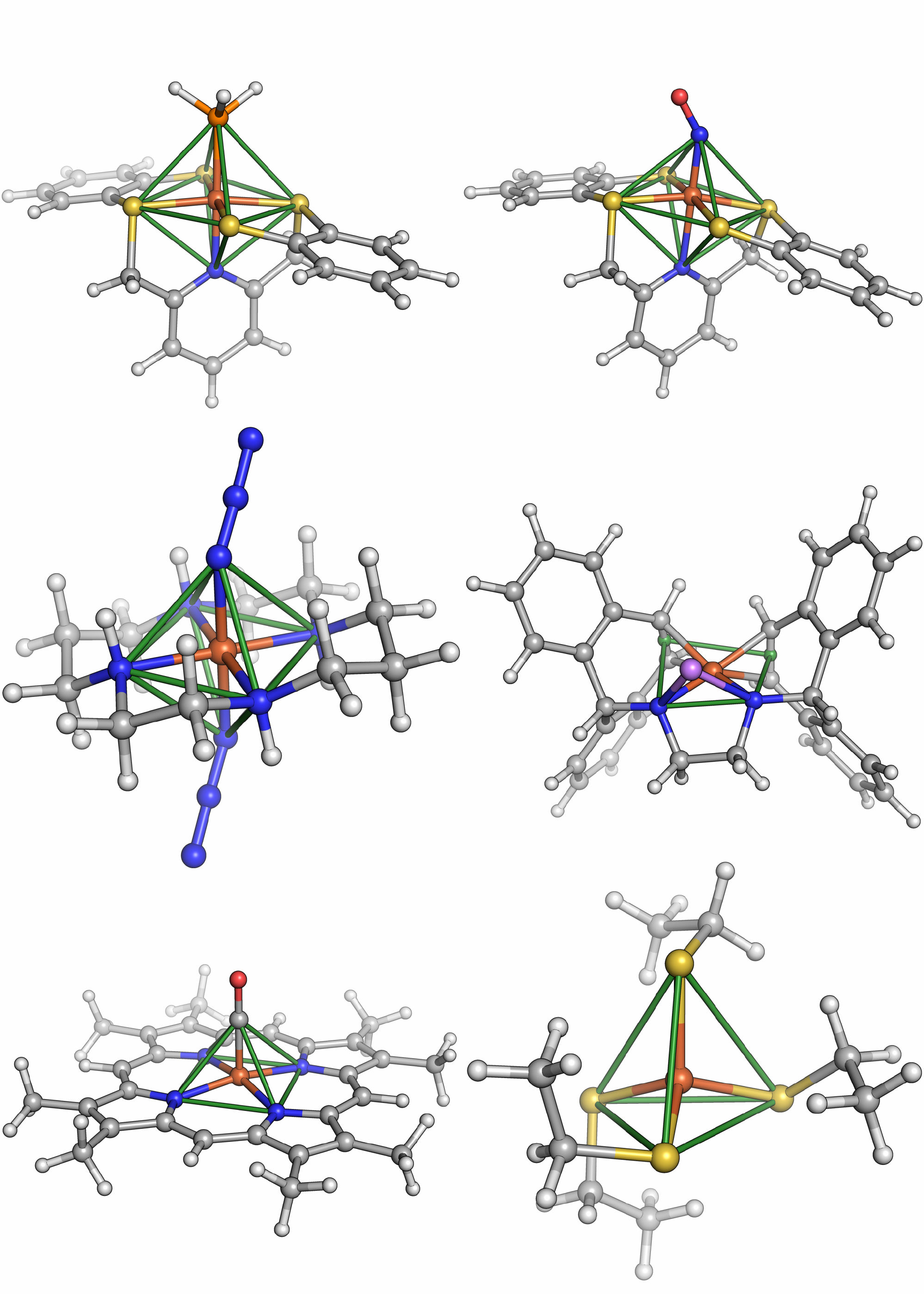}
  \caption{Iron compounds \#19--24 arranged left-to-right, then row-wise, with
  interpreted polyhedral shapes at iron atoms highlighted in green.
  }\label{fig:proppe_compounds_4}
\end{figure}
\begin{figure}
  \centering
  \includegraphics[width=\linewidth]{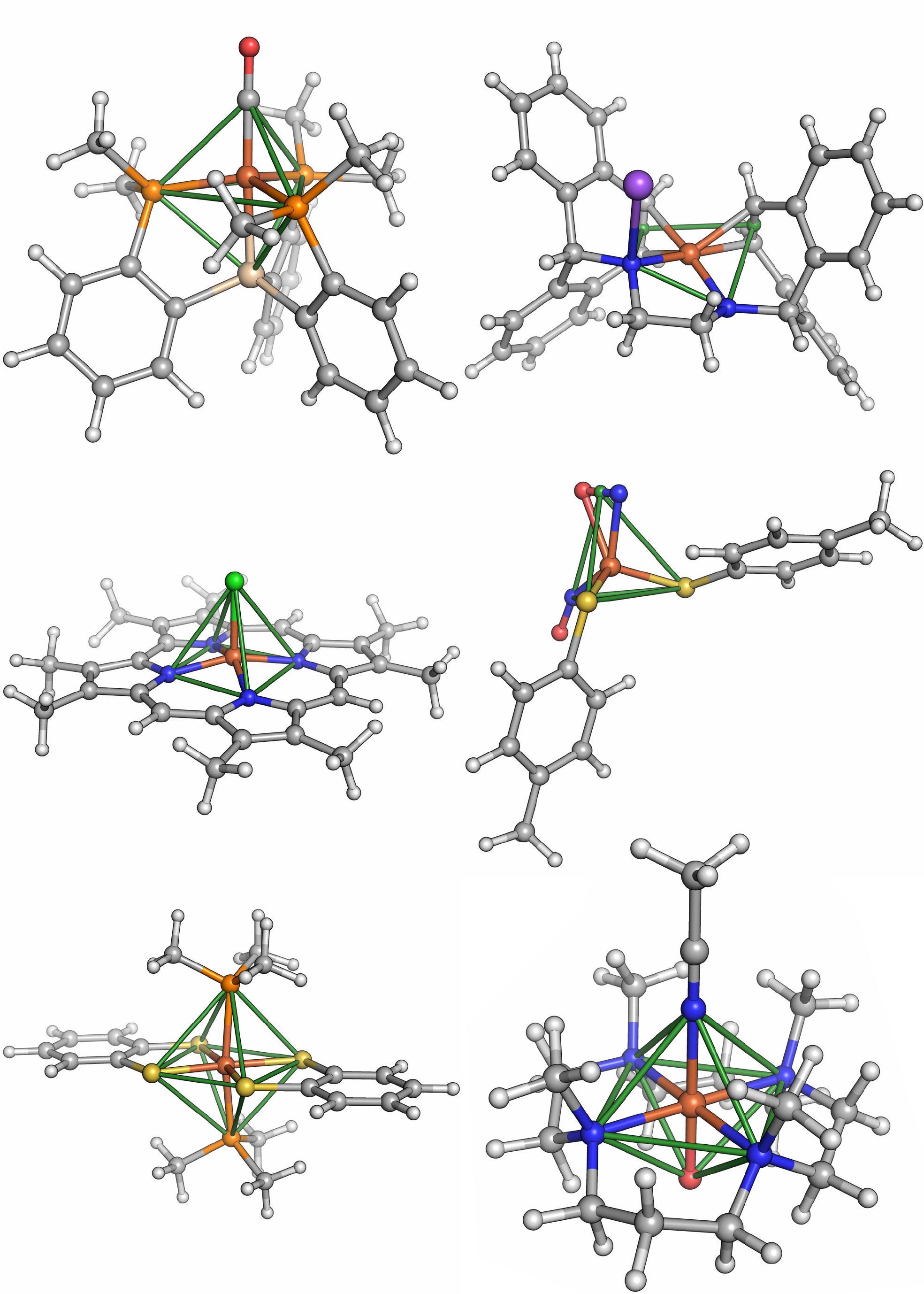}
  \caption{Iron compounds \#25--30 arranged left-to-right, then row-wise, with
  interpreted polyhedral shapes at iron atoms highlighted in green.
  }\label{fig:proppe_compounds_5}
\end{figure}
\begin{figure}
  \centering
  \includegraphics[width=\linewidth]{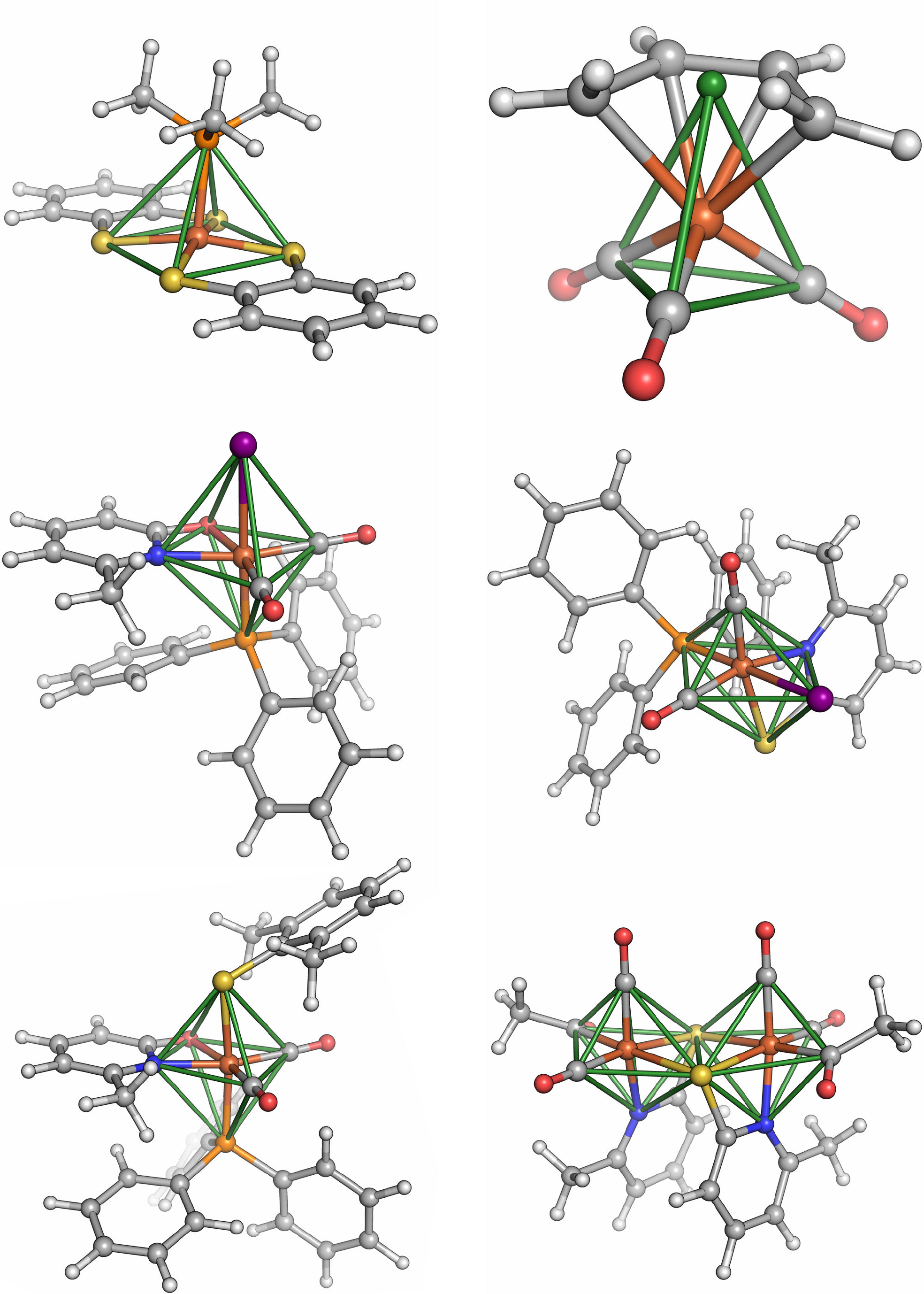}
  \caption{Iron compounds \#31--36 arranged left-to-right, then row-wise, with
  interpreted polyhedral shapes at iron atoms highlighted in green.
  }\label{fig:proppe_compounds_6}
\end{figure}
\begin{figure}
  \centering
  \includegraphics[width=\linewidth]{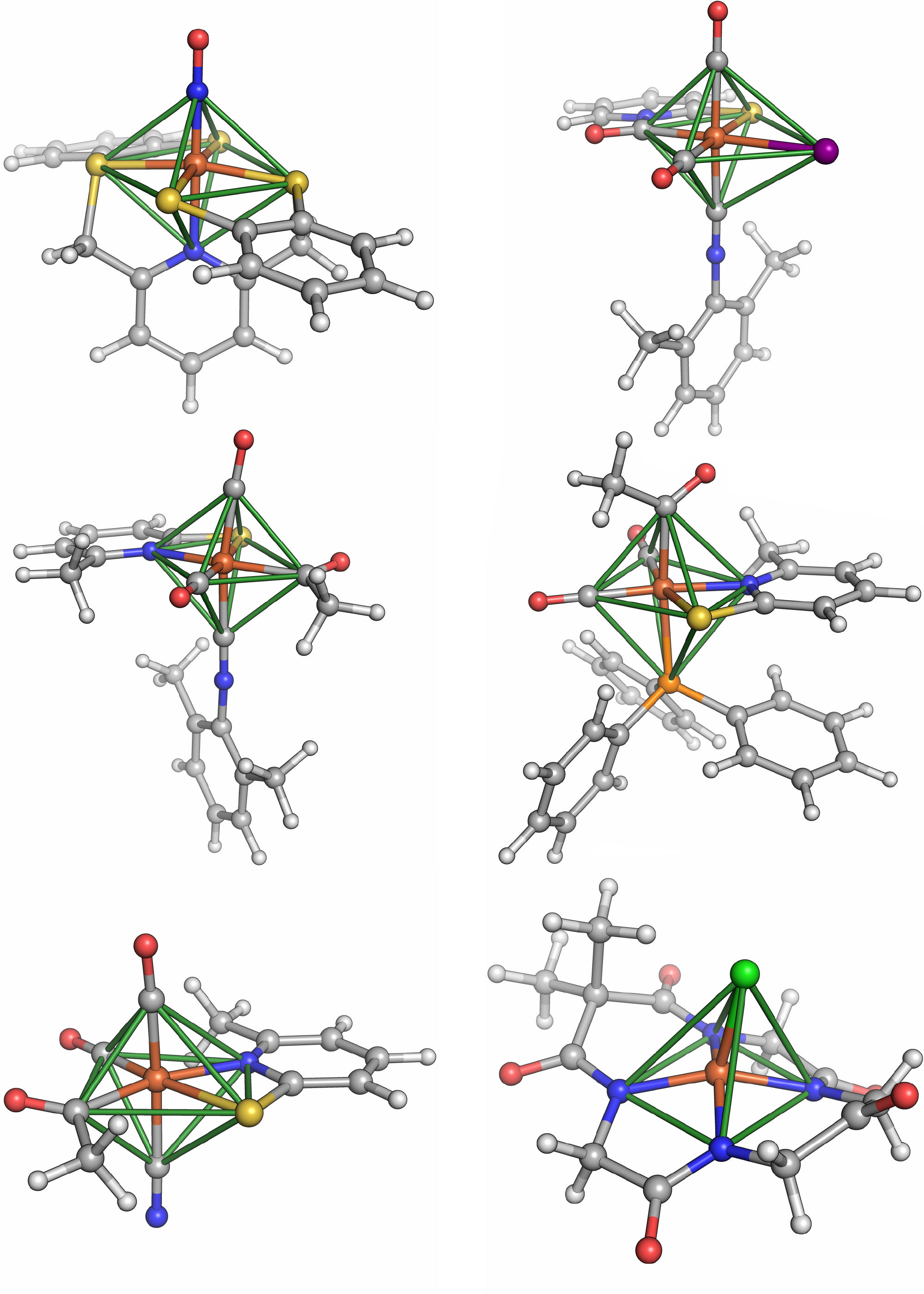}
  \caption{Iron compounds \#37--42 arranged left-to-right, then row-wise, with
  interpreted polyhedral shapes at iron atoms highlighted in green.
  }\label{fig:proppe_compounds_7}
\end{figure}
\begin{figure}
  \centering
  \includegraphics[width=\linewidth]{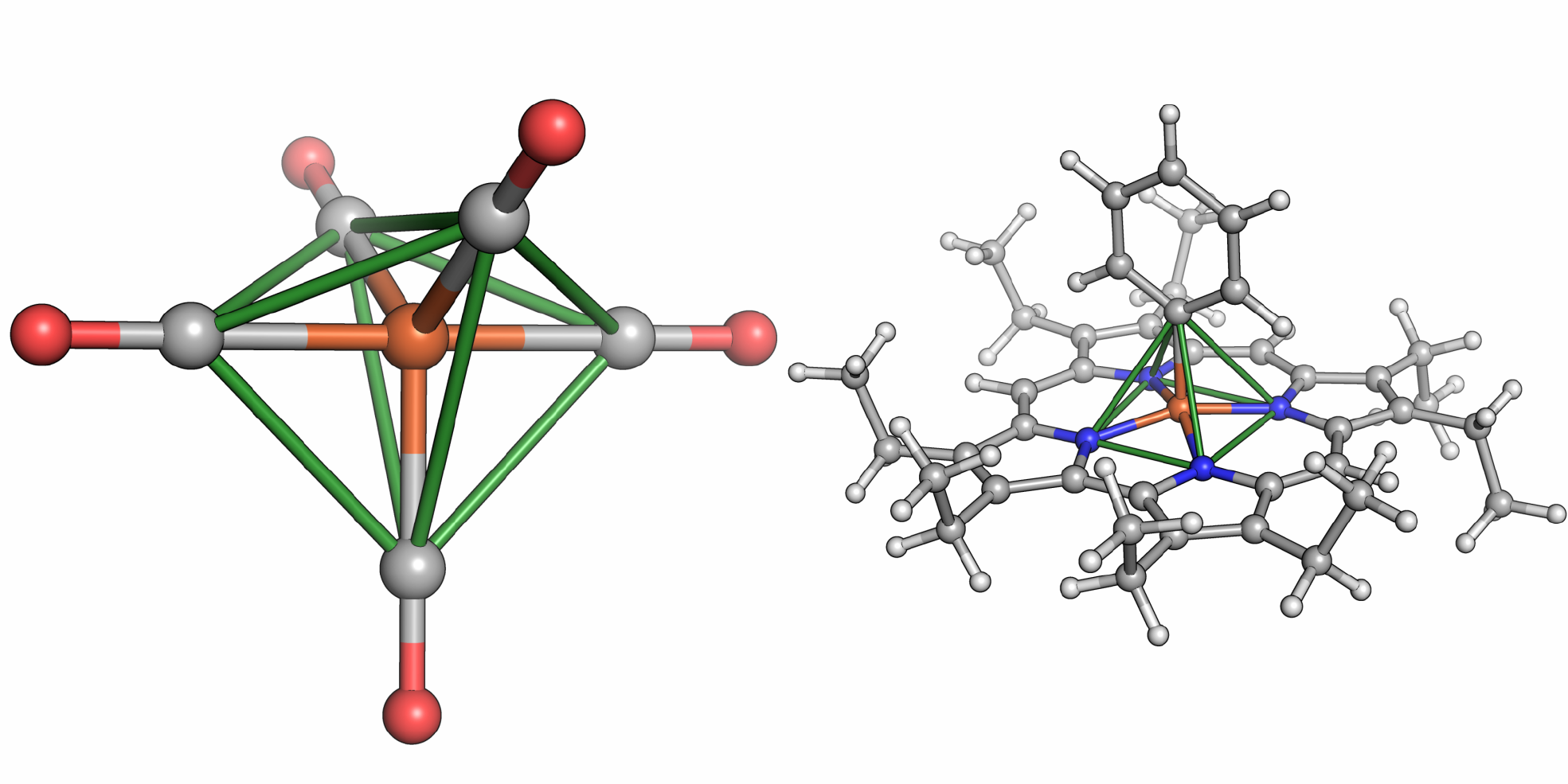}
  \caption{Iron compounds \#41 and 42, with
  interpreted polyhedral shapes at iron atoms highlighted in green.
  }\label{fig:proppe_compounds_8}
\end{figure}

\clearpage

Lastly, in order to demonstrate a part of the program interface for the
generation of conformers, we generate a conformer of each feasible
stereopermutation of a structure:

\begin{lstlisting}[caption=Dataset molecule conformer generation]
import sys
import scine_molassembler as masm

filename = sys.argv[1]
assert filename.endswith(".mol")  # We want full topologies here
mol = masm.io.read(filename)

# The first atom of each structure in the dataset is an iron atom:
iron_atom = 0
stereopermutator = mol.stereopermutators[iron_atom]
assert stereopermutator is not None

conformer_seed = 6604

for i in range(stereopermutator.num_assignments):
    print("Trying to generate conformer for assignment {}".format(i))
    # Change the stereopermutator
    mol.assign_stereopermutator(iron_atom, i)

    # Distance Geometry can fail, so we allow several attempts
    xyz = "Failure reasons are communicated by string"
    attempts = 0
    while isinstance(xyz, str):
        if attempts == 10:
            break

        print("Attempt {}".format(attempts))
        # Reuse attempts for seeding to get different reproducible conformers
        xyz = masm.dg.generate_conformation(mol, conformer_seed + attempts)
        attempts = attempts + 1

    if isinstance(xyz, str):
        print("Could not generate a conformer for assignment {} : {}".format(i, xyz))
        continue

    conf_filename = filename.replace(".mol", "-{}.mol".format(i))
    masm.io.write(conf_filename, mol, xyz)
\end{lstlisting}

\clearpage

\begin{figure}
  \centering
  \includegraphics[width=\linewidth]{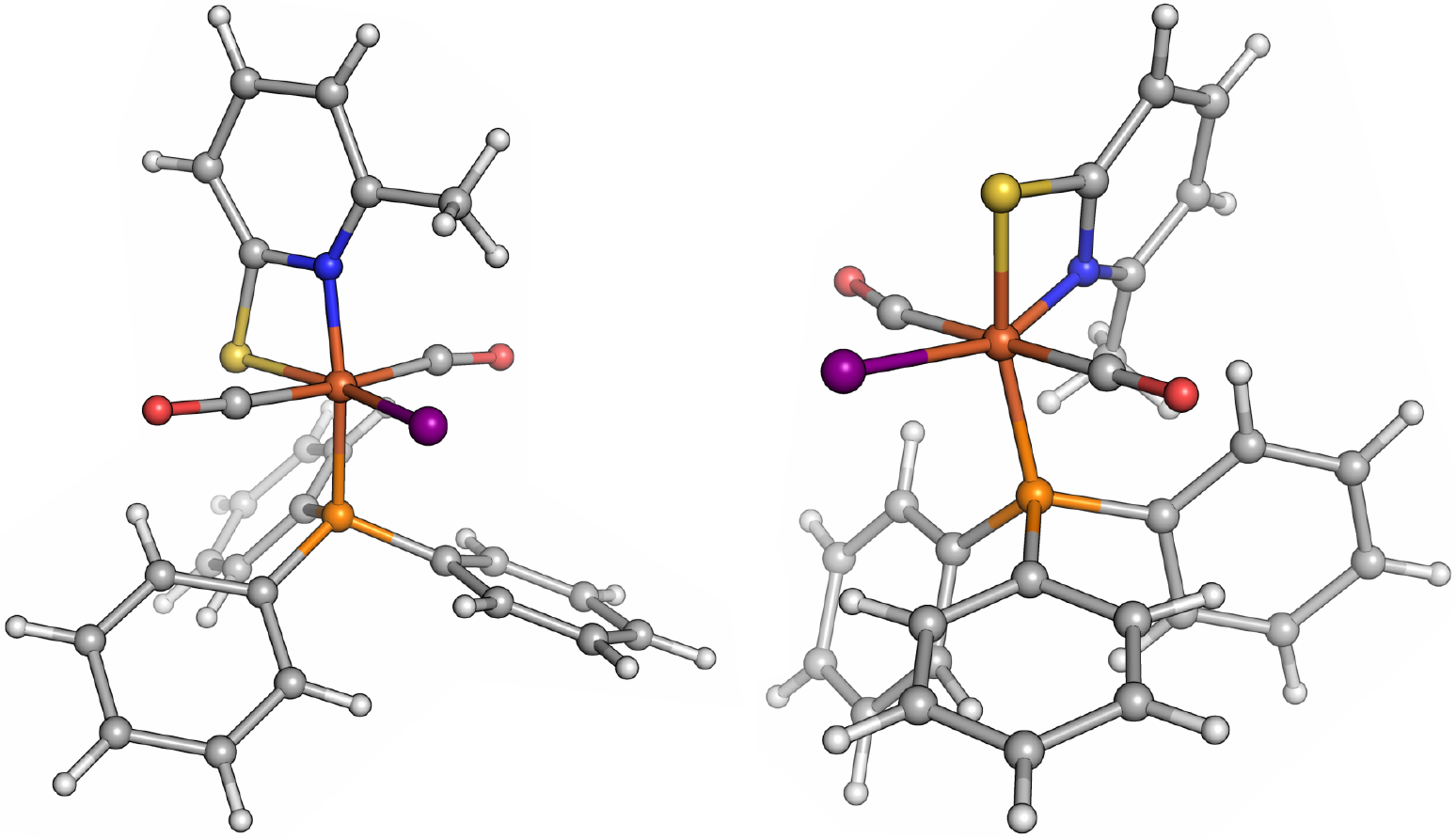}
  \caption{Conformations of stereoisomers of compound \#34 of the feasible
  stereopermutations indexed as 10 and 11 of the iron atom-centered
  stereopermutator as generated by \textsc{Molassembler}.
  }\label{fig:proppe_34_stereoisomers}
\end{figure}

Conformations of two stereoisomers of compound \#34 are displayed in
Figure~\ref{fig:proppe_34_stereoisomers}.

\end{appendix}

\newpage
\providecommand{\refin}[1]{\\ \textbf{Referenced in:} #1}

\end{document}